%% file: main.tex
\documentclass[twocolumn,floatfix]{aastex631}
\usepackage{paralist}
\usepackage{enumitem}
\usepackage[caption=false]{subfig}
\usepackage{mathpazo}
\usepackage{amsmath}
\usepackage{overpic}
\usepackage{upgreek}
\usepackage{amssymb}
\usepackage{graphicx}
\usepackage{amssymb}
\usepackage{amsmath}
\usepackage{nicefrac}
\usepackage{xspace}
\usepackage{bm}
\usepackage[hang,flushmargin]{footmisc}
\usepackage{multirow}
\usepackage{url}
\usepackage{aas_macros}
\usepackage{rviewport}
\usepackage{mathtools}
\usepackage{rotating}
\usepackage{natbib}
\usepackage[capitalise]{cleveref}
\usepackage{booktabs}
\usepackage{array,etoolbox}
\usepackage{xargs}
\usepackage{soul}
\usepackage{tikz}
\usepackage{balance}
\usetikzlibrary{shapes}
\usepackage{mathptmx}

\preto\tabular{\setcounter{magicrownumbers}{0}}
\newcounter{magicrownumbers}

\defcitealias{2017A&A...600A..29T}{TF17}
\defcitealias{JF12coh}{JF12}
\defcitealias{dataAndCode}{this link}

\newif\ifall
\alltrue

\input{abbreviations.tex}

\usepackage{hyperref}
\hypersetup{
  bookmarksopen=true, bookmarksopenlevel=1,
  unicode=false, pdftoolbar=true, pdfmenubar=true,
  colorlinks=true, linkcolor=darkBlue, citecolor=darkBlue,
  filecolor=darkBlue, urlcolor=darkBlue
}

\shorttitle{The Coherent Magnetic Field of the Milky Way}
\shortauthors{M.~Unger and G.R.~Farrar}

\begin{document}
\title{The Coherent Magnetic Field of the Milky Way}
\author[0000-0002-7651-0272]{Michael Unger}
\email{michael.unger@kit.edu}
\affiliation{Institute for Astroparticle Physics (IAP),
  Karlsruhe Institute of Technology (KIT), Karlsruhe, Germany}
\affiliation{Institutt for fysikk, Norwegian University of Science and Technology (NTNU), Trondheim, Norway}
\author[0000-0003-2417-5975]{Glennys R. Farrar}
\email{gf25@nyu.edu}
\affiliation{Center for Cosmology and Particle Physics, Department of Physics,
             New York University, New York, NY 10003, USA}

\begin{abstract}
We present a suite of models of the coherent magnetic field of the
 Galaxy (GMF) based on new divergence-free parametric functions describing the global structure of the field.  The model parameters are fit to the latest full-sky Faraday rotation measures of extragalactic sources (\RMs) and polarized synchrotron intensity  (\PI) maps from WMAP and Planck. We employ multiple models for the density of thermal and cosmic-ray electrons in the Galaxy, needed to predict the skymaps of \RMs and \PI for a given GMF model.  The robustness of the inferred properties of the GMF is gauged by studying many combinations of parametric field models and electron density models.
 We determine the pitch angle of the local magnetic field ($(11\pm1)^\circ$), explore the evidence for a grand-design spiral coherent magnetic field (inconclusive), determine the strength of the toroidal and poloidal magnetic halo fields below and above the disk (magnitudes the same for both hemispheres within $\approx$10\%), set constraints on the half-height of the cosmic-ray diffusion volume ($\geq2.9$~kpc), investigate the compatibility of \RM- and \PI-derived magnetic field strengths (compatible under certain assumptions) and check if the toroidal halo field could be created by the shear of the poloidal halo field due to the differential rotation of the Galaxy (possibly).
 A set of eight models is identified to help quantify the present uncertainties in the coherent GMF spanning different functional forms, data products and auxiliary input. We present the corresponding skymaps of rates for axion-photon conversion in the Galaxy, and deflections of ultra-high energy cosmic rays.
\end{abstract}

\section{Introduction}
\ifall
\input{introduction.tex}
\fi
\section{Observables}
\label{sec:observables}
\ifall
\input{observables.tex}
\fi
\section{Data Products}
\label{sec:data}
\ifall
\input{rm.tex}
\input{synchrotron.tex}
\fi
\section{Auxiliary Models}
\label{sec:auxmodels}
\ifall
\input{thermalElectrons.tex}
\input{cosmicRayElectrons.tex}
\fi
\section{Magnetic Field Models}
\label{sec:magModels}
\ifall
\input{magModels.tex}

\input{glossary.tex}
\fi

\section{Model Optimization}
\label{sec:modelopt}
\ifall
\input{analysis.tex}
\fi

\section{Results}
\label{sec:results}
\ifall
\input{results.tex}
\fi
\section{Applications}
\label{sec:applications}
\ifall
\input{applications.tex}
\fi
\section{Summary}
\label{sec:summary}
\ifall
\input{summary.tex}
\fi
\balance
\newpage
\input{ackn.tex}
\bibliography{mf.bib}
\ifall
\appendix
\input{appendix.tex}
\fi
\end{document}

%% file: abbreviations.tex
\DeclareMathAlphabet{\mathcal}{OMS}{cmsy}{m}{n}
\newcommand{\R}{\ensuremath{{\mathcal R}\xspace}}

\DeclareMathAlphabet\mathbfcal{OMS}{cmsy}{b}{n}

\DeclareMathOperator{\sign}{sgn}

\DeclareMathOperator{\sech}{sech}
\DeclareMathOperator{\sigmoid}{\mbox{\(\sigma\)}}

\newcommand{\muG}{\ensuremath{\upmu}\textrm{G}\xspace}

\newcommand{\Q}{\ensuremath{Q}\xspace}
\newcommand{\U}{\ensuremath{U}\xspace}
\newcommand{\RM}{\ensuremath{\text{RM}}\xspace}
\newcommand{\RMs}{{\RM}s\xspace}
\newcommand{\radm}{\ensuremath{\text{rad}/\text{m}^2}\xspace}
\newcommand{\DM}{\ensuremath{\text{DM}}\xspace}

\newcommand{\EM}{\ensuremath{\text{EM}}\xspace}
\newcommand{\PI}{\ensuremath{\text{PI}}\xspace}
\newcommand{\dd}{{\text{d}}}
\newcommand{\NE}{{\scshape Ne2001}\xspace}
\newcommand{\YMW}{{\scshape Ymw16}\xspace}
\newcommand{\DRAGON}{{\scshape Dragon}\xspace}
\newcommand{\HEALPix}{{\scshape HEALPix}\xspace}
\newcommand{\GALPROP}{{\scshape Galprop}\xspace}
\newcommand{\ROOT}{{\scshape Root}\xspace}
\newcommand{\JFTwelve}{{\scshape JF12}\xspace}
\newcommand{\ncre}{\ensuremath{n_\text{cre}}\xspace}
\newcommand{\nel}{\ensuremath{n_\text{e}}\xspace}
\newcommand{\Planck}{{\scshape Planck}\xspace}
\newcommand{\WMAP}{{\scshape Wmap}\xspace}
\newcommand{\CosmoGlobe}{{\scshape CosmoGlobe}\xspace}

\newcommand{\JFTor}{\ensuremath{{\sigmoid}_{r}\,{\exp}_{z}}}
\newcommand{\creuf}{{\scshape DPD}}
\newcommand{\toro}{{explicit}}

\definecolor{rossoCP3}{cmyk}{0,.88,.77,.40}
\definecolor{darkBlue}{rgb}{0, 0, 0.8}
\definecolor{darkGreen}{rgb}{0, 0.5, 0}

\def\figRes{lowres}
\def\figType{png}
\newcommand{\fitPlot}[1]{
  \raggedright
  masked model\\
  {\centering
  \includegraphics[width=\mpfigw\linewidth]{#11_\figRes_c.\figType}\quad%
  \includegraphics[width=\mpfigw\linewidth]{#14_\figRes_c.\figType}\quad%
  \includegraphics[width=\mpfigw\linewidth]{#17_\figRes_c.\figType}\\}
  data\\
  {\centering
  \includegraphics[width=\mpfigw\linewidth]{#10_\figRes_c.\figType}\quad%
  \includegraphics[width=\mpfigw\linewidth]{#13_\figRes_c.\figType}\quad%
  \includegraphics[width=\mpfigw\linewidth]{#16_\figRes_c.\figType}\\}
  pull\\
  {\centering
  \includegraphics[width=\mpfigw\linewidth]{#12_\figRes_c.\figType}\quad%
  \includegraphics[width=\mpfigw\linewidth]{#15_\figRes_c.\figType}\quad%
  \includegraphics[width=\mpfigw\linewidth]{#18_\figRes_c.\figType}\\}
}
\newcommand{\modelPlot}[4]{
  \raggedright disk\\
  {\centering
  \includegraphics[width=\mpfigw\linewidth]{#1_#2_RM_\figRes.\figType}\quad%
  \includegraphics[width=\mpfigw\linewidth]{#1_#2_Q_\figRes.\figType}\quad%
  \includegraphics[width=\mpfigw\linewidth]{#1_#2_U_\figRes.\figType}\\}
  toroidal halo\\
  {\centering
  \includegraphics[width=\mpfigw\linewidth]{#1_#3_RM_\figRes.\figType}\quad%
  \includegraphics[width=\mpfigw\linewidth]{#1_#3_Q_\figRes.\figType}\quad%
  \includegraphics[width=\mpfigw\linewidth]{#1_#3_U_\figRes.\figType}\\}
  poloidal halo\\
  {\centering
  \includegraphics[width=\mpfigw\linewidth]{#1_#4_RM_\figRes.\figType}\quad%
  \includegraphics[width=\mpfigw\linewidth]{#1_#4_Q_\figRes.\figType}\quad%
  \includegraphics[width=\mpfigw\linewidth]{#1_#4_U_\figRes.\figType}\\}
  full model\\
  {\centering
  \includegraphics[width=\mpfigw\linewidth]{#1_RM_\figRes.\figType}\quad%
  \includegraphics[width=\mpfigw\linewidth]{#1_Q_\figRes.\figType}\quad%
  \includegraphics[width=\mpfigw\linewidth]{#1_U_\figRes.\figType}\\}
}

\newcommand{\fitPlotCaption}[1]{
    \caption{Data and {#1 model} prediction for \RM, \Q and \U.
    The predictions from the individual model components are shown
    in the top rows, followed by full model prediction. The bottom
    three rows show the masked model, the data and the pull, i.e.\ the
    difference between model and data in units of the standard deviation
    of the data.}
}

\makeatletter
\newcommand*{\whiten}[1]{\llap{\textcolor{white}{{\the\SOUL@token}}\hspace{#1pt}}}
\DeclareRobustCommand*\myul{%
    \def\SOUL@everyspace{\underline{\space}\kern\z@}%
    \def\SOUL@everytoken{%
     \setbox0=\hbox{\the\SOUL@token}%
     \ifdim\dp0>\z@
        \raisebox{\dp0}{\underline{\phantom{\the\SOUL@token}}}%
        \whiten{1}\whiten{0}%
        \whiten{-1}\whiten{-2}%
        \llap{\the\SOUL@token}%
     \else
        \underline{\the\SOUL@token}%
     \fi}%
\SOUL@}
\makeatother

\newcommand{\inlineeqnum}{\refstepcounter{equation}~~ \hspace*{\fill} \mbox{(\theequation)}}

\def\Offline{\mbox{$\overline{\textrm%
{Off}}$\hspace{.05em}\protect\raisebox{.4ex}%
{$\protect\underline{\textrm{line}}$}}\xspace}

\makeatletter
\newcommand\mathcircled[1]{%
  \mathpalette\@mathcircled{#1}%
}
\newcommand\@mathcircled[2]{%
  \tikz[baseline=(math.base)] \node[draw,ellipse,inner sep=1pt] (math) {$\m@th#1#2$};%
}
\makeatother

\newcommand{\fszfunc}{\ensuremath{h_\text{d}}\xspace}
\newcommand{\fsrfunc}{\ensuremath{g_\text{d}}\xspace}
\newcommand{\fsz}{\ensuremath{z_\text{d}}\xspace}
\newcommand{\fswz}{\ensuremath{w_\text{d}}\xspace}
\newcommand{\fsri}{\ensuremath{r_\text{1}}\xspace}
\newcommand{\fswri}{\ensuremath{w_\text{1}}\xspace}
\newcommand{\fsro}{\ensuremath{r_\text{2}}\xspace}
\newcommand{\fswro}{\ensuremath{w_\text{2}}\xspace}
\newcommand{\Bdm}{\ensuremath{B_m}\xspace}
\newcommand{\Phidm}{\ensuremath{\phi_m}\xspace}
\newcommand{\rRefd}{\ensuremath{r_0}\xspace}
\newcommand{\pitch}{\ensuremath{\alpha}\xspace}

\newcommand{\BLS}{\ensuremath{B_\text{1}}\xspace}
\newcommand{\phiS}{\ensuremath{\phi_\text{1}}\xspace}
\newcommand{\wS}{\ensuremath{w_\text{S}}\xspace}
\newcommand{\phiCS}{\ensuremath{\phi_\text{c}}\xspace}
\newcommand{\wCS}{\ensuremath{w_\text{c}}\xspace}
\newcommand{\lCS}{\ensuremath{L_\text{c}}\xspace}

\newcommand{\BNT}{\ensuremath{B_\text{N}}\xspace}
\newcommand{\BST}{\ensuremath{B_\text{S}}\xspace}
\newcommand{\zT}{\ensuremath{z_\text{t}}\xspace}
\newcommand{\rT}{\ensuremath{r_\text{t}}\xspace}
\newcommand{\wT}{\ensuremath{w_\text{t}}\xspace}

\newcommand{\BP}{\ensuremath{B_\text{p}}\xspace}
\newcommand{\zP}{\ensuremath{z_\text{p}}\xspace}
\newcommand{\rP}{\ensuremath{r_\text{p}}\xspace}
\newcommand{\wP}{\ensuremath{w_\text{p}}\xspace}
\newcommand{\aP}{\ensuremath{a_\text{c}}\xspace}

\newcommand{\vZero}{\ensuremath{v_0}\xspace}
\newcommand{\rV}{\ensuremath{r_v}\xspace}
\newcommand{\zV}{\ensuremath{z_v}\xspace}

\newcommand{\rSun}{\ensuremath{r_\odot}\xspace}
\newcommand{\hDiff}{\ensuremath{h_{D}}\xspace}

\newcommand{\rotRadialFunc}{\ensuremath{f_r}\xspace}
\newcommand{\rotVerticalFunc}{\ensuremath{g_z}\xspace}

\newcommand{\modelNe}{\textnormal{\texttt{neCL}}\xspace}
\newcommand{\modelXr}{\textnormal{\texttt{expX}}\xspace}
\newcommand{\modelBase}{\textnormal{\texttt{base}}\xspace}
\newcommand{\modelSpur}{\textnormal{\texttt{spur}}\xspace}
\newcommand{\modelCre}{\textnormal{\texttt{cre10}}\xspace}
\newcommand{\modelSyn}{\textnormal{\texttt{synCG}}\xspace}
\newcommand{\modelTwist}{\textnormal{\texttt{twistX}}\xspace}
\newcommand{\modelKappa}{\textnormal{\texttt{nebCor}}\xspace}

%% file: introduction.tex
Spiral galaxies are known to be permeated by large-scale magnetic
fields, with energy densities comparable to the turbulent and thermal
energy densities of the interstellar medium; see
e.g.\ \citet{2016A&ARv..24....4B} for a recent review. A good
knowledge of the global structure of these fields is important for
understanding their origin, inferring their effect on galactic
dynamics, estimating the properties of the diffuse motion of low-energy
Galactic cosmic rays, and studying the impact of magnetic deflections
on the arrival directions of extragalactic ultrahigh-energy cosmic
rays. The GMF is also important for new physics studies, for instance
axion-photon conversion in the GMF or the interpretation of possible
signatures of astrophysical dark matter annihilation.

The determination of the large-scale structure of the magnetic field
of our Galaxy is particularly challenging since one must infer it from
the vantage point of Earth, located inside the field. Previous
attempts to model the Galactic magnetic field (GMF) are summarized
by~\citet{Jaffe:2019iuk}. In this paper, we focus on the coherent
magnetic field of the Galaxy, leaving the study of its turbulent
component to the near future. Following \citet{JF12coh}
(hereafter \citetalias{JF12coh}), we derive the GMF by fitting
suitably general parametric models of its structure to the two
astrophysical data sets which are the most constraining of the
coherent magnetic fields: the {\itshape rotation measures} ({\RM}s) of
extragalactic polarized radio sources and the {\itshape polarized
intensity} (\PI) of the synchrotron emission of cosmic-ray electrons
in the Galaxy.

The relation of these two astrophysical observables to the magnetic
field is detailed in
Sec.~\ref{sec:observables}, followed by a description of the \RM and
\PI data in Sec.~\ref{sec:data}. The interpretation of these
data relies on the knowledge of the three-dimensional density of
thermal electrons and cosmic-ray electrons in the Galaxy. We discuss
these {\itshape auxiliary models} in Sec.~\ref{sec:auxmodels}. The
parametric models of the GMF investigated in this paper are introduced
in Sec.~\ref{sec:magModels} and the model optimization is described in
Sec.~\ref{sec:modelopt}.

The combination of different data sets, auxiliary models and
parametric functions yields an ensemble of GMF models that reflect
the uncertainties and degeneracies inherent in the inference of the
global field structure from the limited information provided by the
\RM and \PI data. In Sec.~\ref{sec:results} we  narrow down these
model variants to a few benchmark models that encompass the largest
differences within the ensemble.

A typical application of the global magnetic field models is the
inference of the arrival direction of cosmic rays at the edge of the
Galaxy and the determination of the conversion probability of axions
in the magnetic field of the Galaxy. In Sec.~\ref{sec:applications} we
briefly comment on the implcations of our study on these topics.

We conclude this paper by giving a brief summary of our findings in
Sec.~\ref{sec:summary} and addressing the question: What is known and not
known about the coherent magnetic field of the Galaxy?

%% file: observables.tex
In this work, we derive the global structure of the GMF from two
astrophysical observations, Faraday rotation measures and polarized
synchrotron intensity. These two quantities provide complementary
information on the parallel and perpendicular components of the
coherent magnetic field along the line of sight from the observer at
Earth through the Galaxy.\\

The {\itshape rotation measure} (\RM) relates the Faraday-rotated
polarization angle $\chi$ of an astrophysical source observed at a
wavelength $\lambda$ to its intrinsic polarization angle $\chi_0$
after passing through the magnetized plasma of the Galaxy via
     \begin{equation}
       \chi(\lambda) = \chi_0 + \RM \, \lambda^2.
       \label{eq:chiRot}
     \end{equation}
We follow the IAU convention based
on \citet{1972ApJ...172...43M} \citep[see also][]{2021MNRAS.507.4968F}
in which the RM is positive for a photon traveling in the same
direction as the magnetic field points, through an ambient medium with
free electrons.  In our notation, taking the observer to be at the
origin and the positive unit vector $\bm{u}_r$ pointing away from the
observer, RM is given by
\begin{equation*}
      \RM = - C_\RM \int_0^\infty \nel(\bm{x}(r))\, \bm{B}(\bm{x}(r)) \, \bm{u}_r \, \text{d}r,
       \label{eq:RM}
\end{equation*}
where the position $\bm{x}$ at distance $r$ from an observer located
at $\bm{x}_0$ is $\bm{x}(r) = \bm{x}_0
+ \bm{u}_r\,r$,\footnote{Throughout this work, we assume an observer
at $\bm{x}_0 = \bm{x}_\odot = (-\rSun,\, 0, \, z_\odot)$ in a
right-handed Galactic coordinate system.  The vertical distance of the
Sun from the Galactic plane $z_\odot$ is somewhere between $0.006\pm
0.001$~kpc~\citep{2016A&A...593A.116J} and $0.026 \pm
0.003$~kpc \citep{2009MNRAS.398..263M} and can thus be neglected for
the purpose of this work, i.e., we set $z_\odot \rightarrow 0$.  For
the distance of the Sun to the Galactic center we adopt the precise
measurement of the {\scshape Gravity} collaboration, $\rSun =
8.178 \pm 0.025$~kpc~\citep{2019A&A...625L..10G}.} $\nel$ denotes the
density of the thermal electron plasma in the interstellar medium, and
$\smash[b]{C_\RM = e^3/ (8\,\varepsilon_0\,\pi^2\,m_e^2\,c^3)\approx
0.8119\, (\nicefrac{\text{rad}}{\text{m}^2}) \,
(\nicefrac{\text{cm}^3}{\text{pc}\,\upmu\text{G}})}$
\citep[e.g.][]{bradtAstroProc}. Thus, \RM is negative for a magnetic field
oriented away from the observer.

The interpretation of \RM becomes more complicated if one considers
the possibility of a small-scale correlation or anti-correlation of
magnetic fields and thermal electrons. An anti-correlation could arise
if the magnetic field is in pressure equilibrium with the thermal
electron plasma. In this case, the local Faraday rotation would be
systematically diminished and interpreting the \RM assuming no
anti-correlation would underestimate the magnitude of the integrated
line-of-sight magnetic field.  A positive correlation could be caused
by compression enhancing both the magnetic field and electron
density. These effects were studied by~\citet{2003A&A...411...99B}
(see also~\citep{2021PhRvF...6j3701S}) who provide the following
approximate relation between the rotation measure for the uncorrelated
case, $\RM_{0}$, and the general case with a correlation coefficient
$\kappa$,
\begin{equation}
\RM = \RM_{0}
\left(1+\frac{2}{3}\,\kappa\,\frac{\langle b^2\rangle}{B^2+\langle b^2\rangle}\right),
\label{eq:rmkappa}
\end{equation}
where $B$ denotes the coherent field and $\langle b^2\rangle$ is the
mean squared field strength of the turbulent magnetic field.  In one
of our model variants we will allow for a non-zero $\kappa$.\\

The {\itshape polarized synchrotron intensity (PI)} originates from
cosmic-ray electrons and positrons spiraling in the coherent magnetic
field of the Galaxy. The observed PI depends on the
CR-electron-density-weighted incoherent superposition of synchrotron
emission along the line of sight.  For long-wavelengths, the effect of
Faraday rotation along the path between emission and observation needs
to be taken into account.  At the magnetic field strengths and
frequencies relevant for this analysis (few $\upmu$G and 30~GHz), the
typical cosmic-ray energy responsible for synchrotron emission is of
the order of tens of GeV \citep[e.g.][]{2011hea..book.....L} and the
Faraday depolarization is not significant.

\begin{figure*}[!t]
\centering
\includegraphics[width=0.95\linewidth]{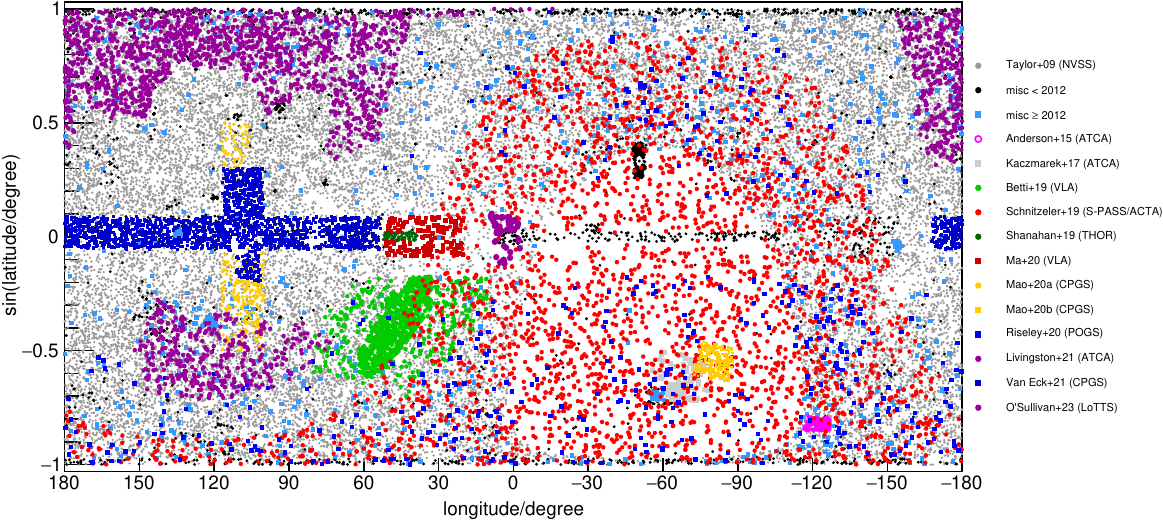}
\caption{Overview of the \RM data used in this analysis. Measurements are shown in Galactic coordinates using a cylindrical projection that visually preserves the point density per solid angle. \RMs published before 2012 (and thus available for the \citetalias{JF12coh} analysis) are shown as gray and black circles.}
\label{fig:rmData}
\end{figure*}

The relationship between the synchrotron volume emissivity and the
magnetic field strength is particularly simple for the case of
electrons with an energy distribution following a power law, $\ncre(E)
= n_0 \, E^{-p}$:
\begin{equation}
  j_\nu \varpropto  n_0\, \nu^\frac{-(p-1)}{2} \, B_\perp^\frac{p+1}{2} \overset{p = 3}{=} n_0 \, \nu^{-1} \, B_\perp^2,
\label{eq:J_coherent}
\end{equation}
where in the last step the approximation of $p\approx3$ was used,
which is applicable for a canonical $E^{-2}$ spectrum at the source
softened to $E^{-3}$ due to energy losses from synchrotron cooling and
inverse Compton scattering.

However, the cosmic-ray electron spectrum is not expected to be an
exact power law \cite[e.g.][]{2013MNRAS.436.2127O}. To obtain the
local volume emissivity at the position $\bm{x}$ we therefore
integrate the single-electron emissivity over energy,
\begin{equation}
  j_\nu(\bm{x}) = \int_0^\infty j(E, B_\perp(\bm{x}))\, n_{\rm cre}(\bm{x}, E)\, \dd E,
\label{eq:J}
\end{equation}
where the emissivity $j(E)$ of an electron of energy $E$ is given in
Eqs.~(8.56) and (8.57) of \citet{2011hea..book.....L} and can be
conveniently evaluated using the synchrotron functions provided by
the \href{https://www.gnu.org/software/gsl/doc/html/specfunc.html#synchrotron-functions}{GSL
library}.  The integral over the emissivities along the line of sight
given the three-dimensional distribution of $\ncre(\bm{x}, E)$ and
$B_\perp(\bm{x})$ yields the Stokes parameters
\Q and \U of the polarized synchrotron intensity. \Q and \U add up quadratically to the total polarized intensity\footnote{Note that instrumental noise
on \Q and \U leads to a positive bias in \PI when estimated via
Eq.~(\ref{eq:polint})~\citep[e.g.][]{1974ApJ...194..249W}. Since
we will fit the measured \Q and \U data, but not \PI, the bias is not
important for this work.} via
\begin{equation}
 \PI^2 = \Q^2 + \U^2
 \label{eq:polint}
\end{equation}
and their ratio defines the observed polarization angle
\begin{equation}
 \psi_{\rm PA} = \frac{1}{2}\, \arctan(\U/\Q).
\end{equation}
$\psi_{\rm PA}$ is perpendicular to the line-of-sight average of
the \ncre-weighted magnetic field angle in the plane of the sky
\begin{equation}
\langle\psi_{\rm mag}\rangle =  \psi_{\rm PA} + \pi/2,
\end{equation}
i.e.\ the analysis of \Q and \U is sensitive to both the strength and
the orientation of the perpendicular projection of the magnetic field
along the line of sight.

Isotropic random magnetic fields do not contribute to the polarized
intensity, but an anisotropic random field results in a net-polarized
intensity, because \Q and \U define only the plane of polarization,
and are thus invariant under a 180$^\circ$ rotation of the magnetic
field. Anisotropic random magnetic fields can originate from
magneto-hydrodynamic turbulence~\citep{1995ApJ...438..763G} or from
the stretching or compression of a flux-frozen isotropic random
magnetic field~\citep{1980MNRAS.193..439L}. The latter can lead to
basically one-dimensional fluctuations along a preferred orientation,
termed ``ordered random'' by \citet{2010MNRAS.401.1013J} or
``striated'' by \citet{JF12coh}.  A striated random field does not
make a net contribution to the RM since the contributions of
opposite-sign regions cancel out. However, it does contribute to the
polarized synchrotron emission since that depends only on the
orientation and not the direction of the field.  \citet{JF12coh}
considered different values of the striation factor for different
components of the field, but found them to be all the same. Here we model
the possible striation via a simple spatially independent
multiplicative factor to the coherent field,
\begin{equation}
B^\prime = (1+\xi) B.
\label{eq:striation}
\end{equation}
The striation factor $\xi$ is related to the striation factor $\beta$
used in~\citetalias{JF12coh}, for which a cosmic-ray electron energy
spectrum with a spectral index of $p=3$ was assumed, via $(1+\xi)
= \sqrt{1+\beta}$.

%% file: rm.tex
\subsection{Rotation Measures}
\label{sec:rmData}
We assemble a sky map of extragalactic \RMs based on individual
measurements from various catalogues and surveys as illustrated
in \cref{fig:rmData}.
Most of the currently known extragalactic \RMs were derived
by~\citet{tss09} from the two-band polarization data of the NVSS
survey \citep{1998AJ....115.1693C} leading to 37\,543 \RMs with a
declination of $>-40^\circ$. Since only two frequencies are available
in this data set, the derived \RMs are susceptible to $n\pi$
ambiguities in \cref{eq:chiRot} for high values of $|\RM|$. We replace
the values of 20 of these high-\RM sources by the ones obtained by the
broadband follow-up observations of~\citet{2019MNRAS.487.3432M}, three
are discarded based on these re-observations.

For measurements predating the year 2014, we use the catalogue of 4553
high-quality \RMs curated by~\citet{2014RAA....14..942X}. Notable
surveys with $N_\RM>100$ collected in this catalogue are from
\citet{skb81}, \citet{bmv88}, \citet{btj03}, \citet{kmg+03},
\citet{bbh+07}, \citet{bhg+07}, \citet{hbe09}, \citet{fem+09},
\citet{mgh+10}, \citet{vbs+11}, \citet{mmg+12a}, and \citet{mmg+12b}.

3220 RMs in the Southern equatorial hemisphere are taken from the
S-PASS/ATCA wide-band radio polarimetry survey (version 0.9)
of~\citet{2019MNRAS.485.1293S} (applying the quality cuts given in
Sec.~4 of that paper and requiring emission at a single Faraday depth
as well as model fits with $\chi^2/\text{ndf}<10$). This survey is of
particular significance, since it complements the sky coverage of the
NVSS survey. GMF model fits predating the S-PASS/ATCA release,
e.g.~\citetalias{JF12coh}, were constrained by only a very small
amount of \RMs in the 18\% southernmost equatorial sky.

Further 5999 \RMs published in the years 2014-2022 are taken from the
data compilation
of \citet{2023ApJS..267...28V} \citep[v1.1.0 at][]{van_eck_cameron_l_2022_7894467}. Notable surveys with $N_\RM>100$ in this compilation are from
\citet{2014ApJS..212...15F},
\citet{2015ApJ...815...49A}, \citet{2017MNRAS.469.4034O},
\citet{2017MNRAS.467.1776K}, \citet{2019ApJ...871..215B},
\citet{2020PASA...37...29R}, \citet{2020MNRAS.497.3097M}, and
\citet{2021ApJS..253...48V}.
In addition, we include the recent 2461 extragalactic \RMs
of \citet{2023MNRAS.519.5723O}.

\begin{figure*}[!t]
  \centering
  \subfloat[Sky map of extragalactic rotation measures (44857 \RMs averaged
  over $N_\text{side}=32$ \HEALPix pixels). The color scale is saturated at $|\RM|\geq\,150\,\radm$)]{
  \includegraphics[clip,rviewport=0 0 1.01 1,width=0.5\linewidth]{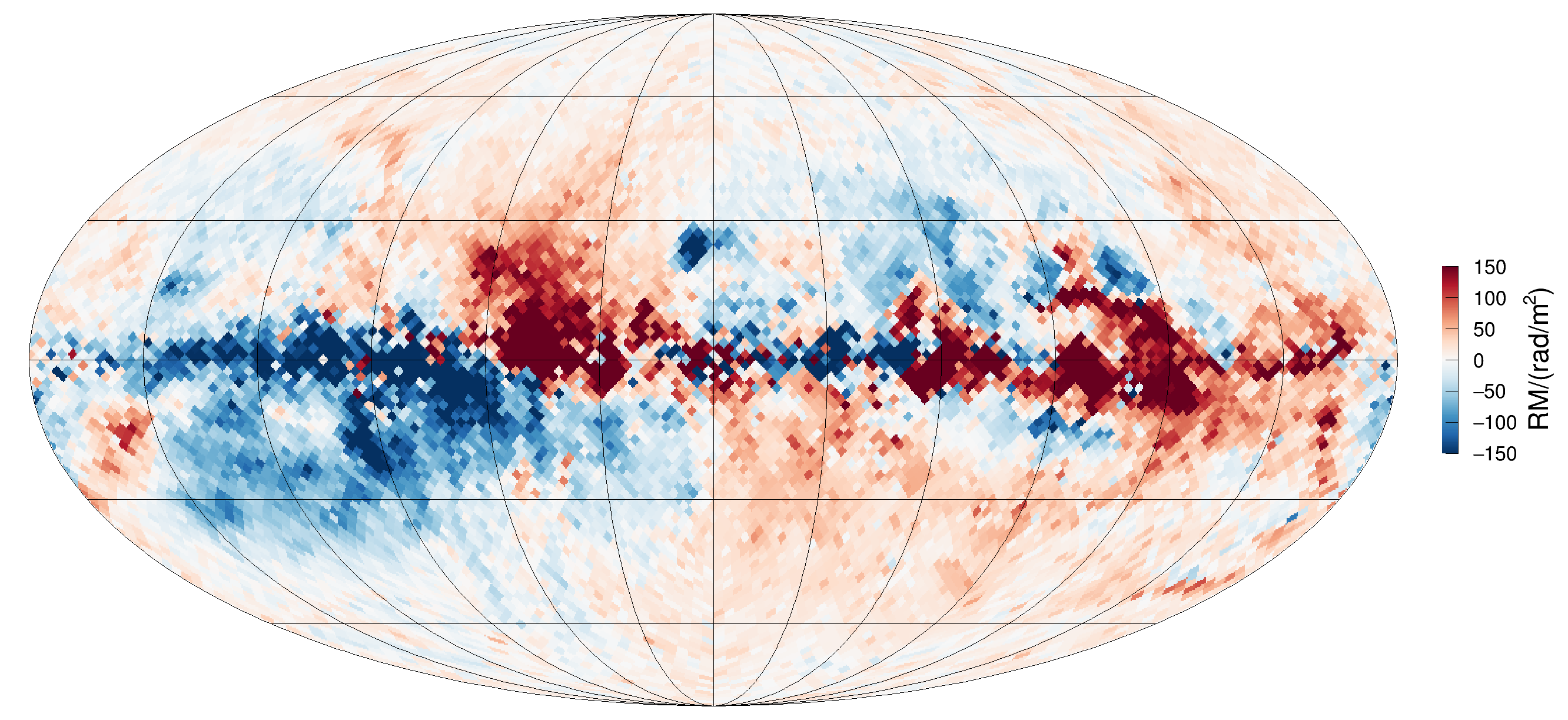}
  \label{fig:maskMap1}
  }%
  \subfloat[Masked region for the \RM analysis. Selected astrophysical objects (red) and HII regions (blue). The gray-scale background shows the  sky map of emission measures.]{
\includegraphics[clip,rviewport=-0.01 0 1 1,width=0.5\linewidth]{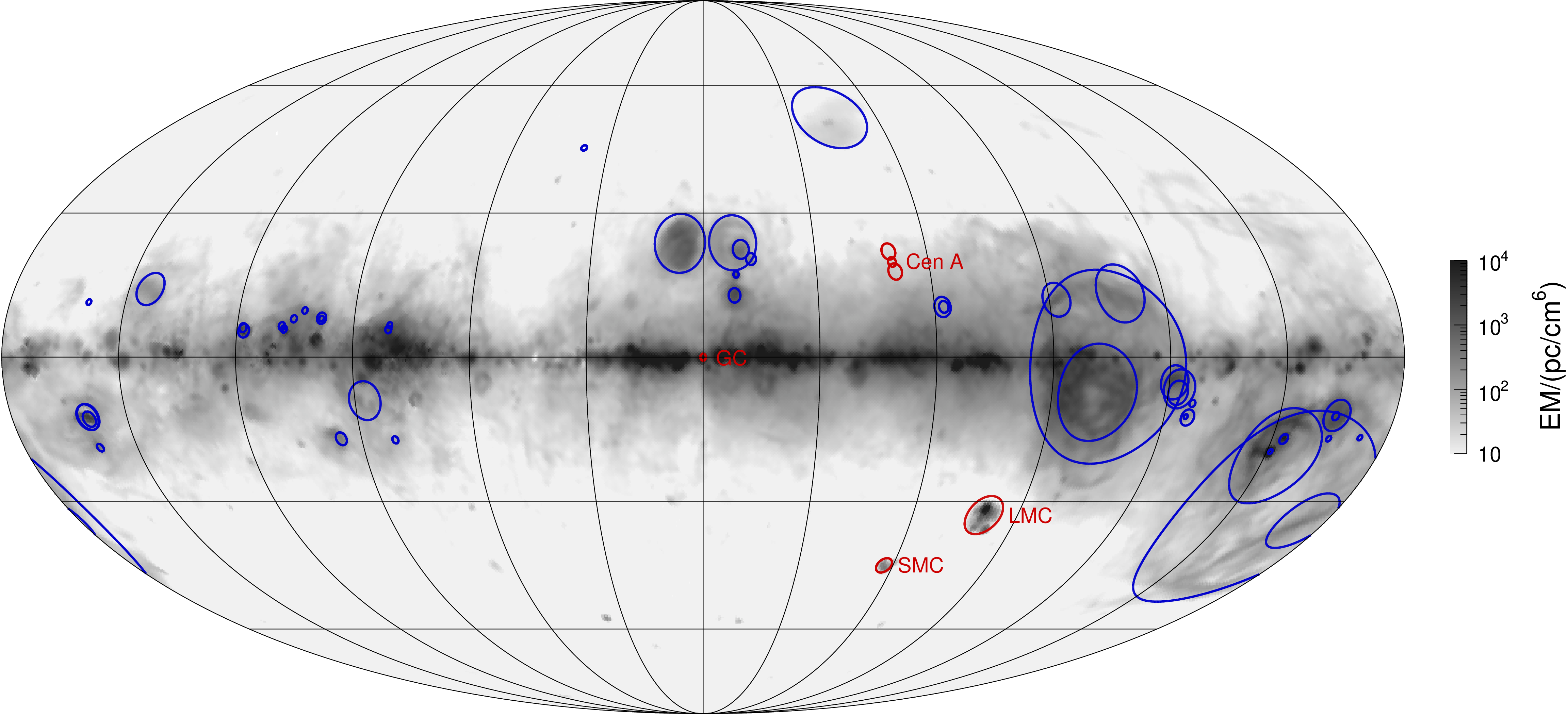}
\label{fig:maskMap2}
}
\caption{Sky map of rotation measures (left) and masked regions (right).}
\label{fig:maskMap}
\end{figure*}

In this data set of 53\,773 \RMs, we identify obvious multiple
measurements of the same extragalactic object if their coordinates
correspond to the same pixel id on a high-resolution \HEALPix sky map
with $N_\text{side}=2048$ (angular pixel width of 1.7 arcminutes).\footnote{The \HEALPix package
  (Hierarchical, Equal Area, and iso-Latitude Pixelation of the
  sphere) is used to subdivide the sky into equal-area
  pixels~\citep{2005ApJ...622..759G} The resolution $N_\text{side}$ of
  a \HEALPix map is related to the number of pixels via $N_\text{pix}
  = \smash[b]{12\,N_\text{side}^2}$.  For a map with
  $N_\text{side}=16$, the standard resolution used in this paper, the
  number of pixels is $N_\text{pix} =3072$ with an angular width of
  approximately $\smash[b]{\theta \approx \sqrt{4\pi/N_\text{pix}} \;
    180^\circ/\pi = \sqrt{3/\pi} \; 60^\circ / N_\text{side} =
    3.66^\circ}$.} All duplicate \RMs but the measurement with the
best frequency coverage are dropped, leaving 47\,054 \RMs of unique
extragalactic objects.

In the next step of data selection, we apply a two-pass algorithm to
reject outliers caused by either a large source-intrinsic \RM or a
wrong resolution of the $n\pi$-ambiguity. The procedure is similar in
spirit but different in detail to that used for the data set
of~\citetalias{JF12coh}.  For each \RM measurement we accumulate
surrounding measurements within an angular collection radius starting
at 1$^\circ$. The collection radius is increased until a sample of at
least 10 independent surrounding \RM measurements are found that
differ by less than three median absolute deviations from the median
RM of that sample. In the second pass, these sources are then used to
calculate the mean and standard deviation of the \RM in the region
surrounding each object. We keep objects if their \RM value is closer
than three standard deviations from the surrounding mean.

The remaining 44\,857 \RMs are displayed in \cref{fig:maskMap1}.  For
this figure, we show the average of \RMs in an $N_\text{side}=32$
\HEALPix map.\footnote{This and subsequent maps shown in this paper
  are area-preserving Mollweide projections of the sky in Galactic
  coordinates with the Galactic center at the origin. The longitude
  increases toward the left from $0^\circ$ at the center to
  $+180^\circ$ and decreases toward the right from $+360^\circ$ at
  the center to $+180^\circ$ (or, equivalently, from $0^\circ$ to
  $-180^\circ$).} For the purpose of this figure, 1524 pixels (12\%) without a
measurement were in-painted with the average of their surrounding pixels. At this somewhat higher resolution than the one used for
the analysis (see below), both the large and small-scale features of
the extragalactic rotation measures are visible. The most salient
large-scale features of the \RM sky are known since the early advent
of \RM catalogues, see e.g.\ the discussion in
\citep{1988Ap.....28..247A, 1997A&A...322...98H}: For the inner Galaxy
(longitudes between $-90$ to $90^\circ$) there is an anti-symmetry in
both longitude and latitude of the sign of the average rotation
measures, whereas in the outer Galaxy (longitudes from $90^\circ$ to
$270^\circ$), the average \RMs have the same sign above and below the
plane, as summarized by the following schematic of an \RM sky map,
\begin{equation}
  \RM(\ell,b) \approx \mathcircled{\!\!\frac{-+-+}{--++}\!\!}\,.
  \label{eq:rmantisym}
\end{equation}
In addition, many small-scale features can be identified, some of
which can be attributed to foreground objects. We remove some of the
most obvious regions, indicated by circles in
\cref{fig:maskMap2}. Firstly, we discard lines of sight if they pass
through magnetized objects with a large angular size, i.e.\ M31, the
Small and Large Magellanic Clouds, Centaurus A (core and lobes) and the
Galactic Center. The excluded regions around these objects are shown
in red in \cref{fig:maskMap2}. Secondly, we deselect \RMs if the
thermal electron density along the line of sight is dominated by a
single object. Such local overdensities in the diffuse WIM are caused
by the ultraviolet light of young massive stars ionizing the interstellar
medium around them, creating an HII region.  \RMs in lines of sights passing
through HII regions are often dominated by the product $\nel
B_\parallel \Delta$ inside the region of thickness $\Delta$, see
e.g.\ \citet{1981ApJ...247L..77H} and
\citet{2011ApJ...736...83H}. These \RMs are thus not representative of
the large-scale GMF. Regions of locally enhanced thermal electron
density can e.g.\ be identified by the emission measure, $\EM =
\int_0^\infty \nel\, \dd l$, which is displayed in \cref{fig:maskMap2}
based on the composite of H$\alpha$ surveys (VTSS, SHASSA and WHAM)
from \citet{2003ApJS..146..407F}. More than 8000 Galactic HII regions
are known in the Galaxy \citep{2014ApJS..212....1A}, but most of them
are either far away, and thus of small angular extent, and/or at low
Galactic latitudes, where the overall \RM integral through the plane
is large enough that it is not overshadowed by the contribution of a
single HII region. We therefore deselect only the lines of sight that
overlap with an HII region if it is either of large angular extent
($r{>}10^\circ$) or at high latitudes, $|b|{>}5^\circ$. These regions are
displayed in blue in \cref{fig:maskMap2} if their size is
${>}0.5^\circ$.

The final 41\,686 \RMs are then binned in 3072 angular pixels of an
$N_\text{side}=16$ \HEALPix map with an angular diameter of ${\sim}
3.7^\circ$. For each pixel, we calculate the mean and unbiased sample
variance to be used in the model optimization (see \cref{eq:chi2} below). If the
number of \RMs within one pixel is ${<}10$, then the variance (but not
the mean) is calculated using the 10 \RMs with the closest angular
distance to the center direction of the pixel. This leads to 2838
pixels with \RM data, where 46 pixels are excluded because no data was
observed in this direction and another 188 pixels are excluded by the
masks displayed in \cref{fig:maskMap2}. The final masked and binned
\RM data are shown in Fig.~\ref{fig:basemodel} in
Sec.~\ref{sec:results}.

%% file: synchrotron.tex
\subsection{Polarized Synchrotron Emission}
\label{sec:syndata}

The polarized synchrotron emission from the Galaxy is best observed at
high frequencies, where the depolarization of the signal due to
Faraday rotation is negligible. At Faraday depths typical for the
Galaxy, the change in polarization angle becomes negligible in the
tens of GHz range, see~\cref{eq:chiRot}.  The received polarized
intensity in units of {Rayleigh-Jeans antenna temperature}, $T = c^2 /
(2\nu^2\,k_\text{B})\, \PI$, decreases as $\nu^{\beta_\text{s}}$ with
frequency. Here the synchrotron spectral index $\beta_\text{s} \approx
-3$ for a cosmic-ray electron spectrum with index $p=3$,
see~\cref{eq:J_coherent}.
Another source of polarized emission from the Galaxy originates from
thermal dust with a spectral index of $\beta_\text{d} \approx +1.6$
\citep{2015A&A...576A.107P}. Empirically, the cross-over between the
two components is at around 100~GHz.

The ``sweet spot'' for the observation of synchrotron emission is
therefore at high enough frequencies such that Faraday depolarization
is negligible, but not too high, such that the signal is not dominated
by polarized dust emission.  The \WMAP and \Planck satellites operated
in this frequency regime, with a frequency threshold of 20 and 30 GHz
respectively.

From \WMAP we use the final nine-year results (DR5) on the polarized
synchrotron emission at 22.5\,GHz provided at a \HEALPix resolution of
$N_\text{side}=64$~\citep{2013ApJS..208...20B}. Four variants of \Q
and \U sky maps are available, differing mainly by the constraints
placed on $\beta_\text{s}$ during analysis. These four variants of the
derived polarized synchrotron intensity are in good agreement outside
of the Galactic plane, which is anyway masked in our analysis (see
below). We use the ``\modelBase'' model, which is the most data-driven
variant in which the synchrotron spectral index was allowed to float
freely in each sky pixel. We in-paint the pixels that are flagged as
having an erroneous component separation (2.7\%) with the $Q$ and $U$
averages of their eight surrounding neighbor pixels.

\begin{figure}[!t]
  \def\figw{1}
  \includegraphics[width=\figw\linewidth]{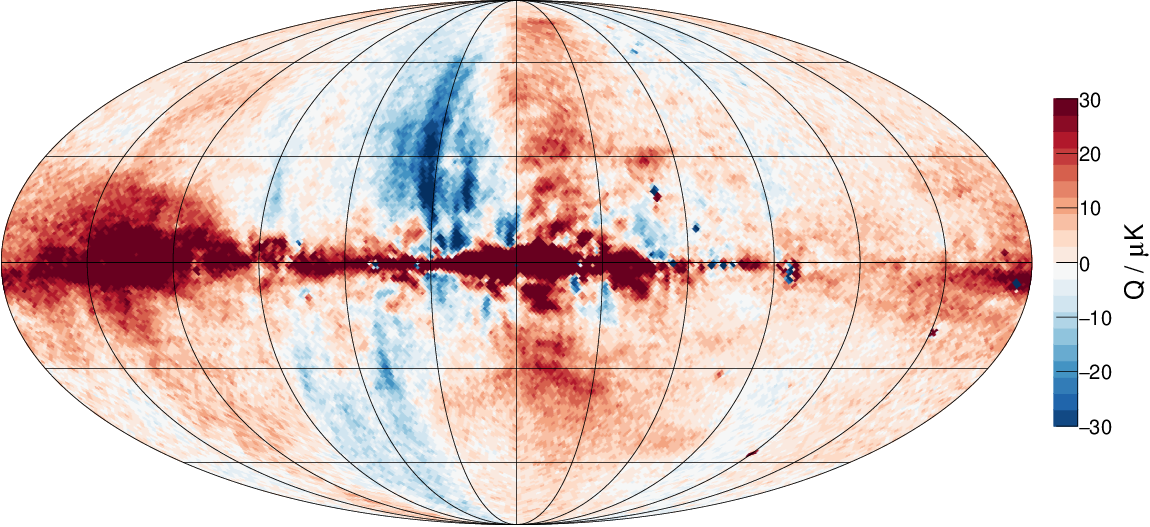}\\[\baselineskip]
  \includegraphics[width=\figw\linewidth]{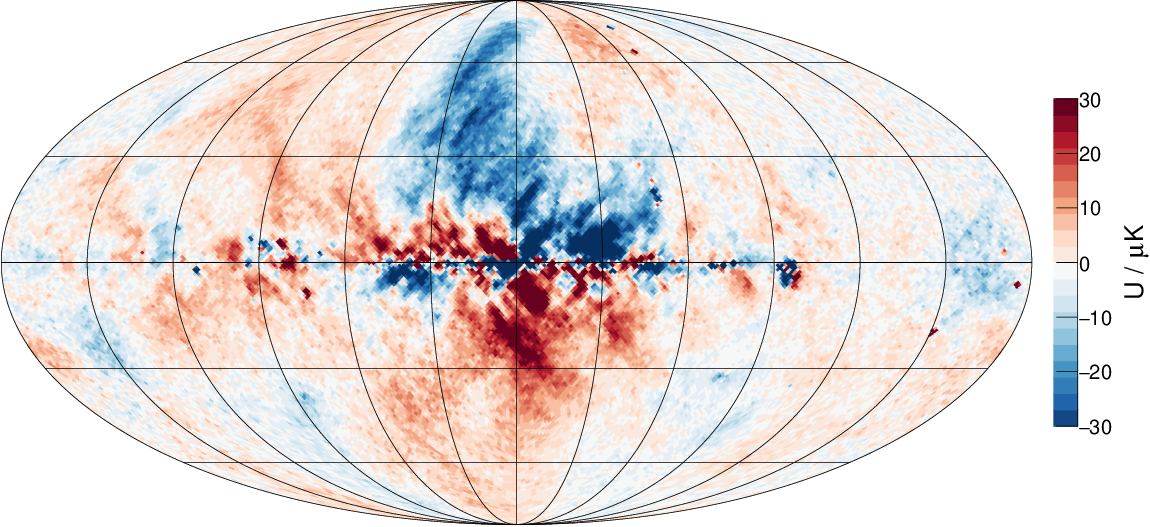}\\[\baselineskip]
  \includegraphics[width=\figw\linewidth]{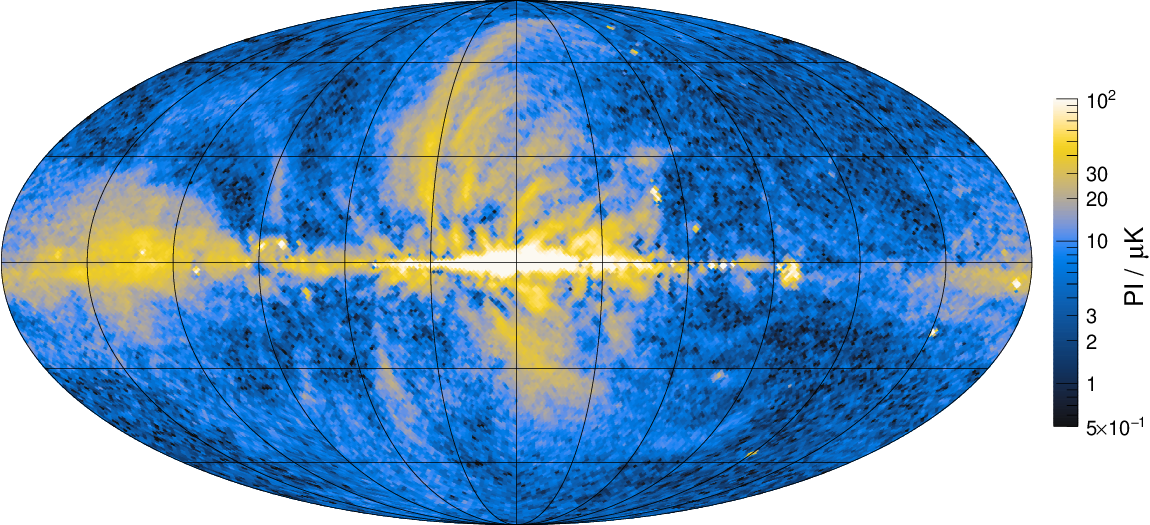}
  \caption{Stokes \Q (top panel) and \U (middle panel) parameters and polarized
    intensity \PI (bottom panel) from the averaged synchrotron maps of
    \Planck and \WMAP at 30~GHz.}
 \label{fig:syndata}
\end{figure}

From \Planck we use the third release (R3.0) of the polarized
synchrotron foreground at 30\,GHz derived from component separation
using the {\scshape Commander}
software~\citep{2020A&A...641A...4P}. These sky maps are provided at
high resolution of $N_\text{side}=2048$ and we average the $Q$ and $U$
values to obtain maps with $N_\text{side}=64$ to match the resolution
of \WMAP. During this averaging, spurious outliers in the
high-resolution maps are identified and discarded if the value is more
than $5\sigma$ away from the median of the values within the
low-resolution pixel. This procedure removes 0.18\% of the
high-resolution data points.

We then combine the \Q and \U sky maps from \WMAP and \Planck into a
lower resolution map at $N_\text{side}=16$ by taking the simple
arithmetic average of all 32 measurements in each pixel (16 from each
experiment). \WMAP intensities are extrapolated to the \Planck
frequency with a spatially-constant synchrotron spectral index of
$\beta_\text{s} = -3.15$ based on the mean value predicted by our
simulations described in Sec.~\ref{sec:ncre}. Note that due to the
proximity of the \WMAP and \Planck frequencies, even a large
difference between the actual and assumed of
$\Delta\beta_\text{s}=0.2$ would introduce a variation in extrapolated
intensity of only
$1-(22.5~\text{GHz}/30~\text{GHz})^{\pm\Delta\beta_\text{s}} = \pm
6\%$. The variance of the 32 data points per pixel is used to
calculate the weights in the $\chi^2$ minimization as discussed in
Sec.~\ref{sec:modelopt}.

\begin{figure}[!t]
  \centering
  \includegraphics[width=\linewidth]{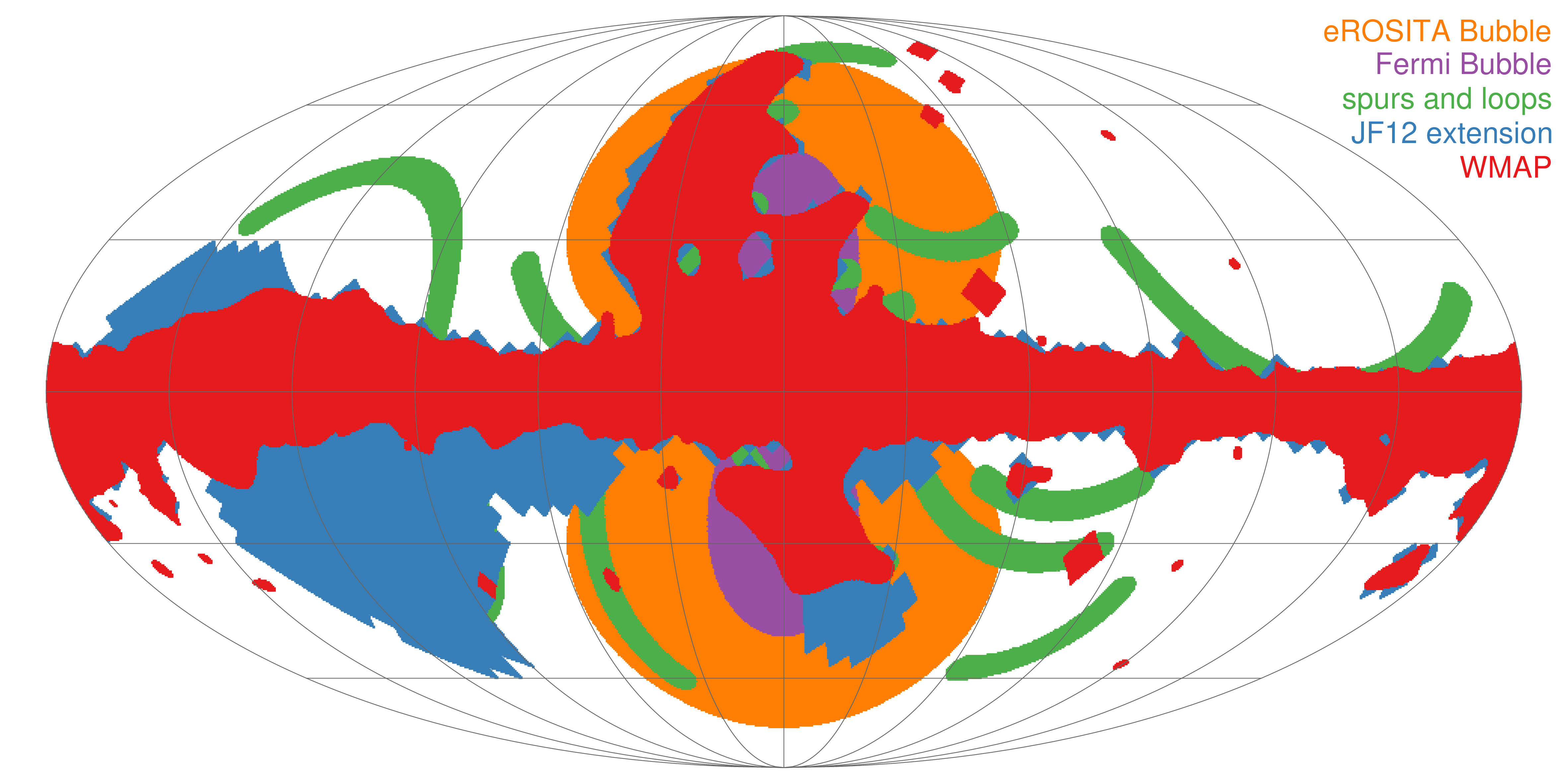}
  \caption{Polarized synchrotron mask used in this analysis. Masked
    regions are shown in red, blue and green (see text). For
    comparison, the outline of the Fermi and eROSITA bubbles are shown
    in violet and orange.}
  \label{fig:synmask}
\end{figure}

The Stokes parameters derived by the \WMAP and \Planck collaborations
exhibit systematic large-scale differences, see \cref{fig:syndatadiff}
in the appendix.  These differences could be attributed either to
large-scale spatial variation of the synchrotron spectral index or to
residual calibration uncertainties in one or both of the data sets. A
combined analysis of \WMAP and \Planck data was recently presented by
the \CosmoGlobe
Collaboration~\citep{2023arXiv230308095W,2023arXiv231013740W},
performing the component separation and the calibration of the data
sets simultaneously. We will use the \CosmoGlobe results as an
alternative to our average and study the effect on the parameters of
GMF models in Sec.~\ref{sec:results}.

Our combined maps of Stokes \Q and \U and polarized intensity \PI of
the Galactic synchrotron emission at 30~GHz are shown in
\cref{fig:syndata}.  As can be seen, both \Q and \U exhibit
large-scale features, which we will interpret in the following  as
imprints of the large-scale features of GMF. However, care must be
taken not to include regions with strong local features that could
bias our fits.  We therefore mask-out some regions of the sky in
fitting to the GMF parameters.

The elements of our mask when fitting the polarized intensity data are
shown in \cref{fig:synmask}.  Red indicates the mask used in the
original \WMAP analysis. It excludes regions of high polarized
intensity along the Galactic plane and the North Polar Spur, and
pixels containing strong extragalactic sources such as the radio lobes
of Centaurus A. The blue region depicts the additional polarized
intensity mask introduced by \citet{JF12coh} mainly to remove a
presumably local, high-latitude polarized emission at
$90^\circ<\ell<180^\circ$. Finally, we also mask the green regions as
an attempt to remove further large circular arcs (called loops, spurs,
or filaments) visible in \PI.  Here we exclude data in the direction
of loops I to IV, as defined by \citet{1971A&A....14..252B} and
further filaments identified by \citet{2015MNRAS.452..656V}. The union
of the above leads to our final PI mask.  The cumulative application
of these three masks leaves 73.1\% (\WMAP), 63.4\% (\WMAP+JF12), and
57.8\% (\WMAP+JF12+loops) of the sky for analysis.

It is worthwhile noting that the exact attribution of features in the
polarized radio sky to local or global phenomena in the Galaxy is
still under debate, see e.g.~\citet{2022arXiv220301312L}.  Some of the
loops and filaments could be caused by local supernova remnants
expanding into the surrounding ambient magnetic
field~\citep{1973A&A....24..149S}, whereas others might be related to
large-scale magnetized outflow from the Galactic
center~\citep{2013Natur.493...66C}, related to the so-called ``Fermi
bubbles'' observed in gamma rays~\citep{2014ApJ...793...64A} and
surrounded by ``eROSITA bubbles'' in
X-rays~\citep{2020Natur.588..227P}. These are shown as violet and
orange regions in \cref{fig:synmask}, demonstrating that our mask
retains some directions which contain contributions from the Fermi and
eROSITA bubbles.

%% file: thermalElectrons.tex
\subsection{Thermal Electrons}
\label{sec:ne}
\begin{figure*}[ht]
\includegraphics[clip,rviewport=0 -0.06 1 1,width=0.7\linewidth]{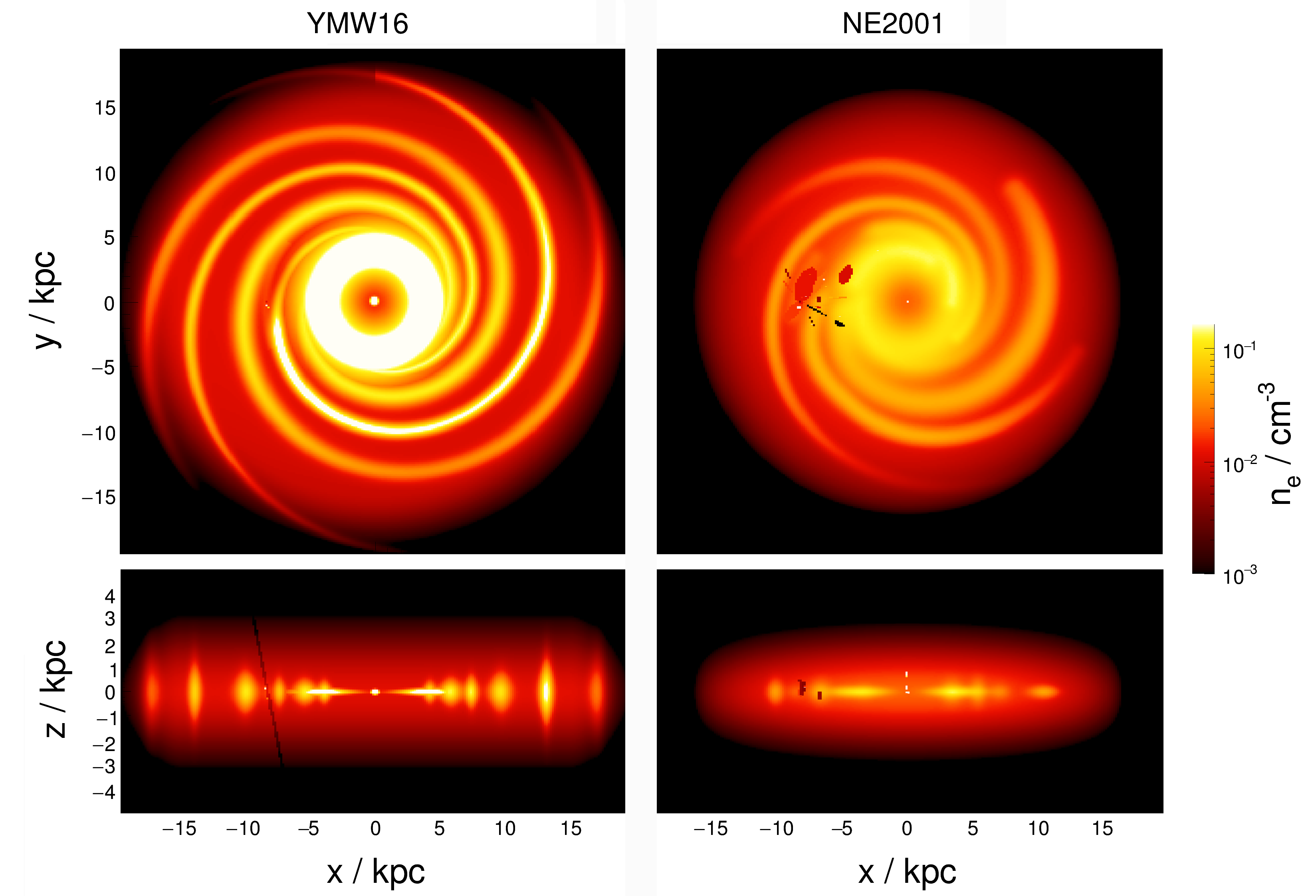}%
\includegraphics[width=0.3\linewidth]{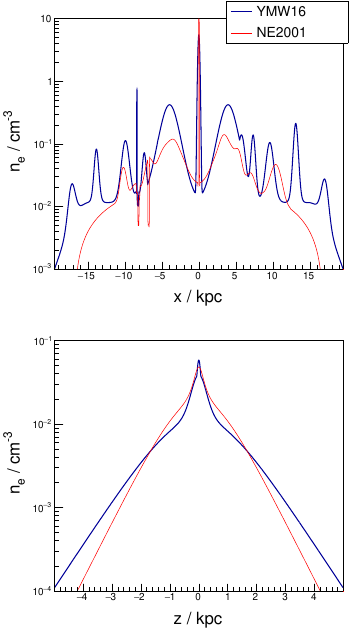}
\caption{{\itshape Left:} Thermal electron density, $\nel$, for the
  \YMW and \NE model. The top panel shows the mid-plane density at
  $z=0$ and the bottom panel gives an edge-on view of the density at
  $y=0$ in the $x-z$ plane. Structures visible at around the position
  of the sun ($\approx (-8.2, 0, 0)$~kpc) are due to the modeled
  under- and over-densities of the local interstellar
  medium. {\itshape Right:} Thermal electron density along the
  Galactic $x$-axis at $z=y=0$ (top) and along the $z$-axis at
  $x=-6$~kpc and $y=0$.}
\label{fig:ne}
\end{figure*}
The magnetized plasma responsible for the Faraday rotation of
extragalactic radio sources resides mostly in the warm ionized medium
(WIM) of the Galaxy \citep[e.g.][]{ryden_pogge_2021}. To interpret
the rotation measures, a three-dimensional model of the density,
$\nel({\bm x})$, of these free thermal electrons is needed. Here we use two
models of $\nel$: \NE of~\citet{2002astro.ph..7156C} and \YMW
of~\citet{2017ApJ...835...29Y}. Both models were tuned to describe the
{\itshape dispersion measure} \DM of Galactic pulsars, which is given
by the line-of-sight integral from Earth to the pulsar at a distance
$d$, $\DM = \int_0^d \nel(\bm{x}(r)) \, \dd r$. More than 3000
pulsars with measured {\DM}s are listed in the current version (1.70)
of the ATNF pulsar catalogue \citep{2005AJ....129.1993M}, but only for
a few is the distance $d$ well known: \NE was tuned to the {\DM}s of
112 pulsars and \YMW used 189. Given the large number of parameters of
these models (e.g.\ the \YMW model has 82 fixed parameters and 32
fitted parameters) and the scarcity of data, they rely, to a
large extent, on astrophysical priors for the geometrical topology of
the thermal electron density.

Particularly important for the
modeling of the large-scale structure of the GMF is the vertical
structure of the Galactic WIM which is relatively well constrained by the
{\DM}s of pulsars in high-latitude globular
clusters~\citet{2008PASA...25..184G}.  Since the fit of \NE pre-dates
most of these data, we replace the original value of the exponential
scale height of the diffuse WIM of $h_\text{WIM} =0.95$~kpc with
$h_\text{WIM} =1.3\pm 0.2$~kpc as derived
by \citet{2012MNRAS.427..664S} for this model. A larger value of
$h_\text{WIM} = 1.67\pm 0.05$~kpc was inferred
by~\citet{2017ApJ...835...29Y} for their \YMW model.

An illustration of the thermal electron densities of \YMW and \NE is
shown in \cref{fig:ne}. As can be seen, the two models differ
substantially, especially regarding the positions and widths of spiral
arms, the density in the molecular ring at $r\approx 5$~kpc and in the
Galactic Center, and the scale-height of the thick disk of
the WIM. Whereas \YMW can predict pulsar {\DM}s with a somewhat higher
fidelity than \NE, the latter model remains a viable
alternative to describe the large-scale features of the
WIM~\citep[e.g.][]{2021PASA...38...38P}. By using both models in our
GMF fits, we can study the systematic effects arising from different
assumptions on the thermal electron density.

%% file: cosmicRayElectrons.tex
\begin{figure*}[t]
  \def\figh{0.29}
  \centering
\includegraphics[height=\figh\textheight]{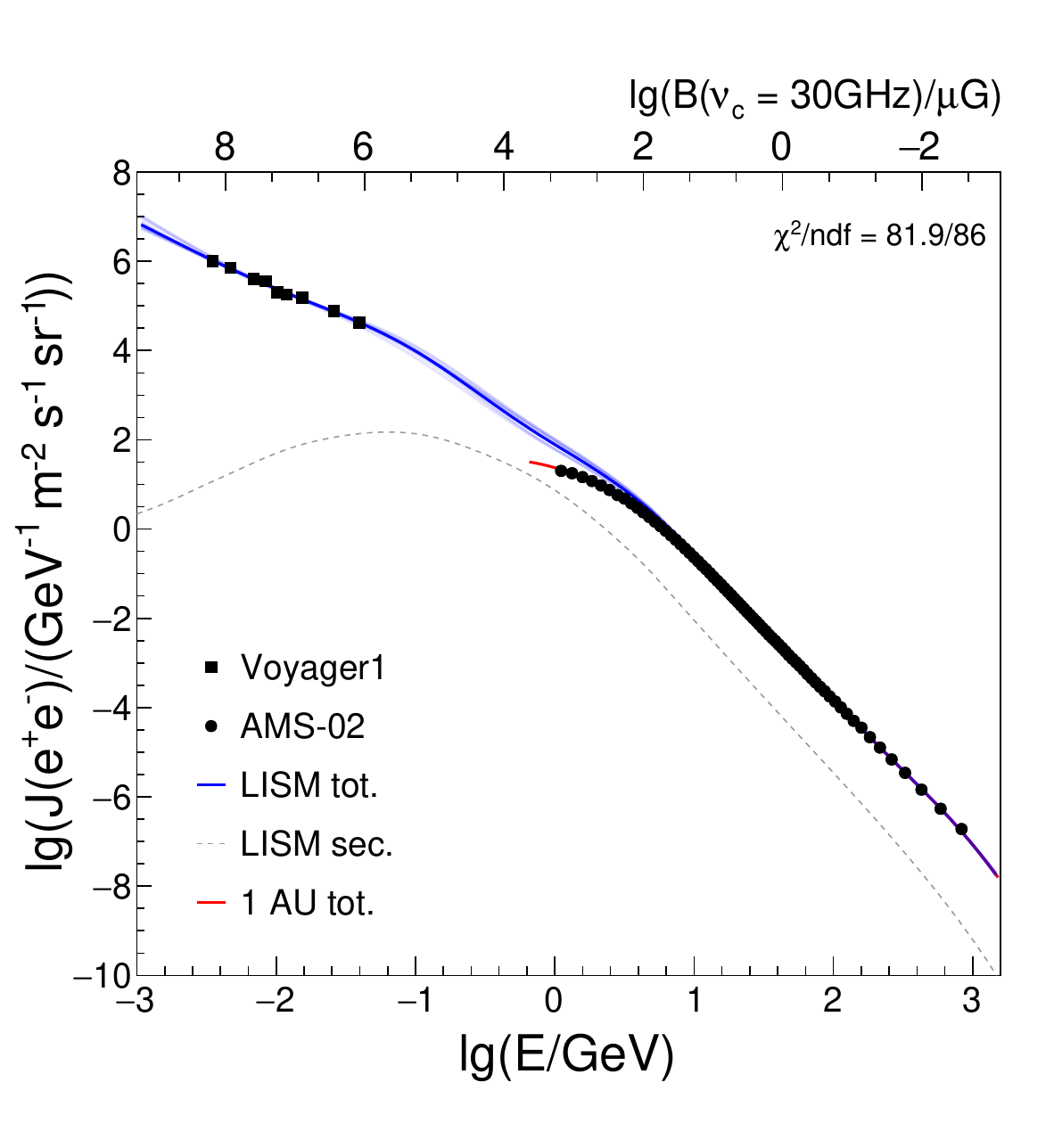}
\includegraphics[clip, rviewport=0 -0.09 1 1.139,height=\figh\textheight]{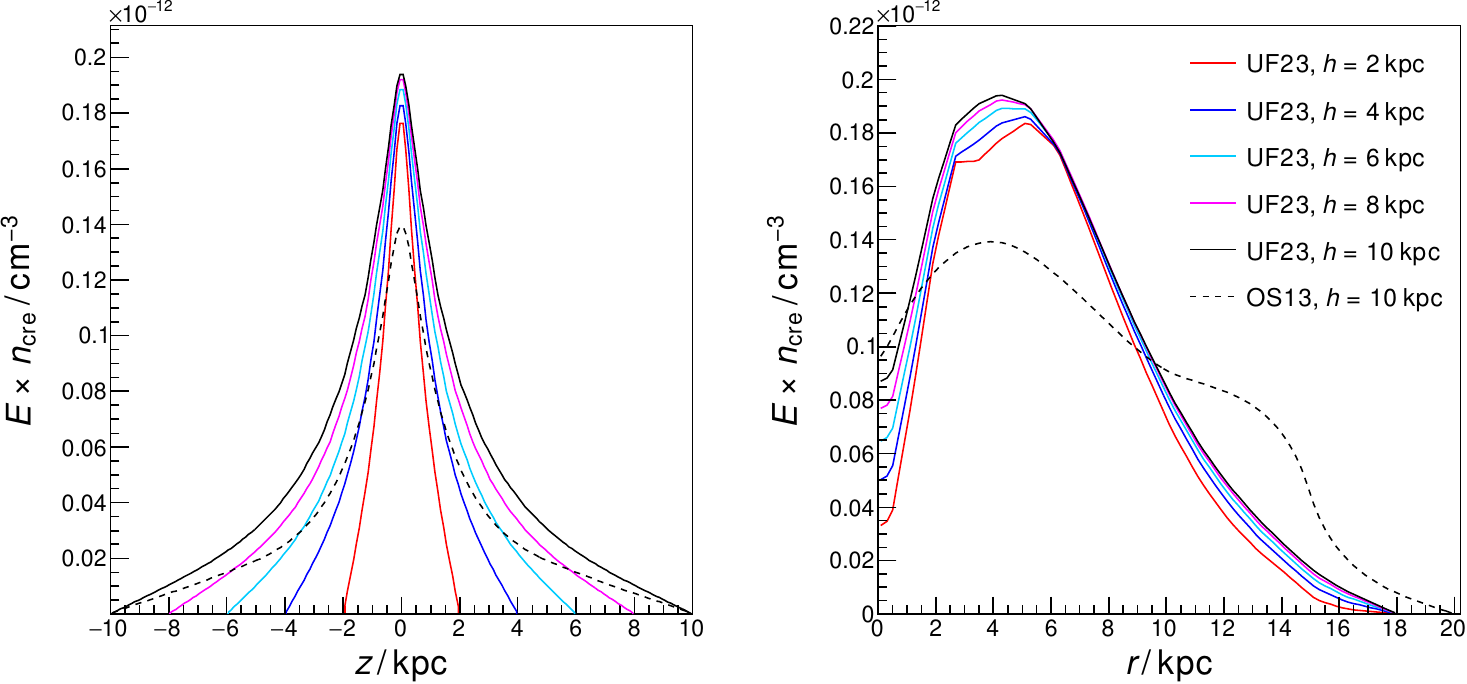}
\caption{{\itshape Left:} Flux of cosmic-ray electrons and positrons
  at Earth. Black points show the measurements.  The calculated flux
  in the local interstellar medium (LISM), for the $\hDiff=6$~kpc
  case, is shown as a blue band and the contribution from secondary
  production ($\text{p}_\text{CR}+\text{p}_\text{ISM}$ interactions)
  is indicated by a dashed line; the solar-modulated flux at Earth is
  shown as a red line. The additional $x$-axis at the top of the
  figure shows the magnetic field strength needed to obtain a
  synchrotron peak at $\nu_c=30$~GHz (i.e.,\ the frequency for which
  the \Planck collaboration reports the polarized synchrotron
  emission used here), for the electron energy given in the $x$-axis at the
  bottom of the figure.  {\itshape Middle:} Density of 10~GV
  cosmic-ray electrons at a galactocentric radius $r=4$~kpc, as a
  function of Galactic height $z$ and {\itshape Right:} as a function
  of radius $r$ at mid-plane ($z=0$). Different choices of the
  half-height \hDiff of the diffusion volume are shown as lines with
  different colors. The dashed line is the result
  of~\citet{2013MNRAS.436.2127O} for $\hDiff=10$~kpc.  }
\label{fig:ncreDens}
\end{figure*}

\subsection{Cosmic Ray Electrons}
\label{sec:ncre}
Calculating Galactic synchrotron emission requires not only a
model of the GMF but also a model of the
three-dimensional density distribution and energy spectrum of cosmic-ray electrons in the Galaxy, $\ncre({\bm x},E)$.

In contrast to the purely phenomenological thermal electron models
discussed in the previous section, predictions of \ncre are based on
detailed modeling of the production and propagation of electrons and
positrons in the Galaxy \citep[e.g.][]{1998ApJ...493..694M}, to obtain
a steady-state solution of the diffusion equation of cosmic rays in
the Galaxy~\citep{1964ocr..book.....G}. In its simplest version
(sometimes referred to as ``plain diffusion''), the defining quantity
of the model is the diffusion coefficient of charged particles in the
interstellar medium (ISM). Most calculations assume a homogeneous and
isotropic Galactic diffusion coefficient $D$ within a ``diffusion
volume'' approximated as a cylinder of half-height \hDiff.

The normalization and rigidity\footnote{The rigidity of a particle
  with charge $Ze$ and momentum $p$ (energy $E$) is $\R=pc/(Ze) \simeq
  E/(Ze)$} dependence of $D$ is determined from measurements of the
fluxes of secondary and primary cosmic-ray nuclei at Earth. The most
precise estimates (in terms of the uncertainty of both cosmic-ray flux
and nuclear cross sections) are derived from the ratio of the flux of
secondary boron nuclei and (mostly) primary carbon nuclei. However,
secondary-to-primary ratios can only constrain the ratio, $D/\hDiff$,
\citep[e.g.][]{2001ApJ...555..585M}. Estimates of $D/\hDiff$ in a
plain diffusion scenario range from 0.03~kpc/Myr
\citep{Genolini:2019ewc,2017PhRvD..95h3007Y} to 0.10~kpc/Myr
\citep{2016ApJ...831...18C} at a reference rigidity of 10~GV.

The degeneracy between \hDiff and $D$ can in principle be broken by
data on ``cosmic clocks'' (e.g. $^{10}$Be/$^{9}$Be or Be$/$C), but due
to the poor quality of current data and uncertainties in the
spallation cross sections, only mild constraints on the halo height
can be derived and the current estimates of \hDiff are in the range 2
to 10~kpc
\citep[e.g.][]{Weinrich:2020ftb,Evoli:2019iih,Maurin:2022gfm}.  We
therefore consider diffusion volumes having $\hDiff = 2, 4 , 6, 8$ and
10~kpc and derive the corresponding \ncre for each, to allow us to
assess the uncertainty in the GMF due to the present uncertainty in
$\hDiff$.  For each of the five different values of the height of the
diffusion volume, $\hDiff$, we solve the cosmic-ray diffusion equation
with the \DRAGON program~\citep{Evoli:2008dv}.\footnote{We choose
  \DRAGON over the other widely used solver
  \GALPROP~\citep{1998ApJ...509..212S} because at the time of setting
  up our analysis chain only \DRAGON supported spatially varying
  diffusion coefficients. In this paper we, however, investigate only
  homogenous diffusion.}

For each value of $\hDiff$ we obtain the diffusion coefficient at
10~GV from $D/\hDiff=0.046$~kpc/Myr and we take the rigidity
dependence of $D(\R)$ to be the one derived for the ``PD2'' model
of~\citet{2016ApJ...831...18C}.  At high rigidities ($\R \gtrsim
5~\text{GV}$) the diffusion coefficient scales as $\smash[b]{D
  \varpropto \R^{0.578\pm 0.073}}$, i.e.\ compatible with the
power-law scaling $\smash[b]{\R^{\nicefrac{1}{2}}}$, typical for a
turbulent magneto-hydrodynamical
cascade~\citep{1964SvA.....7..566I,kraichnan}.  The spatial
distribution of Galactic cosmic-ray sources is taken to follow the
radial distribution of pulsars in the
Galaxy~\citep{2006MNRAS.372..777L}, used as a proxy for the
distribution of supernova remnants, and we use the JF12 magnetic field
model to calculate the cooling of cosmic-ray electrons due to
synchrotron radiation.  The maximum galactocentric radius of the
diffusion volume was set to $R_\text{max}=18$~kpc.

We constrain the solutions of the diffusion equations by the
measurements of cosmic-ray proton and lepton fluxes, in the local
interstellar medium from the Voyager I satellite
\citep{2016ApJ...831...18C}, and inside the heliosphere from AMS-02
orbiting Earth on the International Space Station
\citep{AMS:2014gdf,AMS:2015tnn,AMS:2019rhg}.  We first find the proton
spectrum at the source; this then fixes the contribution to the lepton
flux at Earth coming from proton interactions with cosmic rays and the
ISM ($\text{p}_\text{CR}+\text{p}_\text{ISM}$), accounting for a small
adjustment due to contributions from nuclei.  We then attribute the
remaining lepton flux to the injected lepton spectrum. Thus, in this
simplified ansatz, the well-known ``positron anomaly''
\citep{PAMELA:2008gwm}, i.e.,\ the excess of positron flux beyond
expectations from secondary production at energies $\gtrsim {10}$~GeV,
is attributed to the injected lepton spectrum.  Such an ansatz is
plausible if astrophysical sources of primary positrons have a similar
spatial distribution as the sources of primary electrons, e.g.,\ if
pulsars are the sources of the ``anomalous'' positrons and
electrons~\citep{Hooper:2008kg} and the bulk of electrons is
accelerated in supernova remnants. Then, if the interest is only in
the sum of electrons and positrons, as it is in our case, the two
sources can be lumped together into a single source class. Due to the
fast cooling time of electrons and positrons at high energies, the
high-energy flux at Earth might be dominated by a local source and
thus not representative of the average lepton flux in the Galaxy
\citep[e.g.][]{DiMauro:2014iia,Joshi:2017ogv,Mertsch:2018bqd}.
However, these effects are expected to play a role only above 100~GeV
and are thus not important for the synchrotron frequencies of interest
in this study. An example of our fit of the cosmic-ray lepton flux is
shown in the left panel of \cref{fig:ncreDens}.

  As a further systematic check, we also interpret the synchrotron
  data using the \texttt{z10LMPDE} cosmic-ray electron model, a
  \GALPROP calculation of~\citet{2013MNRAS.436.2127O} (OS13) used
  in~\citep{2016A&A...596A.103P} for GMF
  modeling~\citep{jaffe17}. This model assumes a half-cylinder height
  of $\hDiff=10$~kpc, a maximum radius of $R_\text{max}=20$~kpc and
  the source distribution from~\citet{1998ApJ...509..212S} with a
  radial cutoff at 15~kpc.  Further differences to our fiducial
  \GALPROP calculation include a different magnetic
  field~\citep{2010RAA....10.1287S} for electron cooling and a
  different value of $D/\hDiff = 0.031$ at 10~GV.  We normalized this
  model to match the AMS flux of electrons and positrons at 35~GV.

  The vertical and radial distribution of the cosmic-ray lepton
  density at 10~GV for different values of \hDiff and the OS13 model
  are shown in the middle and right panels of \cref{fig:ncreDens}. As
  expected, a larger half-height of the diffusion volume leads to a
  larger cosmic-ray occupation in the halo, while the density in the
  disk is approximately constant due to the normalization at
  Earth. The different source distribution assumed for OS13 results is
  mainly responsible for the very different radial distribution of
  cosmic-ray leptons.

 Of course, further variations beyond the scale height are possible
 and we studied the effect on synchrotron maps of several other type
 of variations, as described below.  However, only the variation of
 \hDiff makes a significant change in the PI predictions outside of
 the mask we use, so we only include the six \ncre models outlined
 above in our GMF model fitting.\footnote{Note that in the framework
   of a tunable striation parameter, an overall factor in the
   polarized synchrotron intensity does not affect the derived
   coherent magnetic field, but only the striation factor.}.  A spiral
 distribution of sources can affect the \PI up to 25\% in the Galactic
 plane, but has negligible effects outside of the polarization mask
 used in this work. Using the \Planck-tune of the JF12 random
 field~\citep{2016A&A...596A.103P} for synchrotron cooling instead of
 the original model~\citep{JF12rand} affects the intensity by up to
 16\% in the outer Galaxy (where a spiral arm with a large random
 field is present in the original model to describe the intensity from
 the ``fan region''), but the differences are again negligible outside
 the masked region used in this paper. The same holds for \ncre models
 using a three-dimensional model of the interstellar radiation field
 from~\citet{2017ApJ...846...67P}.  Moreover, in this study we did not
 iteratively re-adjust the magnetic field used for the synchrotron
 energy losses in the \ncre calculations. But the resulting refitted
 coherent fields are of the same order of magntitude as the ones of
 JF12 and therefore the electron cooling is dominated by the sum of
 contributions from the random magnetic field and inverse Compton
 scattering.  Finally, it was shown by~\citet{2018MNRAS.475.2724O}
 that the cosmic-ray electron spectrum above a few GeV is insensitive
 to the inclusion of re-acceleration and/or convection in the
 diffusion equation, therefore we do not consider it here.

The six models shown in~\cref{fig:ncreDens} will be used in this paper
to study the systematic effect of the three-dimensional model of \ncre
on the derived structure of the GMF.

We note that future systematic studies should include an investigation
of the effects of a more realistic particle propagation with an
inhomogenous and/or anisotropic diffusion
coefficent~\citep[e.g.][]{DiBernardo:2012zu,Giacinti:2017dgt,Merten:2017mgk,2018MNRAS.477.1258A},
ideally with a self-consistent description of relation between the
magnetic field and the diffusion
coefficient~\citep[e.g.][]{Kuhlen:2022hlo,Blasi:2023quf}.

%% file: magModels.tex
\begin{figure*}[!th]
  \begin{center}
    \def\figh{0.35}
    \includegraphics[clip,rviewport=0 0 0.87 1,height=\figh\textheight]{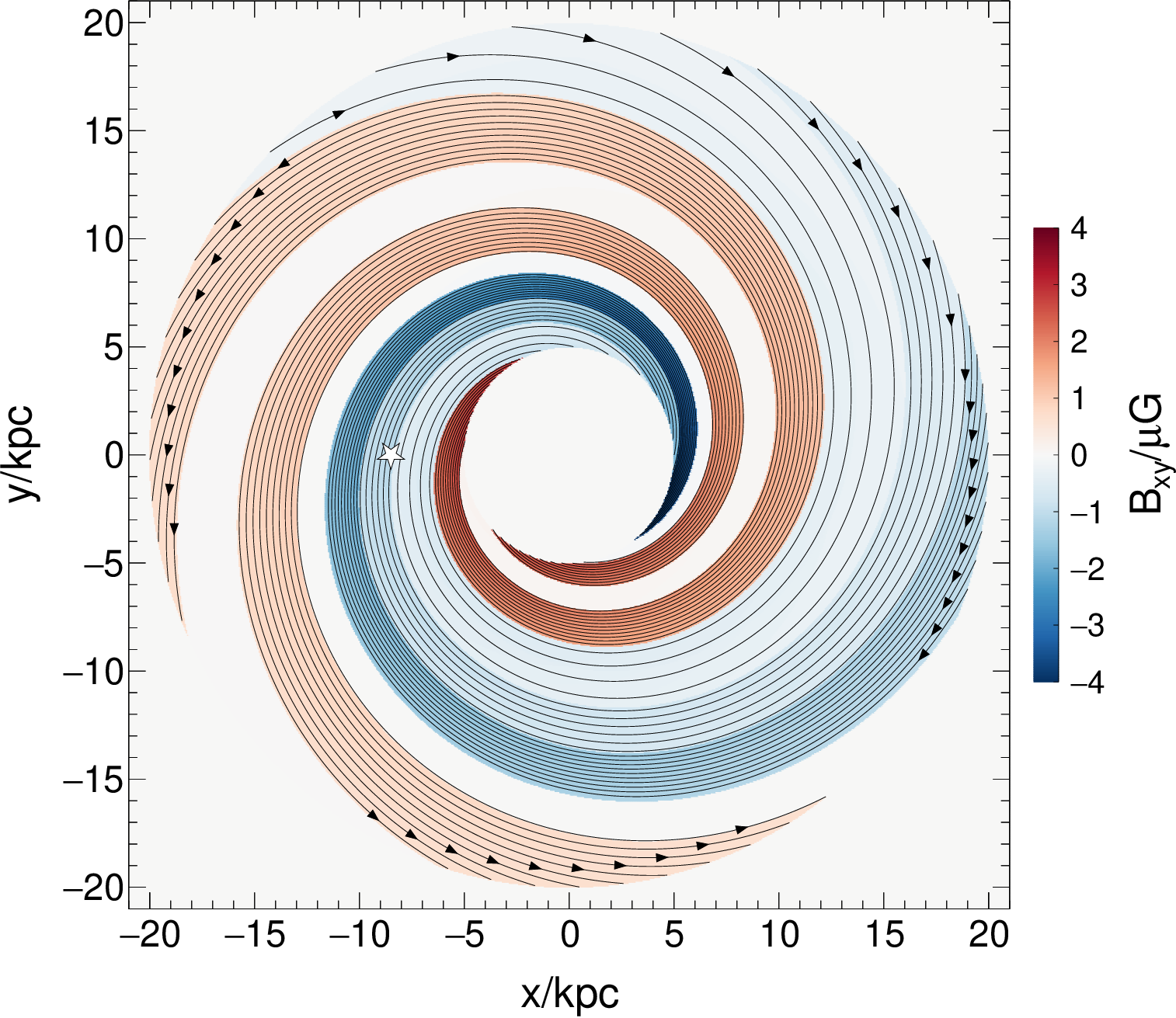}\quad%
    \includegraphics[clip,rviewport=0.105 0 1 1,height=\figh\textheight]{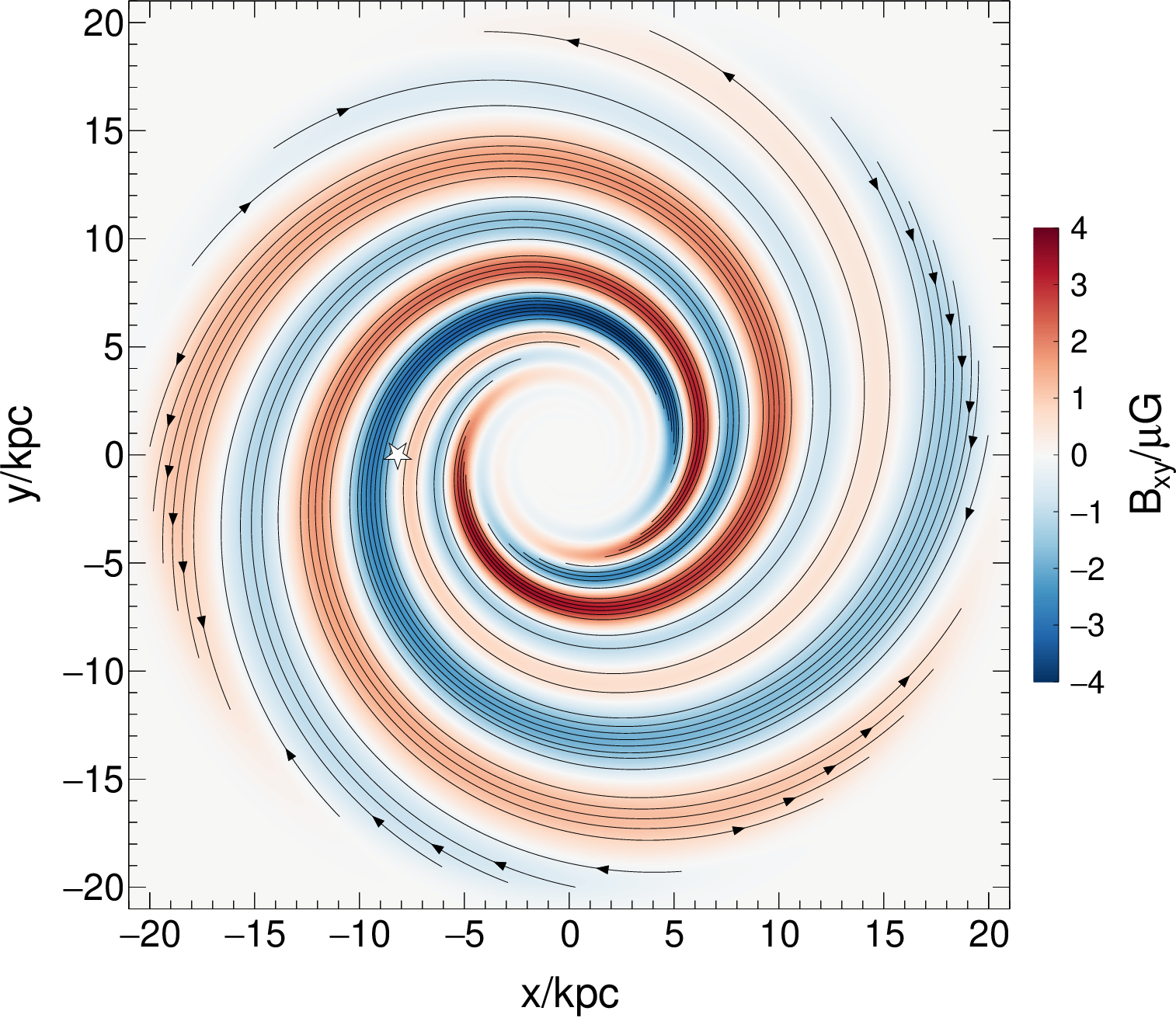}
    \end{center}
 \caption{\emph{Left}: Fixed-arm spiral disk field of the JF12
   model. \emph{Right}: Fourier spiral of the ``\modelNe'' model,
   see Sec.~\ref{sec:results}. The position of the sun is indicated
   with a star.}
 \label{fig:disk}
\end{figure*}

\subsection{General Considerations}
The goal of this work is provide the best possible analytic
approximation to the large scale coherent GMF, given the limitations
of present data and other required input.  To this end, we describe
the field as a superposition of functions that are general enough to
capture the main large scale features of its structure.

Inspired by previous models of the coherent Galactic magnetic field
\citep[e.g.][]{1997ApJ...479..290S,2003A&A...410....1P,2008A&A...477..573S,2011ApJ...738..192P,JF12coh}
we describe the global structure of the GMF as a superposition of a
large-scale halo field and a logarithmic spiral field in the Galactic
disk beyond a minimum radius. The halo field is composed of a toroidal
and a poloidal field as introduced by \citet{JF12coh}.  The need for
each of these components to describe large-scale features of the \RM
and \PI sky is discussed in Sec.~\ref{sec:results}.  In this section,
we give a brief overview of the different parametric descriptions
investigated in this paper and note ways in which the actual coherent
GMF is simpler than the most general case.

\subsection{Disk Field}
\subsubsection{General features}
The observations discussed above of synchrotron polarization patterns
in face-on external spiral galaxies, as well as the \RMs of Galactic
pulsars, suggest that the coherent field of the Milky Way follows a
spiral pattern close to the disk, outside a few kpc from the Galactic
center.  A logarithmic spiral is inherently divergence-free making it
an attractive functional form.

In several previous GMF models the magnetic field was assumed to be
organized in ``arms'' with a pre-defined geometry with the pitch angle
$\alpha$ fixed to a value motivated by that of the spiral matter
density of the Milky Way~\citep[see
  e.g.][]{bhg+07,2010MNRAS.401.1013J,JF12coh,2018ApJS..234...11H}. The
magnitude and sign within each arm was adjusted to match the data,
with the total radial flux being constrained to be zero. Within a
given arm, the field was approximated to be azimuthally constant,
leading to strong discontinuities at the boundaries of the arms.

For a fixed-thickness disk with logarithmic-spiral arms, flux
conservation implies the field $\sim r^{-1}$, leading to an unphysical
singularity at the origin.  Therefore the field in such models was
assumed to vanish or transition to a purely toroidal configuration at
some inner radius.

It should be noted that although the total radial flux vanishes for
both the outer logarithmic spiral and inner toroidal regions, the
radial flux does not vanish locally in the spiral arm region whereas
it does vanish locally in the inner region.  Therefore, the flux lines
must microscopically be reorganized in some transition region at the
boundary.  This structure is physically plausible due the mechanical
action of the Galactic bar, which extends several kpc from the
Galactic center and rotates significantly faster than the disk. The
turbulence in the plasma due to the stirring of the bar entangles
field lines which at larger radius are ordered. In the inner, stirred
region these field lines either contribute to $B_\text{rand}$ or
reconnect, converting magnetic energy to thermal energy.

Global radial flux conservation is compatible with a non-zero toroidal
coherent field in the inner zone, but in the picture above the
toroidal field strength would be small apart from fluctuations
amplified by dynamo action.  We allowed for a possible toroidal disk
field in the inner Galaxy and found it to be consistent with zero
(consistent with previous fits by \citet{JF12coh}) so we dropped it
from the modeling.  In the next subsections we describe our two basic
models for the disk field, and variants we also investigated.

\subsubsection{Fourier Spiral}
\label{ref:fourierspiral}

Here we introduce a new description of the spiral arms of the disk
field, that provides flexibility to fit the geometry of the magnetic
arms and also avoids discontinuities between them.  To specify the
spiral disk field, we need fix the pitch angle $\alpha$, the width and
angle of each arm at some reference radius, and the strength of each
arm.  For this purpose, we decompose the magnetic field strength at
reference radius $\rRefd=5$~kpc, as a function of angle $\phi_0$, into
$n$ modes of strength \Bdm and phase $\phi_m$:
\begin{equation}
  B(\rRefd, \phi_0) = \sum_{m=1}^n \Bdm\, \cos\left(m (\phi_0 - \Phidm)\right).
  \label{eq:fsmodes}
\end{equation}
The magnetic field in cylindrical coordinates at position $(r, \phi, z)$ is given by
\begin{equation}
  \bm{B}_\text{d} = \left(\sin \pitch, \cos \pitch, 0\right) \, \frac{\rRefd}{r}\, B(\rRefd, \phi_0)\,\fszfunc(z) \, \fsrfunc(r),
\label{eq:smoothspiral}
\end{equation}
where the $\phi_0$ corresponding to the given $(r, \phi)$ is found by
following the field line along the logarithmic spiral from $(r, \phi)$
to the reference radius $\rRefd$, the relation being
\begin{equation}
  \phi_0 = \phi - \ln(r/\rRefd) / \tan \pitch.
  \label{eq:phispiral}
\end{equation}
Due to the expansion of $B(\rRefd, \phi_0)$ being in terms of a cosine
series, we refer to this model of the disk field as ``Fourier spiral''
in the following.  The solenoidality of this field model for
$\fsrfunc(r) = 1$ is assured because for each mode in
\cref{eq:fsmodes} the same amount of flux enters and exits along the
circle at $\rRefd$ and, due to the $\frac{\rRefd}{r}$ factor in
\cref{eq:smoothspiral}, the magnetic flux in each arm is constant as a
function of radius.\footnote{Inclusion of an axisymmetric component of
  the $m=0$ component implies a net inward or outward flux of the disk
  field. This excess flux would necessarily flow into the halo and
  imply a net vertical flux, but we do not find evidence of that in
  the data, see Sec.~\ref{sec:poloidal}. Morover, including the
  component does not improve the fit significantly ($\Delta\chi^2 =
  -2.9$). \label{m0comment}} Independently of $\fsrfunc(r)$, the total
radial flux in this disk field model vanishes for every radius.

The functions $h_\text{d}(z)$ and $g_\text{d}(r)$ describe the fade-in
and fade-out of the field in the vertical and radial direction,
respectively. We choose the ansatz
\begin{equation}
  \fszfunc(z) = 1-\sigmoid\left(\frac{|z|-\fsz}{\fswz}\right)
 \label{eq:diskcutoffz}
\end{equation}
and
\begin{equation}
  g_\text{d}(r) =
  \left[1-\sigmoid\Big(\frac{r-\fsro}{\fswro}\Big)\right]\,
  \sigmoid\Big(\frac{r-\fsri}{\fswri}\Big)\left(1-e^{-r^2}\right) ,
  \label{eq:diskcutoffr}
\end{equation}
where $\sigmoid(x)$ denotes the logistic sigmoid function,
\begin{equation}
  \sigmoid(x) = \frac{1}{1+e^{-x}},
  \label{eq:sigmoid}
\end{equation}
such that the disk field is suppressed to half of its value at \fsz,
\fsri and \fsro, and the suppression rate is given by the
corresponding transition widths. The additional factor $(1-e^{-r^2})$ is
needed to assure that the factor $g_\text{d}(r)/r$ in
\cref{eq:smoothspiral} goes to 0 at $r=0$.

Note that mathematically $g_\text{d}(r)$ violates the solenoidality of
the field and is thus to be understood as an effective modeling of the
behavior of the disk field at small and large radii. At large radii we
expect that the coherent magnetic flux gradually spreads out as the
plasma confining it to the disk merges into the circumgalactic medium.
At some point the field is so weak and the electron densities so low
that our observables are not sensitive to it.  We represent this by an
effective outer distance \fsro. In the inner Galaxy, within a radius
designated \fsri, we expect that the coherent log spiral structure is
replaced by a region of low coherent field or possibly a weak toroidal
field, as discussed in the previous subsection.\footnote{Further
  possibilities include that the magnetic flux of the disk field exits
  vertically in the inner Galaxy \citep{2014A&A...561A.100F} (but see
  \footref{m0comment}) or that the inward- and outward-going field
  lines of the disk field connect at \fsri and \fsro as investigated
  by \citet{2019ApJ...877...76K}.  Both of these configurations entail
  regions of very high coherent field which would be energetically
  disfavored, so we did not pursue them.}

For consistency with the outer radius of the molecular ring in the
\YMW model, in this analysis we set \fsri to 5~kpc; we fixed \fsro to
20~kpc having checked that the fit is insensitive to the exact choice.
For the transition widths, we adopt $\fswri=\fswro=0.5$~kpc.  Whether
the details of the inner transition can be constrained by the data is
left for future work.  Without loss of generality, we set the
reference radius to $\rRefd=5$~kpc, such that the coefficients $B_m$
in \cref{eq:fsmodes} denote the amplitude of the modes in the inner
Galaxy.

An example of a Fourier spiral disk field is shown in the right panel
of \cref{fig:disk} and for comparison a fixed-arm spiral field is
displayed in the left panel. Apart from obvious differences in the arm
topologies and field strengths, which are mostly due to different data
sets used to fit these models, it can be seen that the Fourier spiral
results in a smooth disk field without the discontinuities inherent in
the previous fixed-arm models.

\subsubsection{Spiral Spur}
\label{sec:locspure}
As an alternative model to the grand-design magnetic spiral implied by
the Fourier spiral, we investigate if a localized spiral segment can
describe the data equally well. Such an isolated magnetic ``spiral
spur'' in the disk is seen e.g.\ in Fig.\ 2 of the cosmological simulation
of \citet{Pakmor:2013rqa}, due predominantly to a local compression of
field lines.

We model a spur
as a Gaussian of width \wS at reference radius $\rRefd$
\begin{equation}
  B(\rRefd, \phi_0) = \BLS\, \exp\left(-\frac{1}{2} ((\phi_0-\phiS)/\wS)^2\right),
  \label{eq:spurB}
\end{equation}
where \phiS denotes the center of the spur and
 the angle $\phi_0$ follows again from the logarithmic spiral via \cref{eq:phispiral}. The field of the spur in cylindrical coordinates is, similar to \cref{eq:smoothspiral}, given by
\begin{equation}
  \bm{B}_\text{S} = \left(\sin \pitch, \cos \pitch, 0\right) \, \frac{\rRefd}{r}\, B(\rRefd, \phi_0)\,\fszfunc(z) \, g_\text{S}(\phi),
\label{eq:localspur}
\end{equation}
where instead of using the radial attenuation $\fsrfunc(r)$, the size
of the spur is determined from its angular center $\phi_\text{C}$ and
angular half-length \lCS. We attenuate the magnetic field
at $\phi = \phiCS\pm \lCS$ with
\begin{equation}
  g_\text{S}(\phi) = 1 - \sigmoid\left(\frac{\Delta(\phi,\phiCS)-\lCS}{\wCS}\right).
\end{equation}
In the following, we use a fixed attenuation width of $\wCS = 5^\circ$
and, without loss of generality, $\rRefd=8.2$~kpc, such that the
parameter $B_{0,\text{S}}$ is close to the magnetic field strength at
the center of the spur, if it is located in the proximity of the solar
radius, which is the main focus of the spiral spur model.  The
superficial lack of flux conservation is understood as a transition
between locally compressed and a more complex, broadly-distributed
flux distribution not captured in the global model function.

\subsubsection{Further Considerations Regarding the Disk Field}
\label{sec:furtherDisk}
Further variations to the Fourier spiral and spiral spur models were
studied, but not included in the fiducial models presented in this
paper. We investigated a circular ``ring field'' at small
galacto-centric radii. Such a field was introduced
in~\citetalias{JF12coh} at $3~\text{kpc} < r <5~\text{kpc}$ with an
estimated field strength of $0.1\pm0.1$~\muG. Here we confirmed that
a ring field does not significantly improve our fits and
therefore we did not include it in our fiducial models.

We also investigated the sensitivity of the data to the particular
choice of $\fszfunc(z)$ and found no significant changes when
replacing the logistic sigmoid with a Gaussian ($\Delta\chi^2=-3$) and
a slightly worse description when using an exponential
($\Delta\chi^2=+30$).

Furthermore, we studied a flaring disk field, i.e.\ an increase of the
vertical extent of the disk with galacto-centric radius as
e.g.\ observed for the atomic hydrogen of the
disk~\citep{2009ARA&A..47...27K}. We also added the Galactic warp as
determined by~\citet{2006ApJ...643..881L} to the disk field model.
Neither of the two variations improved our preliminary fits reported
in \citep{Unger:2019xct} and we therefore do not include them in this
analysis.

A model variant we did not explore here is to allow a net inward or
outward flux in the disk which is balanced by flux entering the disk
due to an imbalance in the N-S halo fields; this was checked
by~\citet{deepak} for the \citetalias{JF12coh} model and the flux
transfer was found to be consistent with zero, compatible with our
findings on the symmetry of the halo field as reported in
Sec.~\ref{sec:poloidal} below.

\subsection{Halo Field}

Our knowledge of the global structure of the magnetic halo of Milky
Way-like galaxies relies on a combination of edge-on observations of
external galaxies and high-latitude rotation measures and radio
emission~\citep[for a review see, e.g.,][]{2012SSRv..166..133H}.

The large-scale antisymmetric features of extragalactic rotation
measures, cf.~\cref{eq:rmantisym} and \cref{fig:maskMap1}, follow
naturally from a large-scale toroidal
\footnote{Here we use the term toroidal field synonymously with a
  purely azimuthal field, $\bm{B} = (0, B_\varphi, 0)$, while the term
  poloidal field denotes a vector field with a vertical but not an
  azimuthal $\varphi$-component, $\bm{B} = (B_r, 0, B_z)$.} field of
opposite sign above and below the plane, leading to the observed
pattern in RM when superimposed with the local spiral arm of the disk
field, schematically
\begin{equation}
  \underset{\RM_\text{tot}}{\mathcircled{\!\!\frac{-+-+}{--++}\!\!}} = \underset{\text{toroidal halo}}{\mathcircled{\!\!\frac{++--}{--++}\!\!}} + \underset{\text{local arm}}{\mathcircled{\!\!\frac{-\phantom{+-}+}{-\phantom{-+}+}\!\!}}   \,.
  \label{eq:rmcomponents}
\end{equation}
In addition, the large-scale patterns of the \Q and \U parameters, in
particular the tilted nature of the corresponding polarization
vectors, can be explained by the presence of an addition poloidal halo
component, as suggested by \citet{JF12coh} who introduced a coherent
``X-field'' halo component inspired by the X-shaped radio polarization
halos observed in edge-on spiral
galaxies~\citep[e.g.][]{2020A&A...639A.112K}.

Here we will model the magnetic halo of the Galaxy either by a
superposition of a separate toroidal and poloidal field, or by a
unified halo model, as described below.

\begin{figure*}[!th]
  \def\figh{0.25}
  \centering
  \includegraphics[clip,rviewport=0 0 0.87 1, height=\figh\textheight]{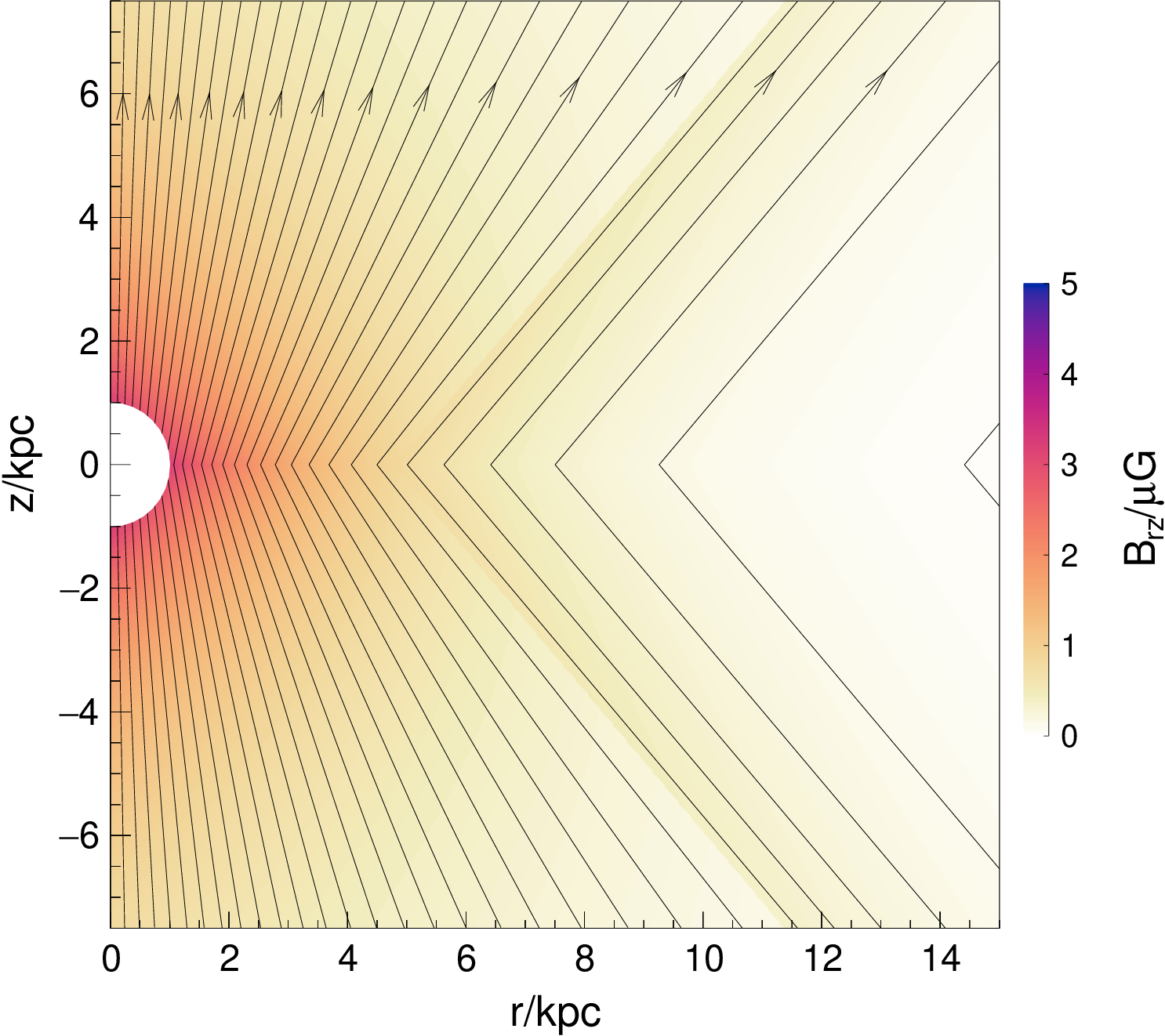}\quad%
  \includegraphics[clip,rviewport=0.085 0 0.87 1,height=\figh\textheight]{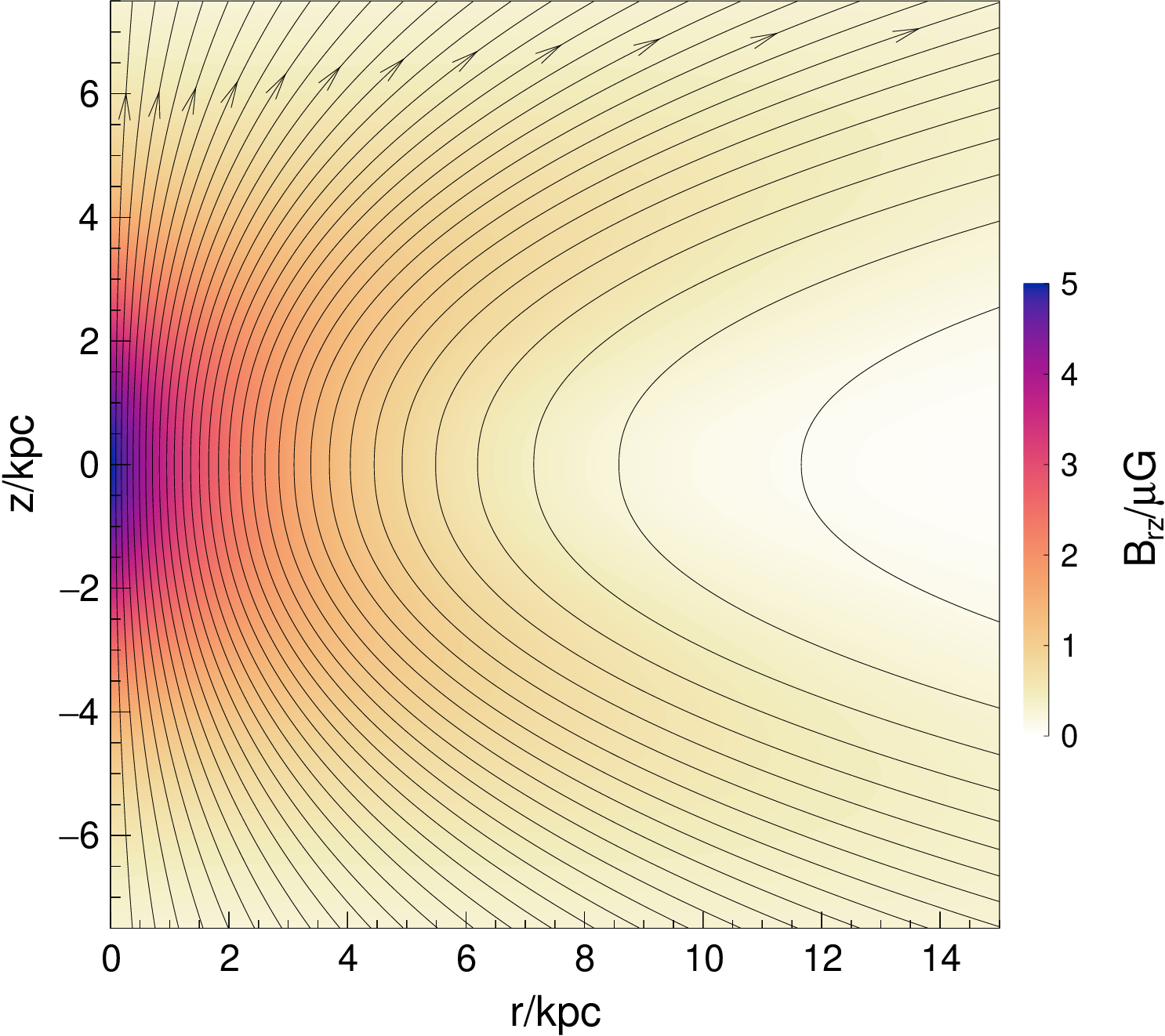}\quad%
  \includegraphics[clip,rviewport=0.085 0 1 1,height=\figh\textheight]{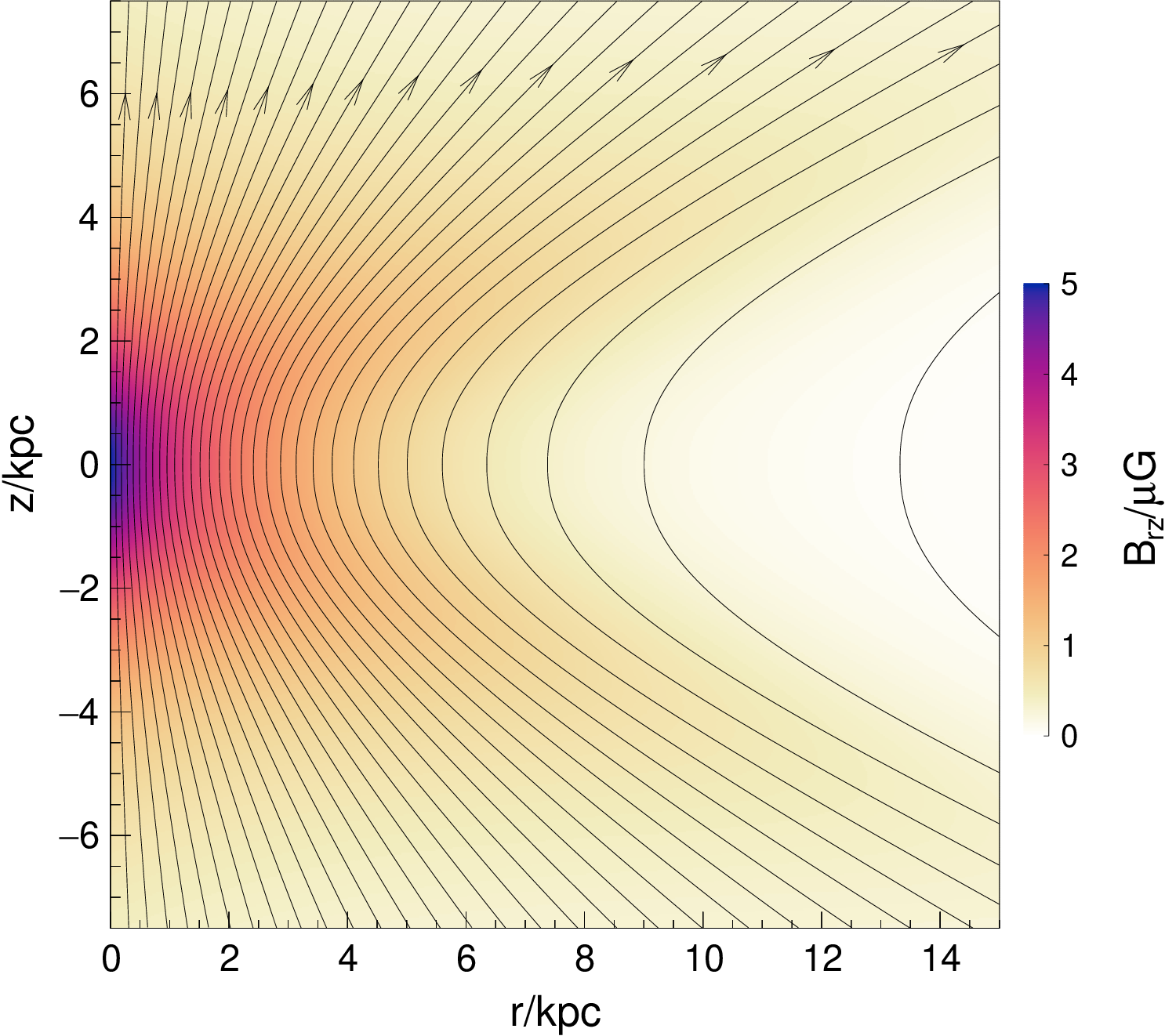}
  \caption{X-field models: JF12 (left), power-function (middle) and coasting X-field (right).}
  \label{fig:xfield}
\end{figure*}
\subsubsection{Toroidal Halo}
For the toroidal field, we adopt the logistic-exponential
ansatz (``\JFTor'') of \citetalias{JF12coh}. It is a purely azimuthal field,
\begin{equation}
  \bm{B}_\text{t} = (0, B_\phi, 0)
\end{equation}
in cylindrical coordinates, with
\begin{multline}
  B_\phi = \left(1-\fszfunc(z)\right) \, e^{-\frac{|z|}{\zT}} \\ \times \left(1-\sigmoid\left(\frac{r-\rT}{\wT}\right)\right)
  \begin{cases}
    \BNT & z > 0 \\
    \BST & \text{otherwise} \\
  \end{cases}
  \label{eq:toroidal}
\end{multline}
where \zT is the exponential scale height of the toroidal
field and the radial cutoff of the toroidal halo is modeled by a
logistic sigmoid function with a transition at \rT and a
width \wT. The maximum field strength above and below the
Galactic plane are given by \BNT and \BST,
respectively; \fszfunc is defined in \cref{eq:diskcutoffz}, i.e.,\
the toroidal halo field is phased in by the complement of the function
that phases out the disk field.

Whereas in~\citetalias{JF12coh} the radial extent of the Northern and
Southern halo was allowed to be different, we use only one value
\rT, since our preliminary fits showed only a minor
deterioration of the fit quality when enforcing a symmetric
toroidal~\citep{Unger:2017wyw}. We also checked whether the data is fit
better with independent functions \fszfunc(z) for the disk and the
halo separately, but found similar transition heights for the disk and
halo, even if they were allowed to take on different values.

\subsubsection{Poloidal Halo}

\citet{JF12coh} introduced a coherent ``X-field'' model that is purely
vertical at $r=0$ and becomes increasingly tilted with increasing
radius until reaching a constant asymptotic angle with respect to the
Galactic plane, $\theta_X \sim 50^\circ$, at a radius of
$r_X\sim5$~kpc. With this model a good description of the polarized
synchrotron data could be achieved. However, this X-field
parametrization has three types of discontinuities: one in the inner
Galaxy, one at the Galactic plane at $z=0$ and one at $r=r_X$, as can
be seen in the left panel of \cref{fig:xfield}. These discontinuities
are avoided by the improved X-field models described in the
following.

\paragraph{A) Power-function X-field}~\citet{2014A&A...561A.100F} employed
a useful method to construct poloidal field models using Euler
potentials. Given the equation of field lines in the form of
$r=f(a,z)$ starting midplane at a radius $a$ (i.e.\ $f^{-1}(r=a) =
0$), the divergence-free cylindrical components of the poloidal field
($B_\phi=0$) are given by
\begin{equation}
  B_r = - \frac{a}{r}\, B_0(a)\, \frac{\partial a}{\partial z}
  \label{eq:ftr}
\end{equation}
and
\begin{equation}
  B_z = \frac{a}{r} \, B_0(a) \, \frac{\partial a}{\partial r}
 \label{eq:ftz}
\end{equation}
where any function $B_0(a)$ of the radial dependence of the
$z$-component of the field at $z=0$ will preserve the solenoidality of
the field.

We extend their parabolic ``model C'' for the
$z$-evolution of the field lines to arbitrary powers $p$,
\begin{equation}
 r = a \, (1 + |z/\zP|^p),
\end{equation}
corresponding to a midplane radius of
\begin{equation}
 a(r, z) = \frac{r}{1+|z/\zP|^p}.
\end{equation}
Using Eqs.~(\ref{eq:ftr}) and (\ref{eq:ftz}), the radial and vertical field
components are
\begin{equation}
   B_r = p\, \frac{z\, a^3}{r^2 \, \zP^2} \,|z/\zP|^{p-2}\, B_0(a)
\end{equation}
and
\begin{equation}
   B_z = \frac{a^2}{r^2}\, B_0(a)
\end{equation}
for $p\geq1$. In Sec.~\ref{sec:results} we will study several
possibilities for the radial midplane dependence of the field
strength,
\begin{equation}
   B_0(a) = \BP\,f_\text{X}(a),
\end{equation}
with normalization constant $\BP$ and one of the following radial functions:
\begin{itemize}
  \item exp: $f_\text{X}(a) = e^{-\frac{a}{\rP}}$, \inlineeqnum \label{eq:bzexp}
  \item gauss: $f_\text{X}(a) = {e^{-\frac{1}{2}(\frac{a}{\rP})^2}}$, \inlineeqnum \label{eq:bzgauss}
  \item sech2: $f_\text{X}(a) = \sech^2(a/\rP)$, \inlineeqnum \label{eq:bzsec}
  \item logistic: $f_\text{X}(a) = 1-\sigmoid((a-\rP)/\wP)$, \inlineeqnum \label{eq:sigm}
\end{itemize}
where $\sech$ denotes the hyperbolic secant, $\sech(x) = 1 / \cosh(x)
= \nicefrac{2}{(e^x + e^{-x})}$ and $\sigma$ is the logistic sigmoid
function defined in~\cref{eq:sigmoid}.

An example of a power-function X-field is shown in the middle panel of
\cref{fig:xfield} (for $p=2.2$ and an exponential radial dependence
with $\rP=2.8$~kpc). As can be seen, the discontinuities present in
the original \citetalias{JF12coh} X-field are avoided, but it differs
qualitatively from the \citetalias{JF12coh} X-field in that the field
lines become more and more parallel to the Galactic plane as the
radius increases rather than reaching an asymptotic angle.

\paragraph{B) Coasting X-field} ~A ``coasting X-field'' with parallel
field lines beyond a certain reference radius $\aP$ can be achieved by
choosing the field-line equation
\begin{equation}
  r = \left(a^n + c \, \frac{a^n}{a^n + \aP^n} \, |z|^p\right)^\frac{1}{n}.
  \label{eq:rcoast}
\end{equation}
The positive solution for $a^n$ is\footnote{Note that $\sqrt{x^2 + k}
  - x$ is of low numerical accuracy if $k\ll x$.  Instead we use
  $(\sqrt{x^2 + k} - x) \frac{\sqrt{x^2 + k} + x}{\sqrt{x^2 + k} + x}
  = \frac{k}{\sqrt{x^2 + k} + x}$ in our numerical implementation.}
\begin{equation}
\begin{aligned}
   a^n = \frac{1}{2}\Big(&\sqrt{\Delta^2 + 4 \aP^n r^n} -\Delta\Big),
\end{aligned}
\end{equation}
where $\Delta = \aP^n + c |z|^p - r^n$. Inserting it into
Eqs.~(\ref{eq:ftr}) and (\ref{eq:ftz}), the radial and vertical field
components are obtained as
\begin{equation}
  B_r = B_0(a) \, \frac{c\,a^2\,p\,z\,|z|^{p-2}}{n\,r\,\sqrt{\Delta^2 + 4 \aP^n r^n}}
  \label{eq:coastingbr}
\end{equation}
and
\begin{equation}
  B_z = B_0(a) \, \left(\frac{r}{a}\right)^{n-2}\frac{a^n + \aP^n}{\sqrt{\Delta^2 + 4 \aP^n r^n}}.
  \label{eq:coastingbz}
\end{equation}
In this paper, we study X-fields for the special case $p=n$ for which \cref{eq:rcoast} becomes
\begin{equation}
  r = a \left[1 + \frac{1}{1 + (a/\aP)^p} \, \left(\frac{|z|}{\zP}\right)^p\right]^\frac{1}{p},
  \label{eq:rcoast2}
\end{equation}
and Eqs.~(\ref{eq:coastingbr}) and (\ref{eq:coastingbz}) simplify
accordingly.  The new parameter $\zP$ is given by $\zP =
\frac{\aP}{c^{1/p}}$.  For large values of the coasting radius $\aP$
($a\ll \aP$)
this three-parameter function simplifies further to the two-parameter
equation
\begin{equation}
  r = a \left[1 + \left(\frac{|z|}{\zP}\right)^p\right]^\frac{1}{p}.
  \label{eq:rcoast3}
\end{equation}
An example of a coasting X-field with $p=nx=2.2$ and $\aP = 7$~kpc is
shown in the right panel of \cref{fig:xfield}.

\begin{figure*}[t!]
  \centering
  \def\figw{0.31}
  \subfloat[$t=0$~Myr]{
  \includegraphics[width=\figw\linewidth]{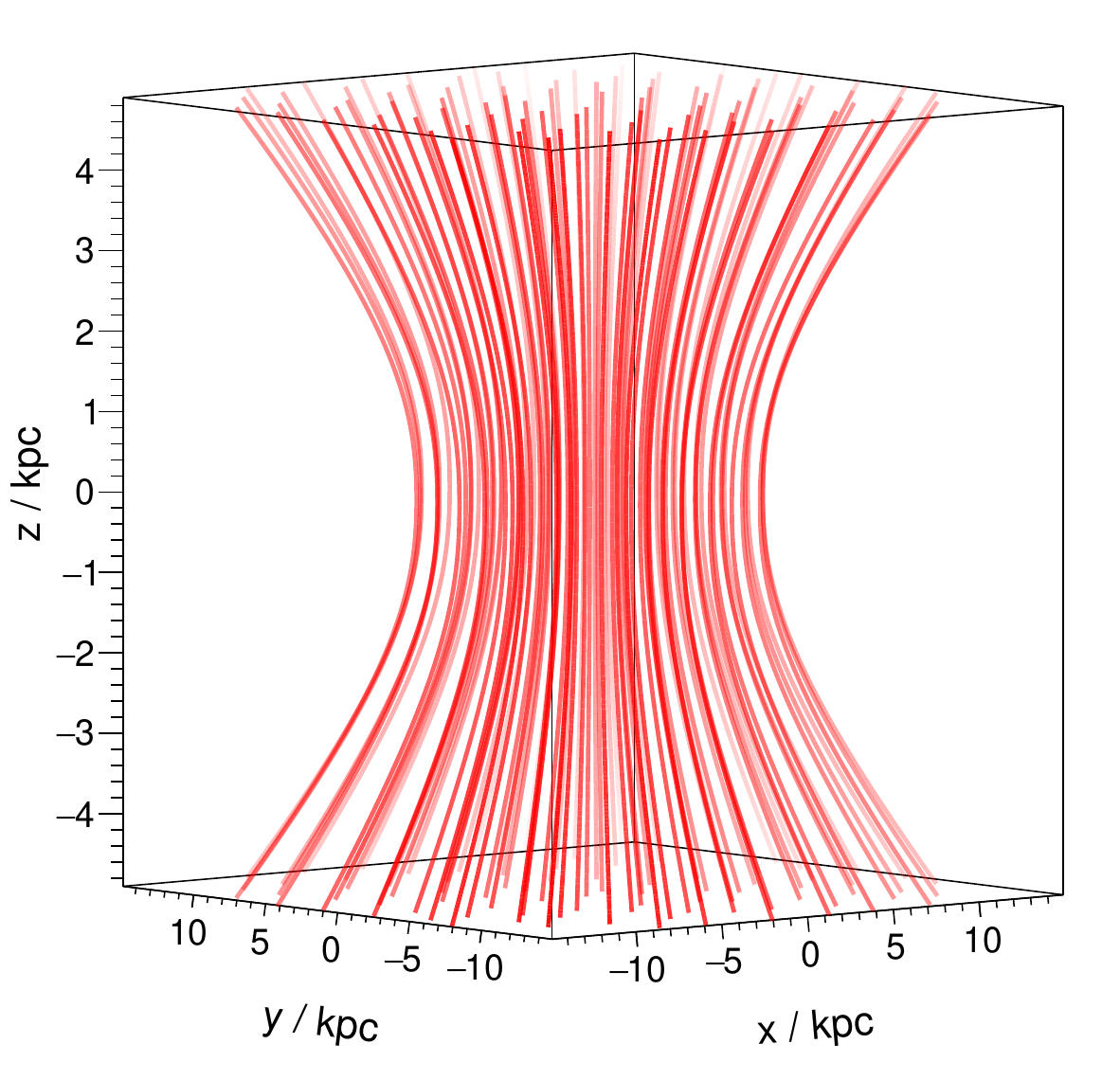}
  \label{fig:twist0}
  }%
  \subfloat[$t=25$~Myr]{
  \includegraphics[width=\figw\linewidth]{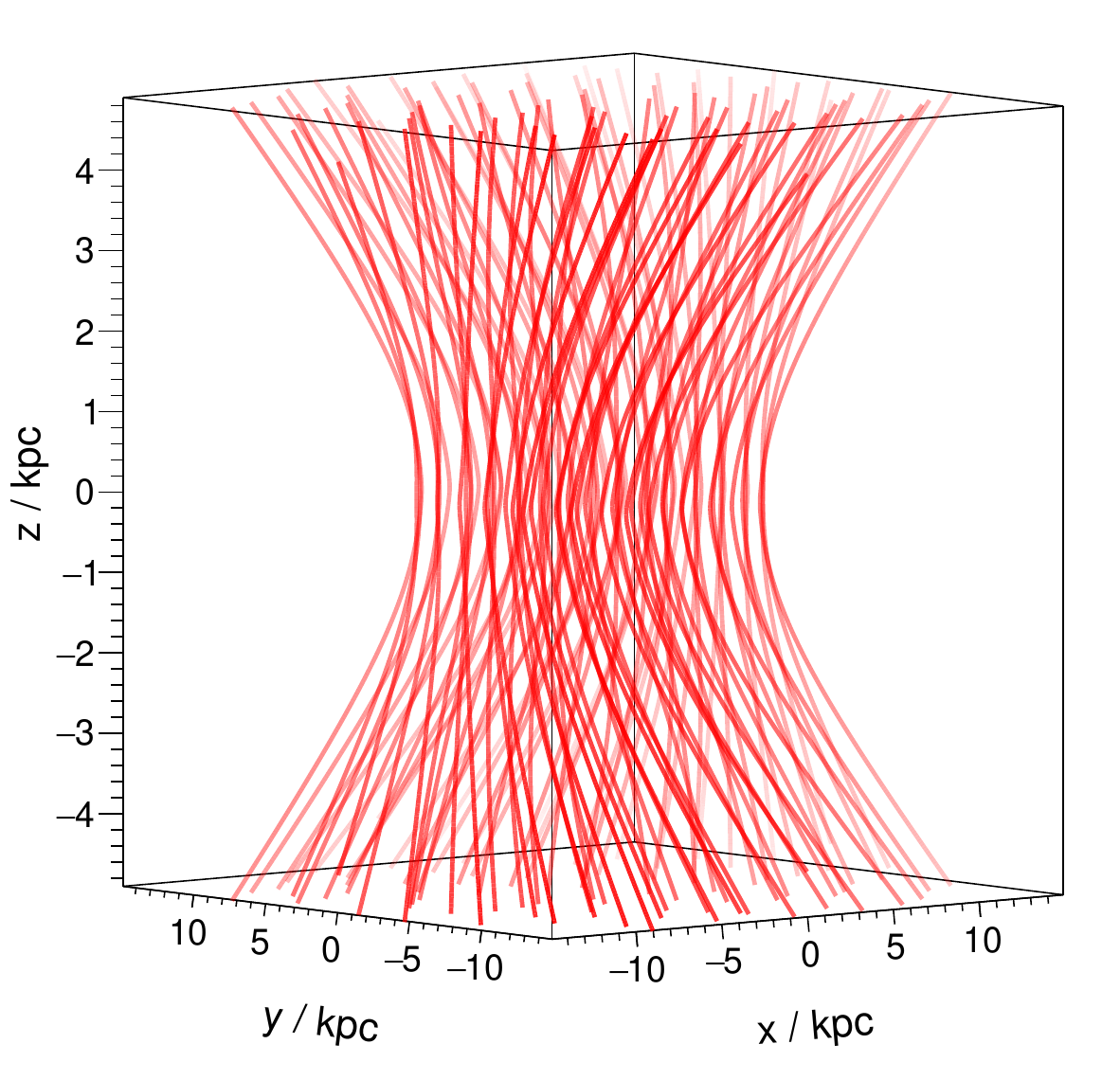}
  \label{fig:twist25}
  }%
  \subfloat[$t=50$~Myr]{
  \includegraphics[width=\figw\linewidth]{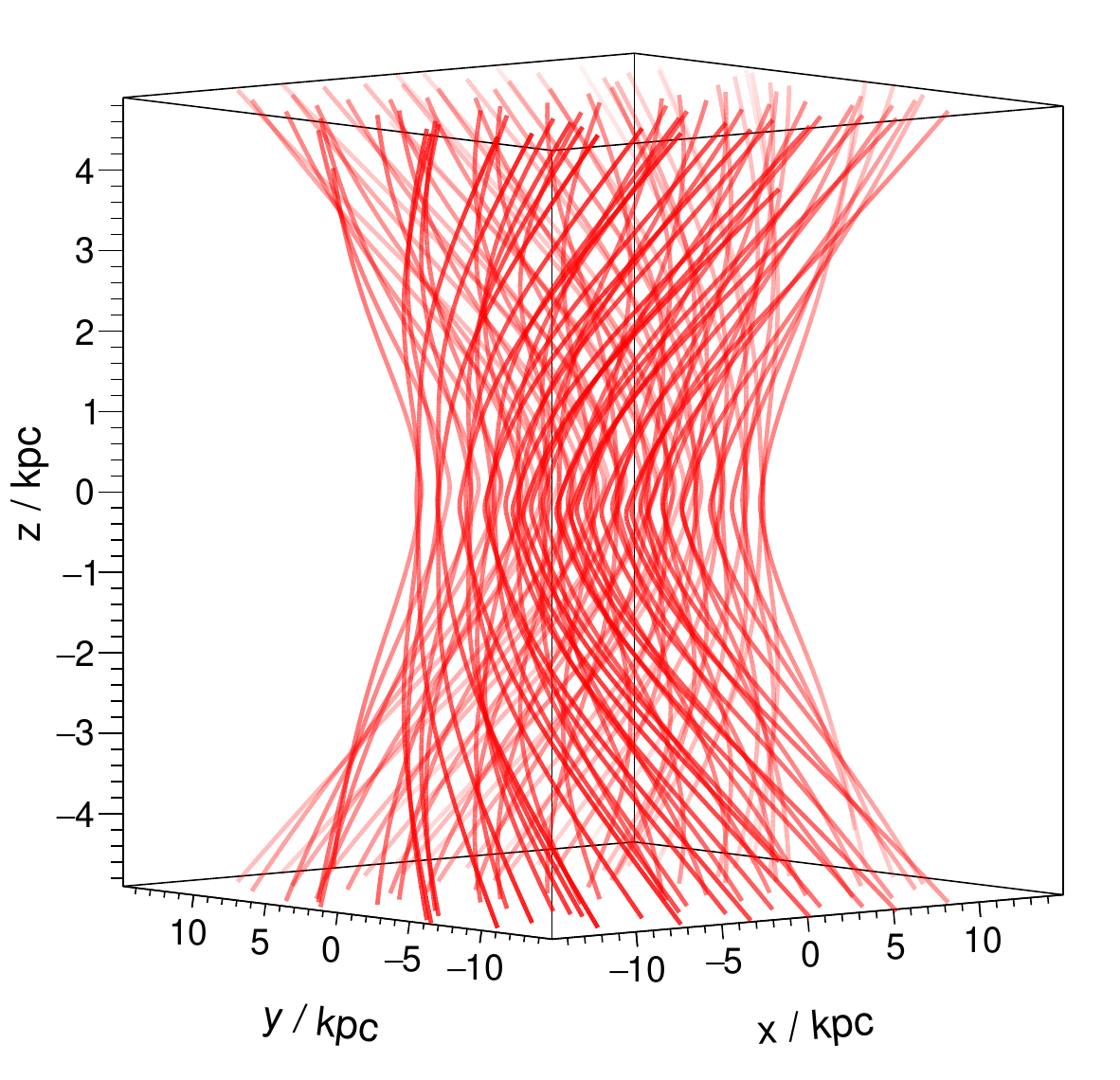}
  \label{fig:twist50}
  }
  \caption{Illustration of the unified halo model at
    different times $t=0$, 25 and 50 Myr.}
\label{fig:twisted}
\end{figure*}

\subsubsection{Unified Halo Model}
\label{sec:xfield}
A toroidal halo field with different directions in the Northern and
Southern hemisphere can be the result of differential rotation of a
poloidal halo
field~\citep[e.g.][]{1988Ap.....28..247A,2003AcASn..44S.151M}.
\citet{Farrar:2014hma} pointed out that -- given the sign of the
dipolar field discovered by~\cite{JF12coh} -- the differential
rotation of the Galaxy would create toroidal halo fields with the
observed directions in the Northern and Southern hemispheres.  However
the question of whether differential rotation can explain the observed
toroidal field strength quantitatively, including its vertical and
radial profiles, was not addressed.  Here, we show that the toroidal
field is in fact remarkably well described by the simplest possible
model based on the observed differential rotation. The success of this
initial simple treatment gives encouragement that a more fully
developed treatment of the effects which limit the buildup of the
toroidal field, canl eventually enable a physics-based model of the
halo field to replace the ad-hoc fitting approach which has been
required up to now.

Consider a poloidal field that is dragged along with the rotation of
the Galaxy and evolves via the induction equation,
\begin{equation}
  \partial_t \bm{B} = \bm{\nabla} \times (\bm{v} \times \bm{B})
      - \underbrace{\bm{\nabla} \times \eta (\bm{\nabla} \times \bm{B})}_{\rightarrow0\;{\rm for}\;\sigma\rightarrow\infty},
\end{equation}
where the ``frozen-in condition'' applies for perfect conductivity,
$\sigma\rightarrow\infty$, for which the magnetic diffusivity
vanishes, $\eta \varpropto 1/\sigma \rightarrow
0$~\citep[e.g.][]{1979cmft.book.....P}. Under these conditions and for
a purely azimuthal rotation velocity~$v$ the induction equation
simplifies to
\begin{equation}
\partial_t \bm{B} =   \bm{\nabla} \times (\bm{v} \times \bm{B}) =  \begin{pmatrix}
   -\frac{v}{r} \,\partial_\phi B_r \\
    \partial_z (v B_z) + \partial_r (v B_r) \\
    -\frac{v}{r} \, \partial_\phi B_z
\end{pmatrix}.
\end{equation}
Thus, for a magnetic field that is poloidal and azimuthally symmetric
at $t=0$ ($\partial_\phi B_r = \partial_\phi B_z = 0$), the poloidal components
are constant and only $B_\phi$ evolves with time,
\begin{equation}
 B_\phi(t) = \left(B_z\, \partial_z v + rB_r \,\partial_r \omega\right)\, t,
\label{eq:evolution}
\end{equation}
where we introduced the angular velocity $\omega = \frac{v}{r}$ and
used the solenoidality of the poloidal field.

Eq.~(\ref{eq:evolution}) can be applied to evolve any type of poloidal
field analytically and it describes the linear increase of $B_\phi$
due to the differential motion of the plasma the magnetic field is
embedded in.
The large-scale motion of the plasma in the Galaxy follows the
Galactic rotation curve, which we take to approximately factorize as
\begin{equation}
  v(r, z) = \vZero \, f(r) \, g(z).
  \label{eq:rotcurve}
\end{equation}
Using
\begin{equation}
  f(r) = 1-e^{-r/\rV} \equiv \rotRadialFunc,
\end{equation}
we adjust \vZero and \rV to match the velocities of high-mass
star-forming regions with parallax measurements
from~\citet{2014ApJ...783..130R}, leading to $\vZero = -240$~km/s
(negative because the Galaxy rotates clock-wise) and
$\rV=1.6$~kpc. For the vertical velocity gradient we use
\begin{equation}
 g(z)  =  2 / (1 + e^{\,2\,|z| / \zV}) \equiv \rotVerticalFunc
\end{equation}
with $\zV=10$~kpc to reproduce $\partial_z v = (-22\pm6)$ (km/s)/kpc
as measured within 100~pc of the Galactic midplane
by~\citet{2008ApJ...679.1288L}.\footnote{The toroidal field in the
  unified model is relatively insensitive to the uncertainty in
  $g(z)$, since the radial and vertical gradients contribute with the
  same sign to the twisting, contrary to the result reported in
  \citep{2003AcASn..44S.151M}. \citet{2017A&A...600A..29T} explored a
  twisted halo model introducing a generic ``winding function''
  $g_\phi(r, z)$ of the field lines. Their Eq.~(27) is equivalent to
  Eq.~(\ref{eq:evolution}) for the choice $g_\phi = t\,
  v(r,z)/r$. However, they did not investigate this explicit
  connection of the winding function to the velocity field $v(r,z)$ of
  the Galaxy, but used an ad-hoc parametric winding function.}

With this particular choice of \rotRadialFunc and \rotVerticalFunc, the
evolution of the azimuthal field via \cref{eq:evolution} is
\begin{equation}
 B_\phi(t) = \left(B_z\, \Delta_z + B_r \, \Delta_r\right)\, t,
\end{equation}
with the two shear terms
\begin{equation}
\begin{aligned}  \Delta_z = \partial_z v = - \sign(z)\, \frac{\vZero}{\zV}\, \rotRadialFunc\,\rotVerticalFunc^2 \, e^{\,2\,|z| / \zV}
\end{aligned}  \label{eq:twistedDeltaZ}
\end{equation}
and
\begin{equation}
   \Delta_r = r \,\partial_r \omega = \vZero\,\left(\frac{1 - \rotRadialFunc}{\rV}- \frac{\rotRadialFunc}{r}  \right)  \rotVerticalFunc~.
  \label{eq:twistedDeltaR}
\end{equation}

An example of a unified halo field is shown in \cref{fig:twisted}
starting at $t=0$ with a poloidal coasting X-field with $\BP>0$ and a
logistic sigmoid cutoff at $r=8$~kpc.  Due to the radial and vertical
shear of the rotation curve, \cref{eq:rotcurve}, an azimuthal field is
created at $t>0$ that has a different sign in the Northern and
Southern hemisphere to which both terms in Eq.~(\ref{eq:evolution})
contribute constructively. Obviously, this process cannot continue in
this na\"ive form over the lifetime of the Galaxy, or it would
overproduce the observed azimuthal field strength.  A steady-state
azimuthal field could be obtained by including a suitable dissipative
term in the induction equation and possibly a source term from the
$\alpha$-effect of dynamo theory~\citep[e.g.][]{2023ARA&A..61..561B},
but for the purpose of this paper, we interpret the best-fit twisting
time as an effective parameter of this simplest version of a unified
halo model.\\

%% file: glossary.tex
In summary, in this section we introduced two variants of the disk
field (grand spiral or spiral spur), two variants of the poloidal field
(power-function or coasting) and two variants for the toroidal halo
(explicit or from twisting).  These magnetic field sub-models as well
as their parameters are listed in Table~\ref{tab:glossary} together with a
few other model parameters.

\begin{deluxetable}{cclcc}[t]
  \tablecaption{List of parameter names of different model components.}
\label{tab:glossary}
\tabletypesize{\footnotesize}
\tablecolumns{5}
\tablewidth{0pt}
\tablehead{\multicolumn{2}{l}{name}&\colhead{explanation}&\colhead{unit}&\colhead{value}}
\startdata
\multicolumn{5}{l}{{\bf disk field}}\\
\multicolumn{5}{l}{{common parameters}}\\
&\pitch  & pitch angle & deg. & free \\
&\fsz    & transition height & kpc & free \\
&\fswz   & vertical transition width & kpc & free  \\
\multicolumn{5}{l}{{a) grand-design spiral}, Eqs.~(\ref{eq:smoothspiral}) and (\ref{eq:fsmodes})} \\
&\Bdm    & magnetic field strength of mode $m$ & \muG & free \\
&\Phidm  & phase of mode $m$ & deg. & free \\
&\rRefd  & reference radius & kpc & 5 \\
&\fsri   & inner radius & kpc & 5 \\
&\fswri  & inner radial transition width & kpc & 0.5 \\
&\fsro   & outer radius & kpc & 20\\
&\fswro  & outer radial transition width & kpc & 0.5 \\
\multicolumn{5}{l}{{b) spiral spur}, Eq.~(\ref{eq:spurB})}\\
&\BLS    & magnetic field strength at \rRefd & \muG & free \\
&\phiS   & azimuth at \rRefd & deg. & free \\
&\wS     & Gaussian width  & deg. & free \\
&\phiCS  & central azimuth & deg. & free \\
&\lCS    & half angular length  & deg. & free \\
&\rRefd  & reference radius & kpc & 8.2 \\
&\wCS    & transition width & deg. & 5 \\ \hline
\multicolumn{5}{l}{{\bf toroidal halo}}\\
\multicolumn{5}{l}{{a) explicit}, Eq.~(\ref{eq:toroidal})}\\
&\BNT    & Northern magnetic field strength & \muG & free \\
&\BST    & Southern magnetic field strength & \muG & free \\
&\zT     & vertical scale height & kpc & free \\
&\rT     & transition radius & kpc & free \\
&\wT     & radial transition width & kpc & free \\
\multicolumn{5}{l}{b) twisted, Eq.~(\ref{eq:evolution})}\\
&$t$& twisting time & Myr  & free \\
&\vZero & Galactic rotation velocity & km/s & -240 \\
&\rV & scale radius of rotation curve & kpc & 1.6 \\
&\zV & scale height of rotation curve & kpc & 10 \\\hline
\multicolumn{5}{l}{{\bf poloidal halo}}\\
\multicolumn{5}{l}{{common parameters}}\\
&\BP & magnetic field strength & \muG & free \\
&$p$ & field line exponent & -- & free \\
&\zP & scale height & kpc & free \\
&\rP & radial scale or transition radius & kpc & free \\
&\wP & transition width & kpc & free \\
\multicolumn{5}{l}{{a) power-function}, Eqs.~(\ref{eq:ftr}) and (\ref{eq:ftz})} \\
\multicolumn{5}{l}{{b) coasting}, Eqs.~(\ref{eq:coastingbr}) and (\ref{eq:coastingbz})} \\
&\aP & coasting radius & kpc & free \\ \hline
\multicolumn{5}{l}{{\bf other model parameters}}\\
&$\kappa$ & $\nel$-$B$ correlation coefficient, \cref{eq:rmkappa}  & -- & free  \\
&$\xi$& magnetic striation factor  \cref{eq:striation} & -- & free\\
\enddata
\end{deluxetable}

%% file: analysis.tex
The parameters $\bm{p}$ of a GMF model are optimized by minimizing
the sum of the variance-weighted squared difference between
the modeled $m$ and measured $d$ observables,
\begin{equation}
\begin{aligned}
 \chi^2 =&  \sum_{i = \scalebox{0.6}{\RM,\Q, \U}} \; \sum_{j=1}^{N_i} \frac{(d_{ij} - m_{ij}(\bm{p}))^2}{\sigma_{ij}^2},
 \label{eq:chi2}
\end{aligned}
\end{equation}
where the second sum runs over the $N_j$ lines of sight available for
the RM and synchrotron data. Each line-of-sight datum $d_{ij}$ is
obtained by averaging available measurements over a finite solid angle
around the line-of-sight direction. These angular {\itshape pixels}
are defined by the \HEALPix resolution of the data. The predicted
values $m_{ij}$ of a GMF model, $\bm{B}(\bm{x}; \bm{p})$, are given by
the numerical evaluation of the line-of-sight integrals for \RM, \Q
and \U.  Even for a perfect GMF model, the differences $d_{ij}-m_{ij}$
are expected to be distributed with a variance of $\sigma_{ij}^2$,
since the data is subject to experimental uncertainties and the model
prediction do not include ``Galactic variance'' originating from
random magnetic fields and from fluctuations of the densities of
thermal electrons and cosmic-ray electrons. In general, Galactic
variance will introduce correlations between adjacent pixels if the
size of the perturbations is larger than the angular size of one
pixel. In that case, the full covariance matrix needs to be included
in the calculation of the $\chi^2$. It is, however, non-trivial to
calculate the coefficients of the matrix, because one needs to know
the spatial distribution of the turbulence and coherence length for
the magnetic field and the thermal and cosmic-ray densities. Different
approximations have been applied in previous
analyses. \citet{2010MNRAS.401.1013J} and \citet{2016A&A...596A.103P}
used a model of the three-dimensional random field strength with a
constant coherence length and unperturbed electron densities to
calculate the diagonal elements of the covariance matrix. In a more
data-driven approach, one can measure the sub-pixel variance
$\hat{\sigma}^2$ of the measurements within one pixel and use it as a
weight in the fit. Na\"ively, the average over $N$ observations within
one pixel would then reduce the fluctuations of the mean by a factor
of $1/N$, but due to the aforementioned coherent effects, the
effective variance will typically be larger. Based on a toy model of
coherent cells along the line of sight, \citet{2011ApJ...738..192P}
suggested to use $\hat{\sigma}^2/\delta^2$ as the pixel variance, with
$\delta\approx3$, but the exact value of $\delta$ depends on the
integration distance in units of coherence length over which the GMF
contributes to the observations (see
also~\citep{2017A&A...600A..29T}). Here we follow the procedure
of \citetalias{JF12coh} and simply use $\hat{\sigma}^2$ itself to
weight the data points.

In this paper, the $\chi^2$ as defined in Eq.~(\ref{eq:chi2}) is used
to optimize the parameters of a given model and to assess the relative
quality of different models.  The interpretation of the $\chi^2$ value
in terms of goodness of fit is less meaningful.  A robust
understanding of the statistical fluctuations in the data and their
correlations is needed for that purpose. These can only be assessed
after the random field and its coherence length have been determined,
which will be the subject of near-future work. As an illustration we
apply Eq.~(\ref{eq:chi2}) to simulated data generated with the
coherent \modelBase model derived in this paper and realizations of a
turbulent magnetic field generated with the algorithm
of~\citet{1999ApJ...520..204G} for the field strength of the JF12b
model from \citet{2016A&A...596A.103P}.  For this case, the comparison
of simulated sky maps $d_{ij}$ to the undisturbed model predictions
$m_{ij}$ together with the ``measured'' pixel-wise standard deviation
$\sigma_{ij}$ yields reduced $\chi^2$ values of 0.8, 1.5 and 2.4 for a
random field with a coherence length of 20, 40 and 80 pc,
respectively.

The best set of $n$ model parameters $\bm{p}$ are found performing a
multi-dimensional optimization of \cref{eq:chi2} with the {\scshape
Minuit} program~\citep{1975CoPhC..10..343J} using its {\scshape
Migrad} method that implements a variable metric gradient
descent \citep{fletcher63,fletcher70}. Deterministic gradient methods
like {\scshape Migrad} descend quickly to a local minimum which may
not always coincide with the global minimum. To find this global
optimum we use the heuristic multi-start method, i.e.\ we perform a
number of gradient-descent minimization runs starting at different
positions $\bm{p}_\text{start}$ distributed uniformly in the
$n$-dimensional hypercube of the parameter space. The run with the
smallest local $\chi^2$ is then considered to be the global
minimum. For the model fits performed in this paper we run typically
$\mathcal O(100)$ minimizations. The fact that many runs from
very different starting positions converge to the same minimum local
$\chi^2$, increases our confidence that the $n$-dimensional likelihood
contour of the optimization problem at hand is well-behaved and that
we have indeed identified the global minimum of each model variation
investigated. The main advantage of this simple multi-start method is
that such a repeated gradient-descent can be trivially parallelized on
a computing cluster and thus it provides an efficient method to search
for the global minimum of a model, since each gradient search is very
fast.\footnote{The optimization of the 20-parameter {\modelBase} model
discussed in Sec.~\ref{sec:results} needs, on average, 4000 $\chi^2$
evaluations to converge. Each evaluation takes 4 seconds on an
Intel$^\text{\textregistered}$ Xeon$^\text{\textregistered}$ E5
processor at 2.4 GHz. For each evaluation, $\bm{B}$, $\nel$, $\ncre$
and $j_\nu$ are calculated at $\mathcal{O}(10^6)$ positions throughout
the Galaxy to calculate the adaptive line-of-sight integrals
of \RM, \Q and \U.}

For each minimization run, a good first approximation of the
covariance matrix of the best-fit parameters can be obtained by the
parabolic estimates derived from the Hessian-matrix of the second
derivatives of the $\chi^2$ with respect to the parameters at the
minimum. For the final fiducial models presented
in Sec.~\ref{sec:results}, more precise confidence intervals of the
parameters at the minimum are derived with the profile likelihood
method, i.e.\ by finding for each parameter the two values
$p_\text{up/low}$ at which $\smash{\chi^2(p_\text{up/low})
= \chi_\text{min}^2+m}$, while marginalizing over the other $n-1$
parameters. These estimates are obtained using the {\scshape Minos}
algorithm of {\scshape Minuit} and each evaluation of one
$p_\text{up/low}$ pair is about as computationally expensive as the
overall minimization itself. Unless stated otherwise, we quote the
68\% (``one $\sigma$'') intervals obtained for $m=1$. More
information on the uncertainties and parameter correlations can be
found in Sec.~\ref{sec:cova} in the Appendix.

It is worth noting that our approach differs from that of
previous GMF studies, which used Markov-Chain Monte Carlo (MCMC) to
explore the model parameters,
\citep[e.g.][]{Jansson:2009ip,2010MNRAS.401.1013J,JF12coh,2017A&A...600A..29T}. MCMC
is the preferred method to sample the posterior distribution of a
model given the data, but it is not an efficient optimizer
\citep[e.g.][]{Hogg:2017akh}.  Moreover, for most applications, the
approximate covariance matrix and the confidence intervals derived
with the profile likelihood method provide information equivalent to
the MCMC samples;  see, e.g.\ \citep{Planck:2013nga}, for a comparison of
the two methods in the context of cosmological parameter
estimation. Most importantly, the ability to efficiently optimize the
parameters of many different models is of paramount importance for the
GMF inference, since, as will be shown in Sec.~\ref{sec:results}, the
systematic differences resulting from different model assumptions are
typically much larger than the precision estimated for the parameter
uncertainties of a particular model.

%% file: results.tex
\def\mpfigw{0.27}
\begin{figure*}[t]
    \modelPlot{pics/Models/uf23_base}{FourierDisk}{JF12Toroidal}{CoastingXfield}
    \fitPlot{pics/Fits/uf23_base}
    \fitPlotCaption{\modelBase}
    \label{fig:basemodel}
\end{figure*}

\subsection{Base Model}
\label{ref:sec}
For the results presented in the following, we have optimized the
parameters of more than 200 combinations of magnetic field models and
auxiliary models to obtain an overview of the range of GMF models
attainable in the full set of possible combinations.\footnote{The
combination of two disk field models, eight poloidal field models, two
toroidal field models, two thermal electron models, two synchrotron
products, six cosmic-ray electron models  and including a \nel-$B$
correlation or not results in 1536 possible model variations.}  Many
of these model variations are performed with respect to our fiducial
``\modelBase'' model which consists a three-mode grand-design spiral
disk field, an explicit toroidal halo and a coasting X-field with a
logistic sigmoid radial dependence; there is no correlation between
$n_e$ and $B$ and the striation is a free parameter. The skymaps
of \RM, \Q and
\U resulting from this model are calculated with the \YMW thermal
electron model and a cosmic-ray electron density derived for $\hDiff=6$~kpc.
The model
parameters are adjusted to fit the data yielding an optimum with an
acceptable goodness of fit of $\chi^2/\text{ndf} = 7923 / 6500 =
1.22$, where $\text{ndf}$ denotes the number of degrees of freedom,
i.e.\ the number of data points minus the number of free parameters,
which is $n_\text{par}=20$ in this case. The contribution from \RM
pixels to the $\chi^2$ is 4354 ($n_\RM =2838$) and from \Q and \U
pixels it is 3569 ($n_{\Q}+n_{\U}= 3682$), i.e.\ the model describes the
polarized synchrotron data slightly better than the \RMs.

The contributions of each model component (i.e.\ disk field and
toroidal and poloidal halo) to \RM, \Q and \U is shown separately in
the three top rows of \cref{fig:basemodel} and the predicted sky maps
for the sum of all components is displayed in the fourth row. Here it
is interesting to see how the interplay of different model components
creates the large-scale features observed in data.  As pointed out
above, the sum of the disk and toroidal field produces the large-scale
structure of \RMs, see \cref{eq:rmcomponents}. The polarized intensity
on the other hand, is mostly generated by the toroidal and poloidal
halo. And, whereas the overall \RM can be obtained by summing the \RM
of each component ($\RM \varpropto\bm{B}_\parallel =
(\sum \bm{B}_i)_\parallel =\sum \bm{B}_{i\parallel}$), the sky pattern
of the Stokes parameters of the full model is not the sum of its
components since e.g.\ $\Q,\U\varpropto \bm{B}_\perp^2 = \left((\sum \bm{B}_i)_\perp \right)^2 \ne \left(\sum \bm{B}_{i\perp}\right)^2$.

The masked model, data and ``pull'' are shown in
the three bottom rows. The pull is the difference between the data and
model in units of standard deviation of the data, and the sum of all
squared pulls yields the $\chi^2$, \cref{eq:chi2}. For a perfect match
of data and model, the pull should fluctuate randomly around zero with
a standard-normal distribution, but here several large regions exist
in which the pull has consistently negative or positive values beyond
$\pm 1 \sigma$. Some of these are close to masked regions, e.g.\ at
the edges of the masks for the North Polar spur and loop III for \Q
and \U and close to the mask for the Gum nebula for \RM, implying a
certain amount of leakage of these features beyond the mask. Others,
like the large region of negative pull for
\RM below the Galactic plane at longitudes of $60^\circ<\ell<150^\circ$ could
indicate deficiencies in the modeling of the global structure of the
GMF, e.g.\ an azimuthal variation of the scale height of the toroidal
field. It is, however, plausible that most regions with large-scale
deviations can be attributed to local perturbations that appear as
structures of large angular scale due to their proximity.  Excluding
pixels with a large pull from the fit changes the values of fit
parameters, but the differences are of similar magnitude or smaller
than the differences between the model variations explored below, see
Sec.~B in the Appendix for further details.

Apart from these potentially local structures, the \modelBase model
successfully describes all the large-scale features of the data, in
particular the anti-symmetric structures of the \RM sky and the tilted
large-scale ``lobes'' of negative and positive \Q and \U. The
parameters of the \modelBase model can be found in the first column of
Table~\ref{tab:modelpars}. The correlation matrix of the
parameters is discussed in Sec.~\ref{sec:cova} in the Appendix.

\begin{figure*}[t]
  \centering
  \subfloat[Half-height \hDiff of the cosmic-ray diffusion volume.]{
    \includegraphics[width=0.45\linewidth]{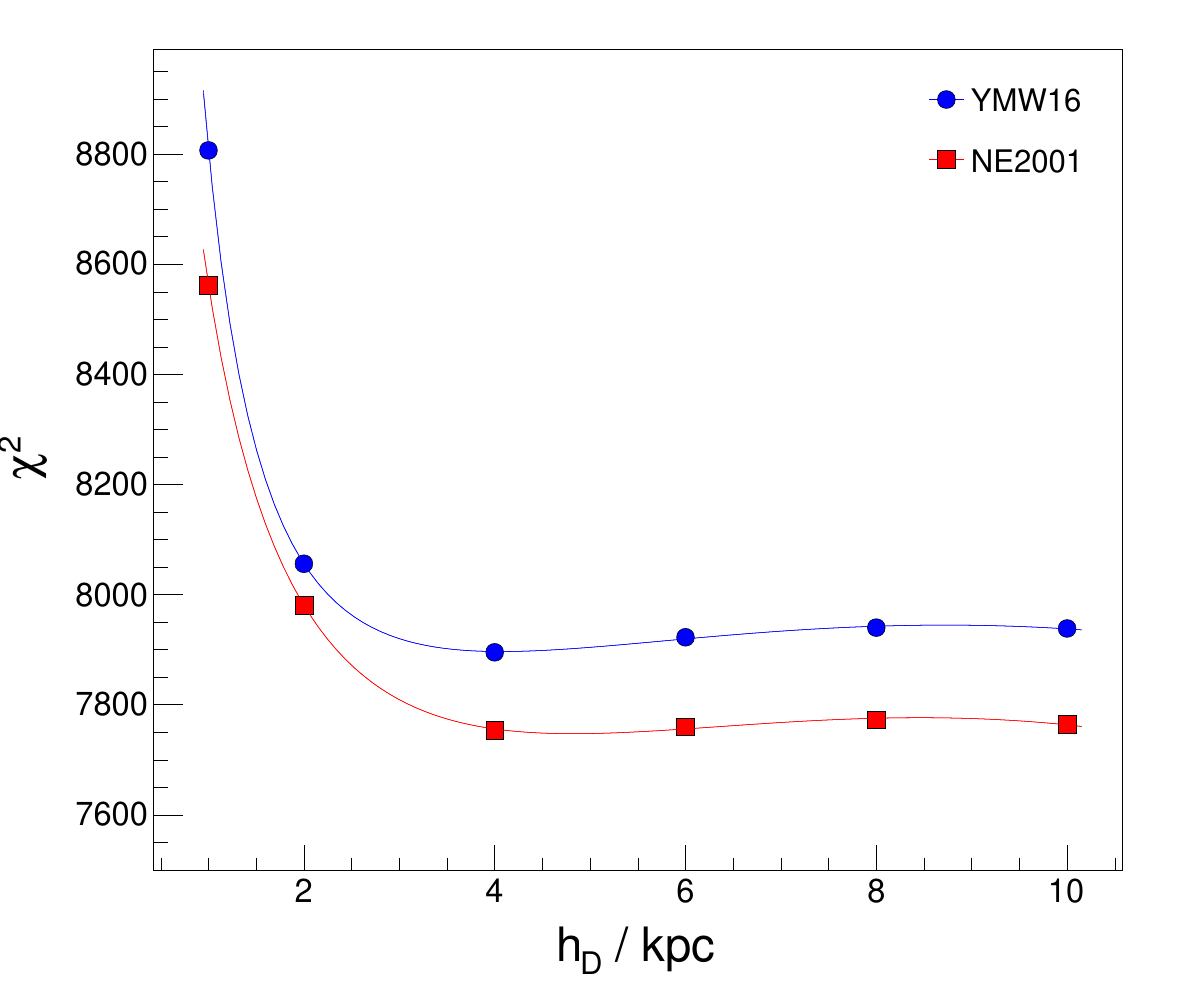}
    \label{fig:necreChi2}
  }\qquad
  \subfloat[Correlation coefficient $\kappa$ between the magnetic
    field and thermal electron density.]{
    \includegraphics[width=0.45\linewidth]{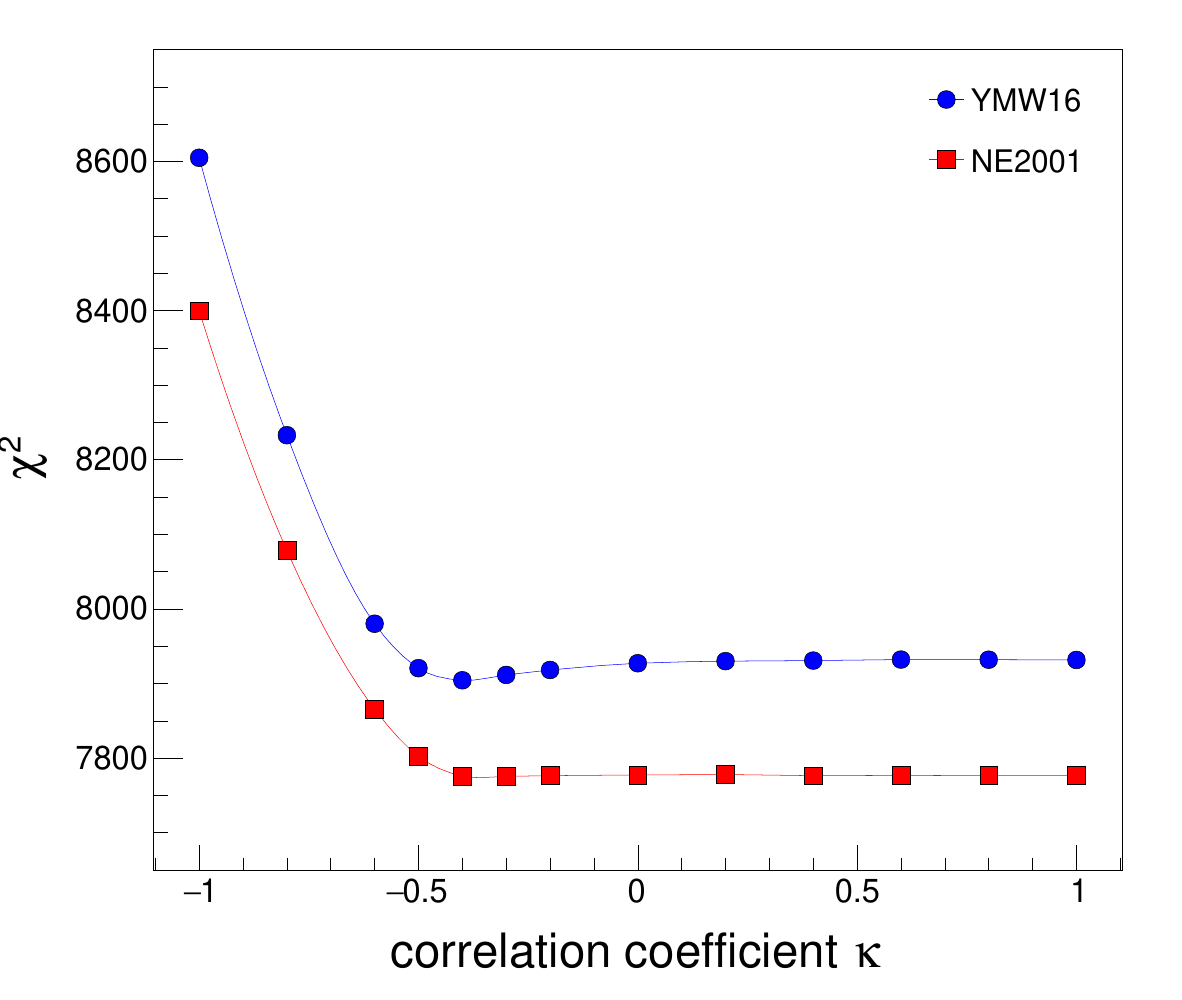}
    \label{fig:kappaChi2}
  }%
  \caption{Goodness of fit, \cref{eq:chi2}, for two
    different models of the thermal electron density, \NE (red) and
    \YMW (blue), and as a function of \hDiff (left) and $\kappa$ (right).}
  \label{fig:fitChi2}
\end{figure*}

\begin{figure*}[t]
  \centering
  \includegraphics[width=0.95\linewidth]{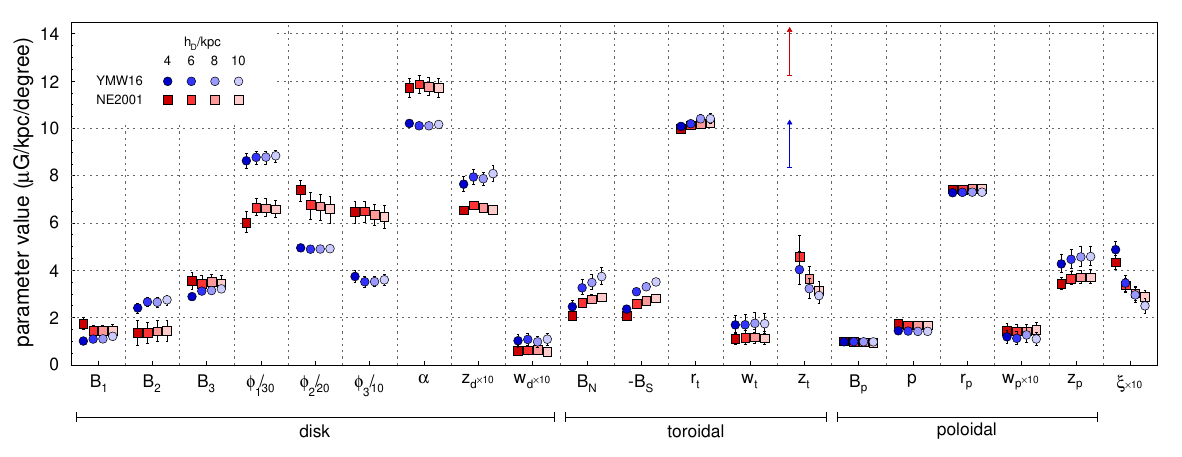}
  \caption{Best-fit parameter values of the disk, toroidal and
    poloidal GMF components for different values of the half-height \hDiff of
    the cosmic-ray diffusion volume, and for two different
    models of the thermal electron density, \NE (red) and \YMW (blue).
    See Table~\ref{tab:glossary} for a short explanation of each
    parameter. Some parameter values have been multiplied by a scale
    factor, as indicated in the axis labels, to fit in one panel with
    a single $y$-axis. One-sigma uncertainties are shown as error bars
    and arrows indicate the 84\% CL lower limits on the parameters}
    \label{fig:necrePars}
\end{figure*}

\begin{figure*}[t]
  \centering
  \includegraphics[width=0.95\linewidth]{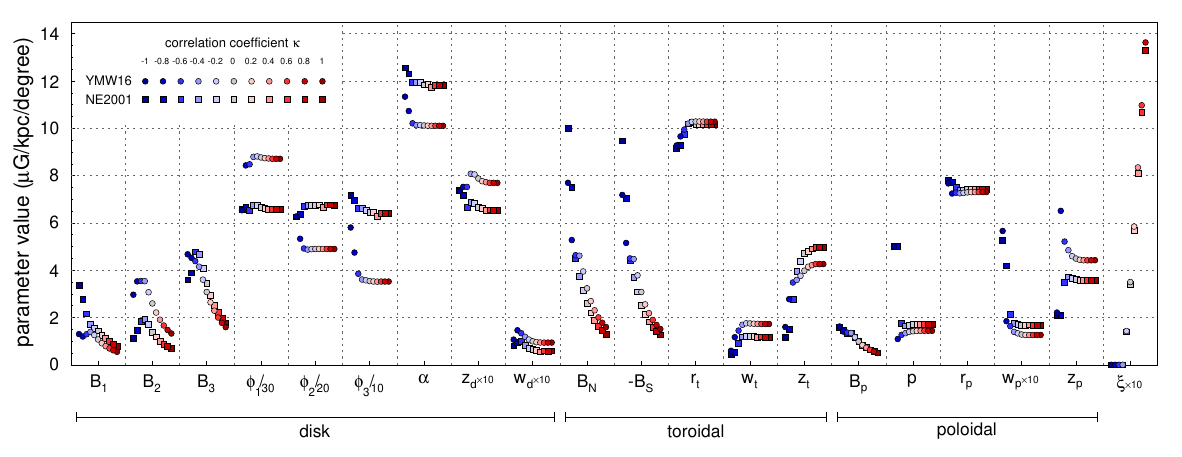}
  \caption{Best-fit parameter values of the disk, toroidal and
    poloidal GMF components for different values of the correlation
    coefficient between the magnetic field and thermal electron
    density $\kappa$ and for two different models of the thermal
    electron density, \NE (red) and \YMW (blue).  See
    Tab.~\ref{tab:glossary} for a short explanation of each
    parameter. Some parameter values have been multiplied by a scale
    factor, as indicated in the axis labels, to fit in one panel with
    a single $y$-axis. No parameter uncertainties are shown here, see
    \cref{fig:necrePars}.}
    \label{fig:kappaPars}
\end{figure*}

\subsection{Thermal and Cosmic-Ray Electrons}
\label{sec:nencre}
To study the dependence of the inferred GMF on the auxiliary models of
the thermal and cosmic-ray electron density, we repeat our fits for
the model variations discussed in Sec.~\ref{sec:auxmodels}. The
best-fit $\chi^2$ values for different half-heights of the cosmic-ray
diffusion volume, \hDiff, and for the \NE and \YMW model are shown in
\cref{fig:necreChi2}. Here the functional forms for the GMF model are
identical to the one used for the \modelBase model.  As can be seen,
the fit quality deteriorates rapidly for small values of the height of
the diffusion volume and it reaches a near-constant value at
$\hDiff\gtrsim 4$~kpc. For small values of \hDiff, the decreasing vertical
extent of the cosmic-ray electron halo cannot be compensated for by a
larger magnetic halo.\footnote{For illustration, assume an exponential
  height dependence of \ncre and $B$ with scale heights $z_\text{cre}$
  and $z_\text{B}$. Then, using \cref{eq:J_coherent}, the
  height-dependence of the synchrotron emissivity is $j_\nu\varpropto
  e^{-|z|/z_\text{syn}}$ with $z_\text{syn}= z_\text{cre} / (1+2
  z_\text{cre}/z_B)$. It follows that for a given $z_\text{syn}$ of
  the data, $z_\text{cre}$ has to be at least $\ge z_\text{syn}$
  ($z_B\rightarrow \infty$).} Interpolating between the fit results
at discrete values of $\hDiff$, we estimate a 5-$\sigma$ lower limit
on the size of the diffusion volume of
\begin{equation}
  \hDiff \geq 2.9~\text{and}~3.5~\text{kpc},
  \label{eq:hdifflimit}
\end{equation}
where the two values correspond to the fits with \NE and \YMW,
respectively. This lower limit is, however, only indicative since
in this analysis, there is no feedback between the derived magnetic field
and the propagation of cosmic-ray electrons, see Sec.~\ref{sec:ncre}.

The interplay of diffusion height and derived magnetic field
parameters can be seen in \cref{fig:necrePars}, where the GMF model
parameters of the different fits for $\hDiff \geq 4$~kpc are
displayed. Especially the fitted scale height \zT of the toroidal field
 depends strongly on \hDiff.  For $\hDiff=4$~kpc only a lower
limit on \zT can be estimated.  In the following, we will therefore
only investigate GMF models derived for \ncre densities within
$6\leq\hDiff/\text{kpc} \leq 10$, for which the best-fit values of \zT
are finite and compatible with current estimates of the size of the
diffusive halo estimated from the analysis of unstable secondary
cosmic-ray nuclei.

Concerning the models of the density of thermal electrons in the
Galaxy, it can be seen in \cref{fig:necreChi2} that the fits with the
old \NE model result in a consistently better fit quality than the
ones performed with the recent \YMW model.  Even though the difference
is statistically significant ($\Delta\chi^2 = -167$ at $\hDiff = 6$)
we consider both models in our analysis, since the latter \nel model
gives a better description of Galactic pulsars {\DM}s than the former.
Switching between these two models has a larger systematic impact on
the values of most GMF parameters than a change in $\hDiff$,
see \cref{fig:necrePars}.

\begin{figure*}[t]
  \def\figh{0.4}
  \centering
  \subfloat[Grand-design spiral]{
    \includegraphics[clip,rviewport=0 -0.05 0.87 1,height=\figh\textheight]{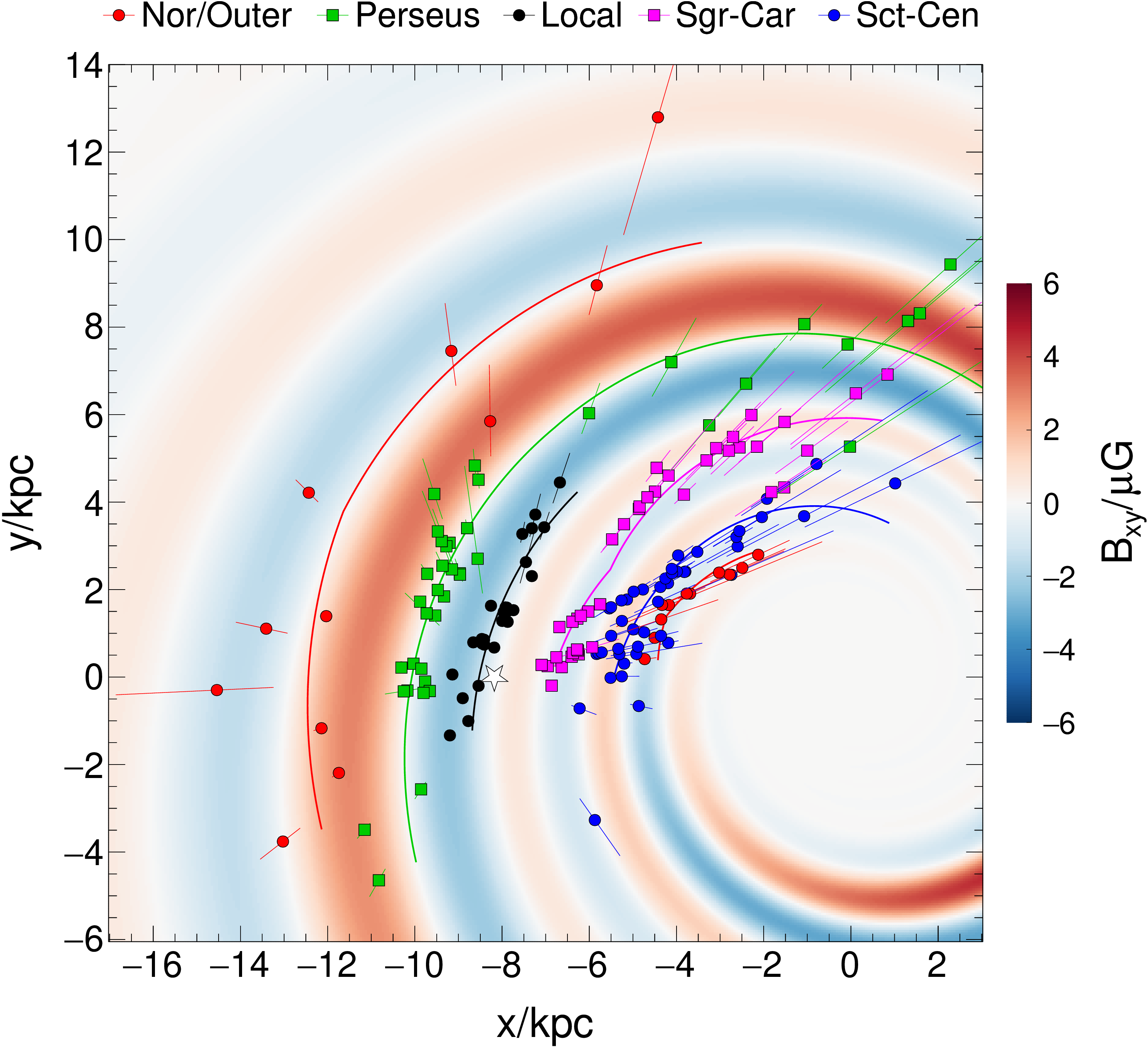}
    \label{fig:grand}
  }
  \subfloat[Spiral spur.]{
    ~\includegraphics[clip,rviewport=0.089 -0.05 1 1.051,height=\figh\textheight]{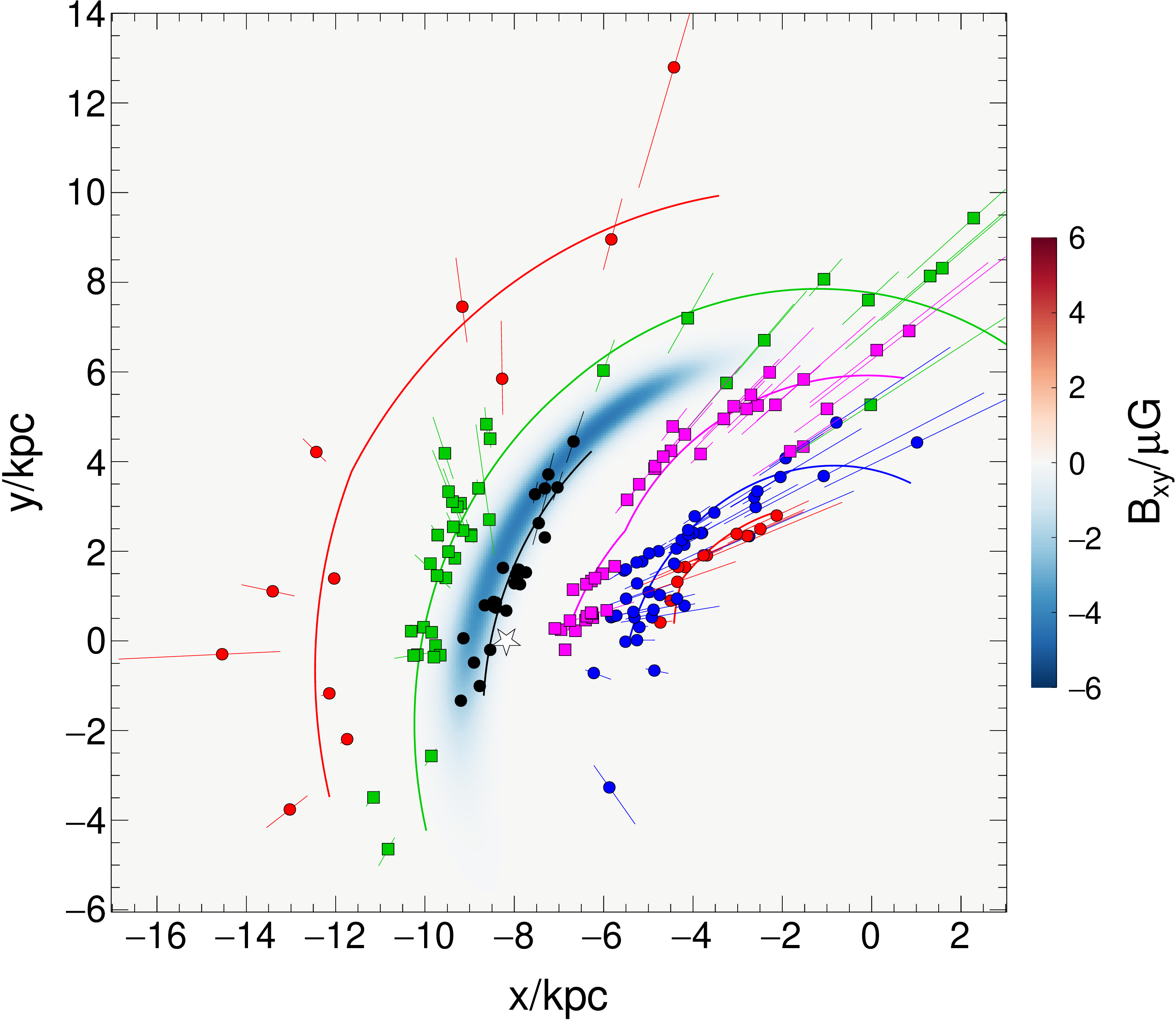}
    \label{fig:spur}
  }
  \caption{Magnetic field of the disk component for the
  grand-design spiral (left) and spiral spur (right) GMF model. The
  horizontal magnetic field strength in the $x$-$y$ plane at $z=0$ is
  displayed with colors ranging from blue (counter-clockwise field) to
  red (clock-wise field). Superimposed points with distance error bars
  are the locations of tracers of the spiral structure of the matter
  density of the Milky Way (high-mass star-forming regions with
  parallax distances) and the curved colored lines are the inferred
  location of spiral arm segments, both
  from \citet{2019ApJ...885..131R}.}
  \label{fig:diskstructure}
\end{figure*}

\subsection{Striation or Correlation?}
\label{sec:striacorr}
It is well known that the magnetic field strength derived from the
observed rotation measures is smaller than the one derived from the
observed Galactic polarized synchrotron emission.  One way to
reconcile the two observables is to postulate the existence of an
anisotropic or striated random field that fluctuates along the
coherent field lines, which leads to an increase in polarized
intensity without changing the rotation measures. For our \modelBase
model, the best-fit value of the striation factor $\xi$,
see \cref{eq:striation}, is
\begin{equation}
  \xi_\text{\modelBase} = 0.35\pm 0.03,
\end{equation}
implying that the energy density in the coherent and striated field
component is almost equal (energy density $u_B \varpropto B^2$,
$B^\prime = (1+\xi) B$, $u_{B^\prime}/u_B = (1+\xi)^2 = 1.8\pm 0.1$).

Another reason for the mismatch between magnetic fields inferred from
\RM and \PI observations could be an anti-correlation of the thermal
electrons and the magnetic field, leading to smaller Faraday rotation
than in the uncorrelated case. Here we perform, to our knowledge for
the first time, an analysis of \RM and \PI allowing for a modified
\RM due to an \nel-$B$ correlation with coefficient $\kappa$, as derived
by~\citet{2003A&A...411...99B} (cf.\ Sec.~\ref{sec:observables}).
Here we use the random field $b$ derived from the total synchrotron
intensity measured by \Planck \citep[JF12b model
of][]{2016A&A...596A.103P} to evaluate \cref{eq:rmkappa} for a fixed
value of $b$, but a coherent field $B$ that is allowed to float freely
during the fit.

The fits are performed for different values of $\kappa$ and the
resulting dependence of the fit quality on the correlation
coefficient is displayed in \cref{fig:kappaChi2}. As can be seen, a
large anti-correlation of \nel and $B$ is disfavored and the
five-sigma lower limit is found to be
\begin{equation}
  \kappa \geq -0.52~\text{and}~-0.49,
  \label{eq:kappalimit}
\end{equation}
where the two values correspond to the fits using the \NE and \YMW
models of \nel, respectively.  The optimal fit to the data is at
$\kappa \approx -0.4$, but the $\chi^2$ minimum is very broad and all
fits above this value fit the data similarly well. The reason for this
degeneracy can be understood by examining the change of fit parameters
with $\kappa$, as shown in \cref{fig:kappaPars}. Given the two degrees
of freedom of rescaling the magnetic field strengths of each model
component and the striation factor, it is always possible to match
the \RM and \PI data as long as $\xi>0$. In the following, we will not
further investigate the fits with $\kappa>0$, since our fiducial fits
with $\kappa=0$ already exhibit a large degree of striation close to
equipartition with the coherent field (see above). Instead, we will
concentrate on the best fit at $\kappa = -0.4$ where the striation
parameter is zero as an alternative to the fiducial model. At this
value of $\kappa$, the magnetic field scale is set by the \PI data,
whereas at $\kappa=0$ the scale is set by the \RM data. In this
extreme scenario, where no striated random fields contribute to the
polarized intensity and the Faraday rotation is diminished by the
anti-correlation of $B$ and \nel, the fitted magnetic field strengths
($B_m$, $B_\text{N/S}$ and $B_\text{p}$) are about a factor 1.4 larger
than in the fiducial case.

\begin{figure*}[t]
\centering
\includegraphics[width=0.87\textwidth]{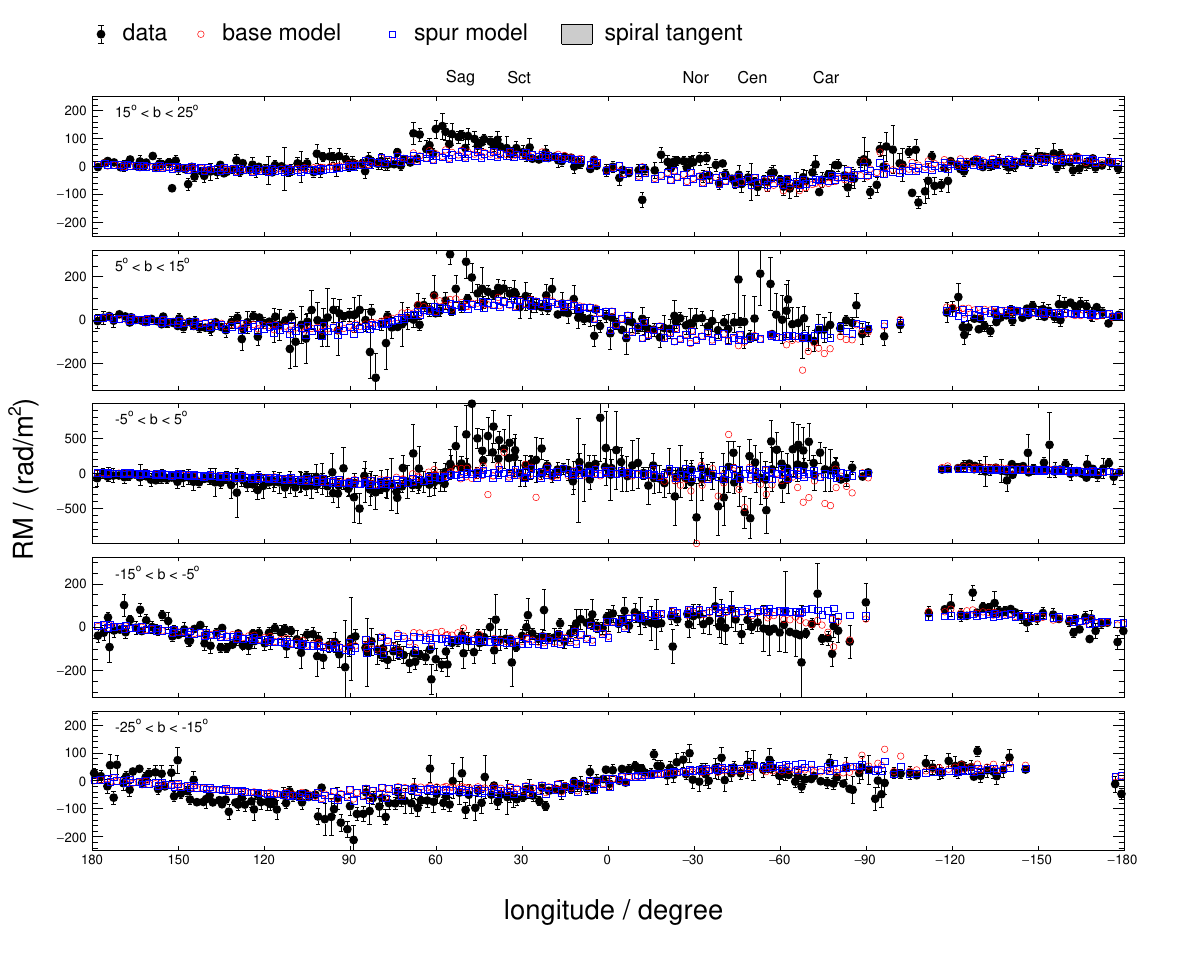}
\caption{Comparison of data (black points with error bars) and models (red open circles: \modelBase model with a grand-design spiral disk field, blue open squares: spiral spur model) in five latitude slices around the Galactic plane. Every point is the RM value detected/predicted in one \HEALPix pixel ($N_\text{side}=16$). Gray bands are drawn at the longitudes of tangent points of the spiral arms of the Galaxy. The width indicates the range of tangent positions in different observables catalogued by \citet{2022NewA...9701896V}.}
\label{fig:spiralAndSpur}
\end{figure*}

\subsection{Synchrotron Data Products}
The choice of data product for the Stokes \Q and \U parameters affects
the GMF fit due to the differences in the large-scale structure of
derived sky maps, see discussion in Sec.~\ref{sec:syndata} and the
comparisons shown in \cref{fig:syndatadiff} in the Appendix. Moreover,
and perhaps more importantly, since these differences are relatively
small, different products differ by the variance of \Q and \U within a
pixel over which we average these values. Since we use these variances
as a weight of the data when optimizing Eq.~\ref{eq:chi2}, they also
affect the outcome of the fit.

For our simple arithmetic average of the \Q and \U values of \Planck
and \WMAP, the variance is calculated for all data points before
averaging, therefore we expect it to be the arithmetic average of
the \Planck and \WMAP variances plus an additional contribution from
the systematic difference between the two products. On average, the
variance derived from the \Planck and \CosmoGlobe maps is about a factor 0.8
smaller than the one of our default synchrotron map, and about the
same for the \WMAP data. Keeping in mind that \WMAP could measure at
lower frequencies where the synchrotron intensity is larger by a
factor of $(22.5~\text{GHz}/30~\text{GHz})^{-3.1}=2.4$, it is not
clear to us to which extent the smaller variance of \Planck (and by
extension \CosmoGlobe) is actually due to an improved measurement of
the synchrotron radiation or due to the use of more aggressive
foreground smoothing priors in the analysis.

As a consequence of smaller variances, the fits using \Planck
and \CosmoGlobe have a $\chi^2$ that is worse than the one of
the \modelBase model by a factor of 1.4 and 1.2, respectively. We have
checked that the worse fit is indeed mostly due to the different
variances (and not the different values of \Q and \U), by performing a
fit of the \Planck data with our default variances, in which case the
fit quality was only worse by a factor of 1.03.

The GMF models obtained with alternative data products are
qualitatively similar, but the parameters obtained in these fits
differ from the one of the \modelBase model. The differences are
mostly not driven by the systematic differences between the sky maps,
but rather caused by the larger relative ``pull'' on the $\chi^2$
of \Q and \U data with respect \RM data. The sensitivity of the
parameter values on the data-driven weights are intrinsic to the
method used in this analysis and needs to be considered when
estimating the GMF model uncertainties, see Sec.~\ref{sec:ensemble}
below.

\subsection{Disk Field}

\subsubsection{Grand-Design Spiral}
Many of the previous attempts to model the global structure of the
disk field enforced a certain number of ``magnetic arms'', often
following the inferred large-scale structure of matter density in the
Galaxy. Due to our more flexible decomposition of the spiral field
into azimuthal Fourier modes, we can let the data decide how much
spiral structure is needed. We have fit the \RM and \PI data using a
different number of Fourier modes ranging from 1 to 5. As expected,
the fit quality continuously improves with the number of modes
$n_\text{mode}$, since each adds two more free parameters (amplitude
and phase) to the fit.  Using Wilks' theorem \citep{Wilks:1938dza}, we
find that the improvements of the fit quality of the \modelBase model
are significant up to $n_\text{mode}=3$, with a $\Delta\chi^2=153.2$
going from 2 to 3 modes and $\Delta\chi^2=12.7$ from 3 to 4 modes. The
same conclusion is reached using the Bayesian information criterion,
$\text{BIC} = \chi^2 +
n_\text{par}\, \ln(n_\text{data})$ \citep{1978AnSta...6..461S}, which
is minimal at $n_\text{mode}=3$. We have also tested that the required
number of modes remains three when changing the height of the
diffusion volume, \hDiff, used to calculate \ncre and when using a
twisted X-field.

Three azimuthal modes correspond to six ``magnetic arms'' of
alternating polarity.  The grand-design configuration of our \modelNe
fit variant (using the \NE thermal electron model) has already been
shown above in the right panel of \cref{fig:disk}. The best-fit disk
field of the \modelBase model is displayed in the left
panel of \cref{fig:diskstructure}, where we also show the location of
tracers of the spiral structure of the matter density of the Milky
Way, given here by the measurements of trigonometric parallaxes of
high-mass star forming regions from~\citet{2019ApJ...885..131R}.  As
can be seen, we find a remarkable alignment of the fitted magnetic
spiral structure and these tracers. Large coherent field strengths are present
in the interarm regions, but the coherent field strength is close to
zero at the location of the spiral matter segments derived
by~\citet{2019ApJ...885..131R}, shown as lines
in \cref{fig:diskstructure}. This result is similar to what is
observed in external galaxies, where the strongest ordered fields are
detected in the interarm regions \citep{2016A&ARv..24....4B}.

The fitted pitch angle, $\pitch$, of the disk field is found to be
nearly independent of the assumed cosmic-ray halo size, \nel-$B$
correlation coefficient ($\kappa\geq -0.4$) and functional form of the
magnetic halo. However, the fits with different models of the thermal
electron density result in pitch angles that are systematically
different with respect to each other by $\pm 1^\circ$,
see \cref{fig:necrePars}.  The average of the value obtained using
the \NE and \YMW thermal electron models is
\begin{equation}
  \pitch = (11.0\pm 0.3\;(\text{stat.})\pm 1.0\;(\text{\nel}))^\circ.
\end{equation}
This value is in good agreement with the pitch angle of the local
(Orion–Cygnus) spiral arm of $(11.4\pm
1.9)^\circ$ \citep{2019ApJ...885..131R} and the pitch angles of
$\alpha = (9.87\dots 11.43)^\circ$ of the grand-design logarithmic
spiral model fitted by \citet{2014A&A...569A.125H} using HII regions
as spiral tracers. Thus, the pitch angles of the spiral structure of
the magnetic field and the matter density in the Milky Way are about
equal, similar to what is observed for external spiral
galaxies~\citep{2015ApJ...799...35V}.

\subsubsection{Local Spur}
What drives the fit of the grand-design spiral disk field? Is it
the \RM values at longitudes where the line of sight is tangential to
a spiral arm?  These magnetic tangents are clearly visible in the
predicted \RMs in the top left plot of \cref{fig:basemodel}, but they
are not so obvious in the \RM of the data, possibly due to the large
variance present in the data at low latitudes. Alternatively, the fit
of the disk field might be mostly determined by the large-scale
``butterfly pattern'' of \RMs, see \cref{eq:rmcomponents}, and
therefore mostly from the disk field in the vicinity of the Sun.

To test this possibility, we fit the data without a grand-design
magnetic pattern, but include only one local magnetic spur, as
introduced in Sec.~\ref{sec:locspure}.  The fitted pitch angle of the
local spur is $(12.1\pm 0.6)^\circ$ and does not depend on the thermal
electron model used. The best-fit local spur is displayed in the right
panel of \cref{fig:diskstructure}. It is located at the edge of the
local Orion–Cygnus spur and has a magnetic field strength of
$4.30~\muG$ at the reference radius of 8.2~kpc. The quality of this
fit is found to be slightly worse than the one of the \modelBase model
(7991 instead of 7923), but close enough to conclude that both models
are approximately equivalent.

For a closer look at the differences between the \modelSpur
and \modelBase model, we show the \RM values in bands of latitude for
data and the two models in \cref{fig:spiralAndSpur}. As can be seen,
both models give the same overall good description of the
longitude-dependence of \RM of the data. The \modelBase model (shown
as red open circles) exhibits distinct \RM features at low latitudes,
shown in the middle panel for $|b|<5^\circ$, but none of these are
visible in the data.\\

The near-equivalence of these two radically different disk field
models is a consequence of the fact that the extragalactic
\RMs used in the fit constrain only the integrated Faraday rotation to the edge of the Galaxy.
The model degeneracy could, in principle, be broken by including \RMs
from Galactic pulsars into the fit to provide cumulative \RMs at
different distances.  However, pulsar \RMs are currently of limited
use, since the distances to most pulsars are not known with sufficient
precision. Some previous GMF studies used the ``\DM-distance'' (i.e.\
the distance $d$ at which $\DM^\prime
= \int_0^d \nel(\bm{x}(r)) \, \dd r$ equals the observed \DM) to
calculate the model \RMs, but this introduces an additional dependence
on thermal electron density models that is difficult to account for
in the GMF model optimization.  Notwithstanding these caveats, it is
interesting to note that \citet{2018ApJS..234...11H} derived a six-arm
grand-design spiral model from the \RMs of pulsars and extragalactic
sources, which is qualitatively very similar to the disk field of
our \modelBase model.  More studies are needed, however, to
unequivocally prove the existence of a grand-design magnetic spiral in
our Galaxy. Until then, a conservative approach is to consider the two
models introduced in this section as extreme possibilities for the
coherent magnetic field in the disk of the Milky Way.

\begin{figure}[t]
\centering
\includegraphics[width=\linewidth]{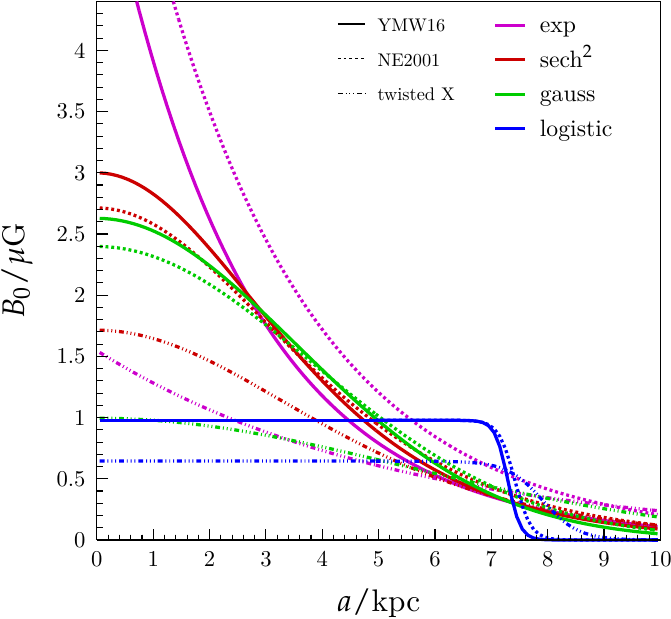}
\caption{Best-fit radial dependence of the $z$-component of the poloidal magnetic field at $z=0$, $B_0(a)$, see Eqs.~(\ref{eq:ftr}) and (\ref{eq:ftz}).
Different radial functions are shown in different colors, see
Eqs.~(\ref{eq:bzexp}) - (\ref{eq:sigm}). Variations in the thermal
electron density and toroidal field type used during the fit are shown
with different line styles, as indicated in the figure legend.}
\label{fig:radialX}
\end{figure}
\subsection{Poloidal Field}
\label{sec:poloidal}
We begin by reassessing the need for a poloidal component to describe
the data, performing a fit without including such a component. The
resulting fit quality deteriorates tremendously, by
$\Delta \chi^2>+2100$ with respect to the \modelBase model,
irrespective of the thermal electron model used in the fit.  As can be
seen in Fig.~\ref{fig:basemodel}, the poloidal component contributes
only little to the \RMs. However it would be incorrect to conclude
that the fit cannot determine whether the X-field is coherent or
striated, since the Stokes parameters of the model components do not
add linearily (see discussion in Sec.\ref{ref:sec}). In fact, if a
striated X-field were equivalent to a coherent one, then the fit
quality would only depend on the orientation and not the direction of
the coherent X-field, but the $\Delta \chi^2$ is $+8679$, when
reversing the direction of the X-field.

We also tried to fit the data with a N-S reflection-symmetric X-field
(characteristic of a S1 dynamo), instead of a dipolar one (A0 dynamo),
however the fit is much worse ($\Delta \chi^2>+1700$). Furthermore, we
checked for the \modelBase model if different strengths of the
magnetic field normalization \BP in the Northern and Southern
hemisphere are preferred by the data. A different value in the two
hemispheres would be required by flux conservation if there were a net
inward or outward flux in the disk in an $m=0$ mode. We find that any
such asymmetry must be small, since the difference of the fitted
values for the two hemispheres is $\Delta \BP = (0.12\pm 0.07)~\muG$.

Inspecting the pixel-by-pixel contributions to the $\chi^2$ we find
that the poloidal fit is mainly driven by the \Q and \U data at
longitudes $|\ell|<60^\circ$. Unless a very peculiar foreground is
responsible for these large-scale features in \Q and \U (see
discussion in Sec.~\ref{ref:fg} of the Appendix), we confirm the
conclusions of \citet{JF12coh} that the polarized synchrotron
intensity cannot be described without the presence of a dipolar
X-field in the Galaxy.

We find that the data is described equally well by a power-function
and a coasting X-field ($\chi^2=7926$ and 7923, respectively), since
the major differences between the field lines of the two models are at
large radii, where the field is small, see \cref{fig:xfield}. There
is, however, a clear preference regarding the choice of the radial
dependence of the mid-plane vertical field,
Eqs.~(\ref{eq:bzexp})-(\ref{eq:sigm}). The best fit to the data is
obtained for the logistic sigmoid function, i.e.\ for a vertical field
strength that is constant with galacto-centric radius and then
vanishes to zero at a certain radius $\rP$ with a transition
width \wP. The tested alternatives of a Gaussian, hyperbolic secant or
exponential radial dependence result in fits that are worse by
$\Delta\chi^2 = +370$, $+452$ and $+508$. This clear preference for a
logistic radial cutoff remains for all model variations tested
(power-function or coasting X-field, twisted X-field and using
different thermal electron models).

The improvement of $\chi^2$ when using a logistic instead of an
exponential function originates mostly from the \RM
($\Delta\chi^2_{\RM} = 315$) and \Q data ($\Delta\chi^2_{\Q} = 155$
and only to a lesser extent from the \U data ($\Delta\chi^2_{\U} =
36$). Much of the improvement in $\chi^2_{\RM}$ is concentrated in one
swath of the \RM sky located at $\ell \approx (-30 \pm 15)^\circ$ and
$ 90^\circ < b < 150^\circ$ and most of the decrease in $\chi^2_{\Q}$
is due to a better description of the data at $15^\circ<\ell<30^\circ$
and $b>0^\circ$.  Since we cannot be certain that these features at
intermediate angular scales are global, we conservatively keep the
exponential radial function as an extreme variation in our model
ensemble.

\begin{figure*}[t]
  \includegraphics[width=0.95\linewidth]{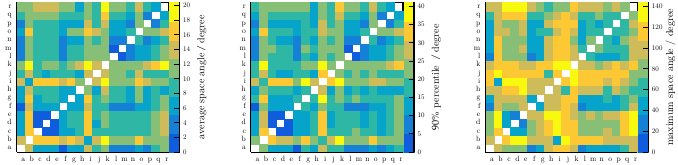}
  \caption{Comparison of models $i=$ a-r and $j=$ a-r
    (see Table~\ref{tab:allmodels}) using the space angle
    $\theta_{ijk} = \arccos(\bm u_{ik}, \bm u_{jk})$ between the direction of
    charged particles ($\R=1\times10^{19}$~V) back-tracked to the edge of the
    Galaxy analyzed for skymaps of arrival directions $k$ at Earth
    with a resolution of $N_\text{side}=32$.  {\itshape Left:} average
    $\theta$, {\itshape middle:} 90\% percentile of $\theta$
    distribution, {\itshape right:} maximum $\theta$.}
    \label{fig:deflComp}
\end{figure*}

The best-fit mid-plane poloidal magnetic field strengths, derived for
different radial functions, are shown in \cref{fig:radialX}. The
largest differences of $B_0(a)$ are at small radii, $a\lesssim 4$~kpc,
where the fit of the poloidal field strength is unconstrained due to
the large number of \Q and \U pixels masked at $|\ell|<30^\circ$.  The
logistic model used in our \modelBase model has a mid-plane vertical
field strength of $\BP = (0.98 \pm 0.03)$~\muG and cuts off inside the
solar circle at $\rP = (7.29\pm 0.06)~\text{kpc}$ with a transition
width of $\wP = (0.112\pm 0.029)~\text{kpc}$. In the case of the
twisted X-field model, the transition is fitted to be at a somewhat
large radius and the transition is broader ($\rP \sim 7.9$~kpc,
$\wP \sim 0.3$~kpc). The \modelXr model leads to a large vertical
magnetic field at the Galactic Center, $\BP = (5.8\pm 0.4)$~\muG that
falls off exponentially with scale length of $\rP = (2.5\pm
0.1)$~kpc. For comparison, the poloidal field of
the \citetalias{JF12coh} model is of similar strength with $\BP =
(4.6\pm 0.3)$~\muG and $\rP= (2.9\pm 0.1)$~kpc.

Thus, the \modelXr and \modelBase models bracket the possible range of
the radial dependence of the strength of the poloidal field. The
corresponding uncertainty of the GMF, quantified here for the first
time, must be taken into account when interpreting the arrival
directions of ultrahigh-energy cosmic rays and when discussing the
radial dependence of cosmic-ray energy spectra derived from gamma-ray
observations~\citep[e.g.][]{Gabici:2019jvz}. A better understanding of
the polarized foregrounds is needed to be able to further constrain
the GMF fits by including \PI data toward the Galactic center.

\subsection{Toroidal Field}

Fitting for an ``explicit'' toroidal halo field, we obtain a radial
extent of $\rT = (10.2\pm 0.2)$~kpc and a vertical scale height of
$\zT = (4.0\pm 0.7)$~kpc within our \modelBase model. The estimate of
the radial extent of the halo is very stable with respect to any of
the model variations we have studied. The vertical scale is less
certain, ranging from 2.9 to 6.1~kpc, where the lowest value is
obtained assuming a large cosmic-ray diffusion volume with $\hDiff
=10$~kpc.

The magnitude of azimuthal field strength in the Northern and Southern
hemispheres was found to vary between about 2 and 5~\muG, depending on
the model variation under study. For the \modelBase model, the
best-fit values are $\BNT = 3.3\pm 0.3$~\muG and $\BST = -3.1\pm
0.3$~\muG.  In all the studied variations, the magnitude of the
Northern and Southern magnetic field strength is found to be
compatible within the estimated uncertainties,
\begin{equation}
  \BNT = -\BST,
  \label{eq:NS}
\end{equation}
which strongly suggests a common origin of the toroidal field in the
two hemispheres.

Our unified halo model (\modelTwist) naturally explains the relationship in Eq.~\cref{eq:NS}, since the
magnetic field in both hemispheres originates from the shearing caused
by the same velocity field. Furthermore, the unified halo model has the practical virtue that its built-in connection between the radial extent of the poloidal and toroidal components should make it less prone to over-fitting local structures in the data.  This suggests that a fitted common radial scale with a large transition width (see the dash-dotted blue line in Fig.~\ref{fig:radialX}) might prove to be a more accurate description
of the global structure of the GMF.

While the fit quality of the twisted X-field is considerably worse
than that of a fit with a separate toroidal and poloidal halo field
($\Delta\chi^2 = +401$ for model a vs.\ b and $\Delta\chi^2 = +462$
for model h vs.\ i) there are only 6 free parameters for the halo,
instead of the 10 parameters needed for the \modelBase model.  On a
technical level, the better $\chi^2$ of the \modelBase model can be
due to the freedom of two different radial cutoffs values, $\rP$ and
$\rT$, for the toroidal and poloidal components that are available in
the fit with an explicit toroidal halo.  A future implementation of
the unified halo model, in which a physically motivated model limits
the build-up of the toroidal field, could introduce additional radial
and vertical dependencies to the amount of twist and hence result in a
potentially even better description of the halo field.

In summary, the unified halo model describes the data well using only
6 free parameters for the halo, instead of the 10 parameters needed
for the \modelBase model. This is the first quantitative demonstration
of earlier conjectures, that the toroidal halo of the Milky Way likely
arose dynamically from differential rotation.

\input{allModelsTable.tex}

\subsection{Model Ensemble}
\label{sec:ensemble}

Based on the results presented in the previous sections, we compiled a
list of viable GMF variations in \cref{tab:allmodels}. Here we already
pre-selected models with $\hDiff\geq 6$~kpc, see
Sec.~\ref{sec:nencre}, and used only the most extreme \nel-$B$
anti-correlation of $\kappa = -0.4$, see
Sec.~\ref{sec:striacorr}. This leads to 18 GMF models, consisting of 7
variations of the parametric models, three variations using \NE rather
than \YMW, one variation including an \nel-$B$ anti-correlation, three
fits with different cosmic-ray electron models and four fits with
different synchrotron products.

How different are these 18 GMF models from each other? One possible
way to measure the model differences is to study the motion of charged
particles in the respective models and calculate the space-angle
difference in their direction after traversing the Galaxy. We
performed such comparisons for many particles (rigidity $\R=
10^{19}$~V) starting in different directions from Earth and determined
the average deflection difference, its 90\% quantile and the maximum
difference for each model combination. The result is shown
in \cref{fig:deflComp}.  Based on these model-to-model comparisons we
can further narrow down the most important variations that encompass
the GMF uncertainties. For instance, it can be seen that of the four
models with different synchrotron products, (a, o, p, q),
the \CosmoGlobe model (q) differs the most from the \modelBase model
(a). The difference is similar when comparing the combination of \NE
and \CosmoGlobe (q) to \NE with default synchrotron (h).

\input{modelParametersTable.tex}

Similar comparisons lead us to select the set of eight GMF models,
identified by the model acronym name in the last column of
Table~\ref{tab:allmodels}.  The parameters of the eight members of
this final GMF model ensemble are given in Table~\ref{tab:modelpars}
and their properties can be summarized as follows:\\

\begin{itemize}[nosep,leftmargin=0em,labelwidth=*,align=left]
\item \modelBase is tuned to the data
  using the \YMW thermal electron model and a cosmic-ray electron
  model calculated for a diffusion volume with a half-height of
  6~kpc. The parametric model of the GMF is the sum of a spiral disk
  field, an explicit toroidal halo and a coasting poloidal X-field,
\item \modelXr uses an exponential dependence of the mid-plane vertical poloidal field
instead of the default logistic radial cutoff,
\item \modelSpur reduces the grand-design spiral disk field to a single local spur (Orion arm),
\item \modelNe replaces thermal electron model \YMW by \NE of \citet{2002astro.ph..7156C},
\item \modelTwist has a 'twisted X-field' resulting from a unified model of the
toroidal and poloidal halo,

\item \modelKappa assumes an anti-correlation between the thermal electron density and the magnetic field strength,

\item \modelCre uses a cosmic-ray electron model in which the half-height of the diffusion halo is increased to 10~kpc, and

\item  \modelSyn swaps the default synchrotron product for the estimate from the \CosmoGlobe analysis.\\
\end{itemize}

The total energy of the coherent magnetic field within a 20 kpc radius
in these eight models, respectively, is \{1.3, 1.5, 1.2, 1.2, 0.6,
2.3, 1.2, 1.0\} $10^{55}$ erg. For the base model, for instance, the
disk, poloidal, and toroidal components contribute \{0.28, 0.26,
0.75\} $10^{55}$ erg to the total.

%% file: allModelsTable.tex
\begin{table*}[t]
\caption{Pre-selection of model variations (GD: grand-design spiral, LS: local spur, CX: coasting X-field, PF: power function, OS13: \citep{2013MNRAS.436.2127O}, \creuf: \DRAGON plain diffusion. W: \WMAP, P: \Planck, CG: \CosmoGlobe). The last column gives the name assigned to the 8 members of the final GMF ensemble, introduced in Sec.~\ref{sec:ensemble}.}
\label{tab:allmodels}
  \begin{tabular}{cccr@{-}lc@{\hspace{1\tabcolsep}}cc@{\hspace{0.5\tabcolsep}}ccrl} \hline\hline
     \multirow{2}{*}{id} &\multirow{2}{*}{disk}&\multicolumn{3}{c}{halo}&\multicolumn{2}{c}{$n_{\rm e}$}&\multicolumn{2}{c}{$n_{\rm cre}$}&\multirow{2}{*}{QU}&\multicolumn{1}{c}{\multirow{2}{*}{$\chi^2/{\rm ndf}$}}&\multicolumn{1}{c}{model}\\
     &    & toroidal & \multicolumn{2}{c}{poloidal}& model & $\kappa$ &model&$h$&  & & \multicolumn{1}{c}{name} \\ \hline
\multicolumn{5}{l}{\bf Parametric models}\\
a &  GD  &   \toro     &  CX &sigm  & YMW16 &  0     & \creuf&6  & (W+P)/2 & $\text{7923  /   6500 = 1.22}$ & \modelBase \\ 
b &  GD  &   twisted & CX &sigm  & YMW16 &  0     & \creuf&6  & (W+P)/2 & $\text{8324  /   6504 = 1.28}$     \\ 
c &  GD  &   \toro     &  CX &gauss     & YMW16 &  0 & \creuf&6  & (W+P)/2  & $\text{8298  /   6500 = 1.28}$ \\ 
d &  GD  &   \toro     &  CX &sech2     & YMW16 &  0  &\creuf&6  & (W+P)/2 & $\text{8381  /   6500 = 1.29}$  \\ 
e &  GD  &   \toro     &  CX &expo      & YMW16 &  0  & \creuf&6  & (W+P)/2 & $\text{8431  /   6500 = 1.30}$ & \modelXr \\ 
f &  GD  &   \toro     &  PF &sigm & YMW16 &  0     & \creuf&6  & (W+P)/2  & $\text{7926  /   6500 = 1.22}$\\ 
g &  LS  &   \toro     &  CX &sigm  & YMW16 &  0     & \creuf&6  & (W+P)/2 & $\text{7991  /   6501 = 1.23}$ & \modelSpur \\ 
\multicolumn{5}{l}{\vspace*{-.2cm}}\\
\multicolumn{5}{l}{\bf Thermal electrons}\\
h &  GD  &   \toro     &  CX &sigm  & NE2001&  0     & \creuf&6  & (W+P)/2&   $\text{7759 / 6500 = 1.19}$  & \modelNe  \\ 
i &  GD  &   twisted & CX &sigm  & NE2001&  0     & \creuf&6  & (W+P)/2&   $\text{8221 / 6504 = 1.26}$     & \modelTwist  \\ 
j &  GD  &   \toro     &  CX &gauss     & NE2001&  0     & \creuf&6  & (W+P)/2&   $\text{8079 / 6500 = 1.24}$\\ 
k &  GD  &   \toro     &  CX &sigm  & YMW16 &  -0.4  & \creuf&6  & (W+P)/2&   $\text{7905 / 6500 = 1.22}$  & \modelKappa  \\ 
\multicolumn{5}{l}{\vspace*{-.2cm}}\\
\multicolumn{5}{l}{\bf Cosmic-ray electrons}\\
l &  GD  &   \toro     &  CX &sigm  & YMW16 &  0     & \creuf&8    & (W+P)/2& $\text{7940  / 6500 = 1.22}$   \\ 
m &  GD  &   \toro     &  CX &sigm  & YMW16 &  0     & \creuf&10   & (W+P)/2& $\text{7939  / 6500 = 1.22}$   & \modelCre \\ 
n &  GD  &   \toro     &  CX &sigm  & YMW16 &  0     & OS13&10  & (W+P)/2& $\text{7965  / 6500 = 1.23}$      \\ 

\multicolumn{5}{l}{\vspace*{-.2cm}}\\
\multicolumn{5}{l}{\bf Synchrotron Data Product}\\
o &  GD  &   \toro     &  CX &sigm  & YMW16 &  0     & \creuf&6  & P &  $\text{11013 / 6500 = 1.69}$         \\ 
p &  GD  &   \toro     &  CX &sigm  & YMW16 &  0     & \creuf&6  & W &  $\text{8845 / 6500 = 1.36}$          \\ 
q &  GD  &   \toro     &  CX &sigm  & YMW16 &  0     & \creuf&6  & CG & $\text{9758 / 6500 = 1.50}$         & \modelSyn  \\ 
r &  GD  &   \toro     &  CX &sigm  & NE2001&  0     & \creuf&6  & CG & $\text{9551 / 6500 = 1.47}$          \\ 
\hline
  \end{tabular}
\end{table*}

%% file: modelParametersTable.tex
\begin{deluxetable*}{ccccccccccc}[t]
 \tablecaption{Parameter values and uncertainties for the eight GMF model variations. See Table~\ref{tab:glossary} for a description of the parameters and Table~\ref{tab:allmodels} for the definition of each model variation.\label{tab:modelpars}}
\tabletypesize{\scriptsize}
\tablecolumns{5}
\tablewidth{0pt}
\tablehead{\multicolumn{2}{l}{name}&\colhead{\modelBase}&\colhead{\modelXr}&\colhead{\modelSpur}&\colhead{\modelNe}&\colhead{\modelTwist}&\colhead{\modelKappa}&\colhead{\modelSyn}&\colhead{\modelCre}&\colhead{unit}}
\startdata
\multicolumn{5}{l}{{\bf disk field}}\\
&\pitch   &$   10.11\pm    0.13$ &$   10.03\pm    0.13$ &$    12.1\pm     0.6$ &$    11.9\pm     0.4$ &$    12.1\pm     0.4$ &$   10.15\pm    0.14$ &$    9.90\pm    0.13$ &$   10.16\pm    0.13$ &  deg. \\
&\fsz     &$   0.794\pm   0.032$ &$   0.715\pm   0.024$ &$   0.750\pm   0.027$ &$   0.674\pm   0.018$ &$    0.94\pm    0.05$ &$   0.812\pm   0.027$ &$   0.622\pm   0.018$ &$   0.808\pm   0.033$ &  kpc  \\
&\fswz    &$   0.107\pm   0.026$ &$   0.099\pm   0.023$ &$   0.123\pm   0.024$ &$   0.061\pm   0.020$ &$    0.15\pm    0.07$ &$   0.119\pm   0.025$ &$   0.067\pm   0.018$ &$   0.108\pm   0.025$ &  kpc  \\
&$B_1$    &$    1.09\pm    0.14$ &$    0.99\pm    0.15$ &$   -4.30\pm    0.18$ &$    1.43\pm    0.26$ &$    1.37\pm    0.17$ &$    1.41\pm    0.19$ &$    0.81\pm    0.12$ &$    1.20\pm    0.14$ & \muG \\
&$B_2$    &$    2.66\pm    0.21$ &$    2.18\pm    0.22$ &          --          &$     1.4\pm     0.4$ &$    2.01\pm    0.30$ &$    3.53\pm    0.27$ &$    2.06\pm    0.20$ &$    2.75\pm    0.21$ & \muG \\
&$B_3$    &$    3.12\pm    0.15$ &$    3.12\pm    0.16$ &          --          &$    3.44\pm    0.34$ &$    1.52\pm    0.26$ &$    4.13\pm    0.21$ &$    2.94\pm    0.14$ &$    3.21\pm    0.15$ & \muG \\
&$\phi_1$ &$     263\pm       9$ &$     247\pm      10$ &$   155.9\pm     1.4$ &$     200\pm      11$ &$     236\pm      11$ &$     264\pm       8$ &$     230\pm      13$ &$     265\pm       8$ & deg.\\
&$\phi_2$ &$    97.8\pm     2.8$ &$    98.6\pm     3.2$ &         --           &$     135\pm      12$ &$     102\pm      10$ &$    97.6\pm     3.2$ &$    97.4\pm     3.3$ &$    98.2\pm     2.8$ & deg.\\
&$\phi_3$ &$    35.1\pm     2.2$ &$    34.9\pm     2.4$ &         --           &$      65\pm       4$ &$      56\pm       6$ &$    36.4\pm     2.5$ &$    32.9\pm     2.4$ &$    35.9\pm     2.2$ & deg.\\
&\wS      &         --           &         --           &$    10.3\pm     0.6$ &         --           &         --           &         --           &         --           &         --           & deg. \\
&\phiCS   &         --           &         --           &$   157.2\pm     3.0$ &         --           &         --           &         --           &         --           &         --           & deg. \\
&\lCS     &         --           &         --           &$    31.8\pm     3.0$ &         --           &         --           &         --           &         --           &         --           & deg.  \\
\multicolumn{5}{l}{{\bf toroidal halo}}\\
&\BNT    &$   3.26\pm    0.31$ &$    2.71\pm    0.19$ &$    2.93\pm    0.23$ &$    2.63\pm    0.17$ &         --           &$     4.6\pm     0.4$ &$    2.40\pm    0.12$ &$     3.7\pm     0.4$ &  \muG  \\
&\BST    &$  -3.09\pm    0.30\phantom{-}$ &$   -2.57\pm    0.18\phantom{-}$ &$   -2.60\pm    0.21\phantom{-}$ &$   -2.57\pm    0.17\phantom{-}$ &         --           &$    -4.5\pm     0.4\phantom{-}$ &$   -2.09\pm    0.11\phantom{-}$ &$   -3.50\pm    0.35\phantom{-}$ &  \muG  \\
&\zT     &$     4.0\pm     0.7$ &$     5.5\pm     0.9$ &$     6.1\pm     1.4$ &$     4.6\pm     0.8$ &         --           &$     3.6\pm     0.6$ &$     5.6\pm     0.8$ &$     2.9\pm     0.4$ &  kpc  \\
&\rT     &$   10.19\pm    0.17$ &$   10.13\pm    0.19$ &$    9.75\pm    0.13$ &$   10.13\pm    0.20$ &         --           &$   10.21\pm    0.17$ &$    9.42\pm    0.08$ &$   10.41\pm    0.20$ &  kpc  \\
&\wT     &$     1.7\pm     0.4$ &$     2.1\pm     0.6$ &$    1.42\pm    0.30$ &$    1.15\pm    0.29$ &         --           &$     1.7\pm     0.4$ &$    0.92\pm    0.16$ &$     1.7\pm     0.4$ &  kpc  \\
&$t$     &         --           &         --           &         --           &         --           &$    54.7\pm     1.1$ &         --           &         --           &         --           & Myr   \\
\multicolumn{5}{l}{{\bf poloidal halo}}\\
&\BP &$   0.978\pm   0.033$ &$     5.8\pm     0.4$ &$    0.99\pm    0.04$ &$   0.984\pm   0.031$ &$   0.628\pm   0.020$ &$    1.35\pm    0.04$ &$   0.809\pm   0.024$ &$   0.969\pm   0.034$ & \muG \\
&$p$ &$    1.43\pm    0.09$ &$    1.95\pm    0.14$ &$    1.40\pm    0.09$ &$    1.68\pm    0.11$ &$    2.33\pm    0.10$ &$    1.34\pm    0.10$ &$    1.58\pm    0.09$ &$    1.42\pm    0.09$ & --  \\
&\zP &$     4.5\pm     0.4$ &$    2.37\pm    0.22$ &$     4.5\pm     0.4$ &$    3.65\pm    0.28$ &$    2.63\pm    0.10$ &$     4.8\pm     0.5$ &$    3.53\pm    0.24$ &$     4.6\pm     0.4$ & kpc \\
&\rP &$    7.29\pm    0.06$ &$    2.50\pm    0.07$ &$    7.20\pm    0.06$ &$    7.41\pm    0.05$ &$    7.92\pm    0.04$ &$    7.25\pm    0.07$ &$    7.46\pm    0.05$ &$    7.30\pm    0.06$ & kpc \\
&\wP &$   0.112\pm   0.029$ &         --           &$   0.123\pm   0.034$ &$   0.142\pm   0.030$ &$   0.291\pm   0.035$ &$   0.143\pm   0.033$ &$   0.150\pm   0.022$ &$   0.109\pm   0.027$ & kpc \\
&\aP &         --           &$     6.2\pm     0.8$ &         --           &         --           &         --           &         --           &         --           &         --           & kpc \\ \hline
\multicolumn{5}{l}{{\bf other model parameters}}\\
&$\kappa$ & 0 & 0 & 0 & 0  & 0 & $-0.4\phantom{-}$ & 0 & 0 & -- \\
&$\xi$ &$   0.346\pm   0.034$ &$    0.51\pm    0.04$ &$   0.330\pm   0.033$ &$   0.336\pm   0.029$ &$    0.78\pm    0.04$ &$       0$ &$    0.63\pm    0.04$ &$   0.250\pm   0.033$ & -- \\
\enddata
\end{deluxetable*}

%% file: applications.tex
Each of the eight GMF model variations introduced in the previous
section can be considered viable estimates of the GMF given the
current \RM, \Q and \U data, and the differences between the models
can be regarded as an estimate of the lower limit on the uncertainty
of our knowledge of the magnetic field of the Galaxy.

In this section, we discuss two applications in which we
propagate the uncertainties of the GMF to uncertainties in the
deflection of ultrahigh-energy cosmic rays and in the conversion
probability of axions in the Galactic magnetic field.

\def\rig{20}

\begin{figure*}[t]
   \def\figw{0.38}
   \def\xlabel{0}
   \def\ylabel{44}
  \centering
\begin{overpic}[clip,rviewport=-0.02 -0.1 1 1,width=\figw\linewidth]{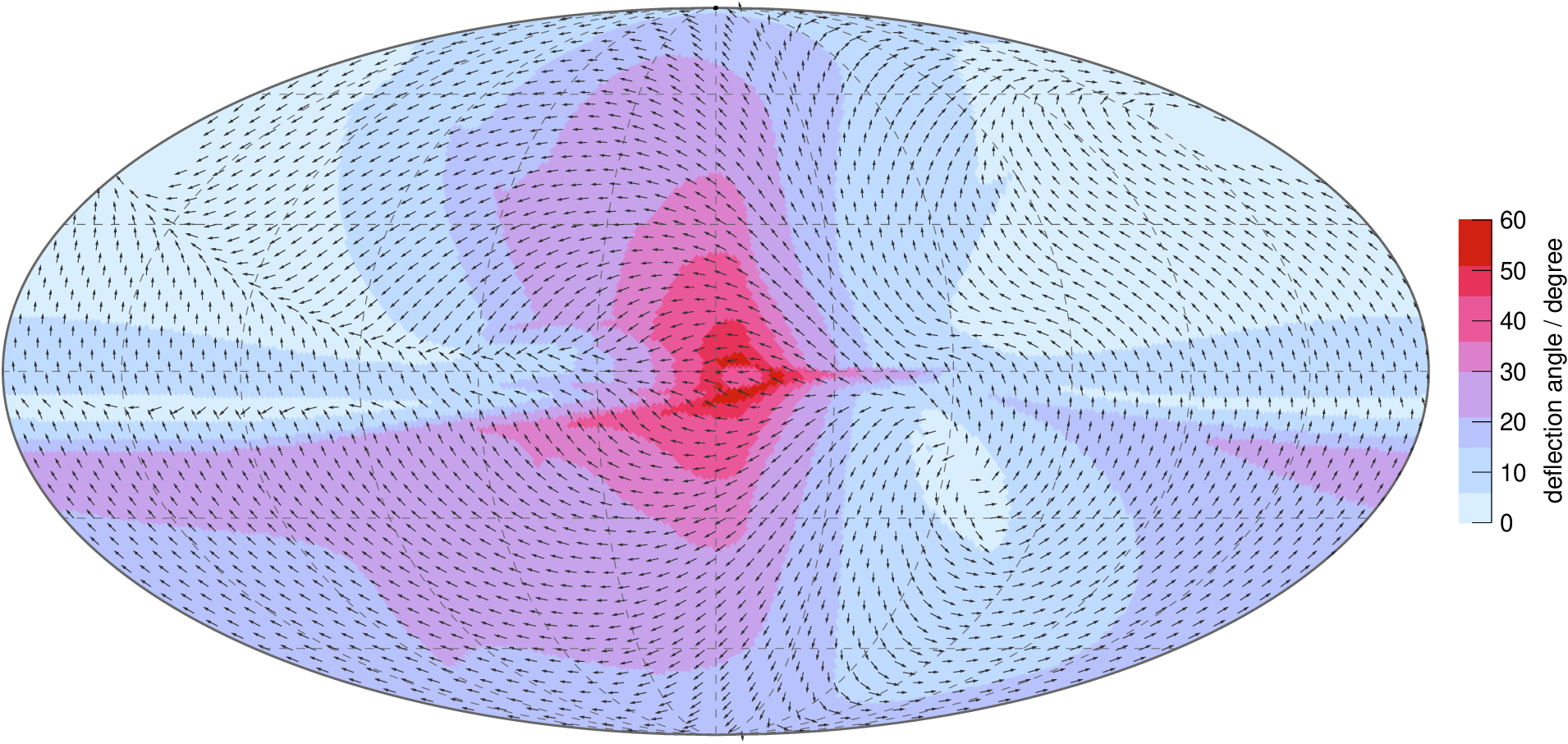}
   \put(\xlabel,\ylabel){\scriptsize \JFTwelve}
\end{overpic}\\
\begin{overpic}[clip,rviewport=-0.02 -0.1 1 1,width=\figw\linewidth]{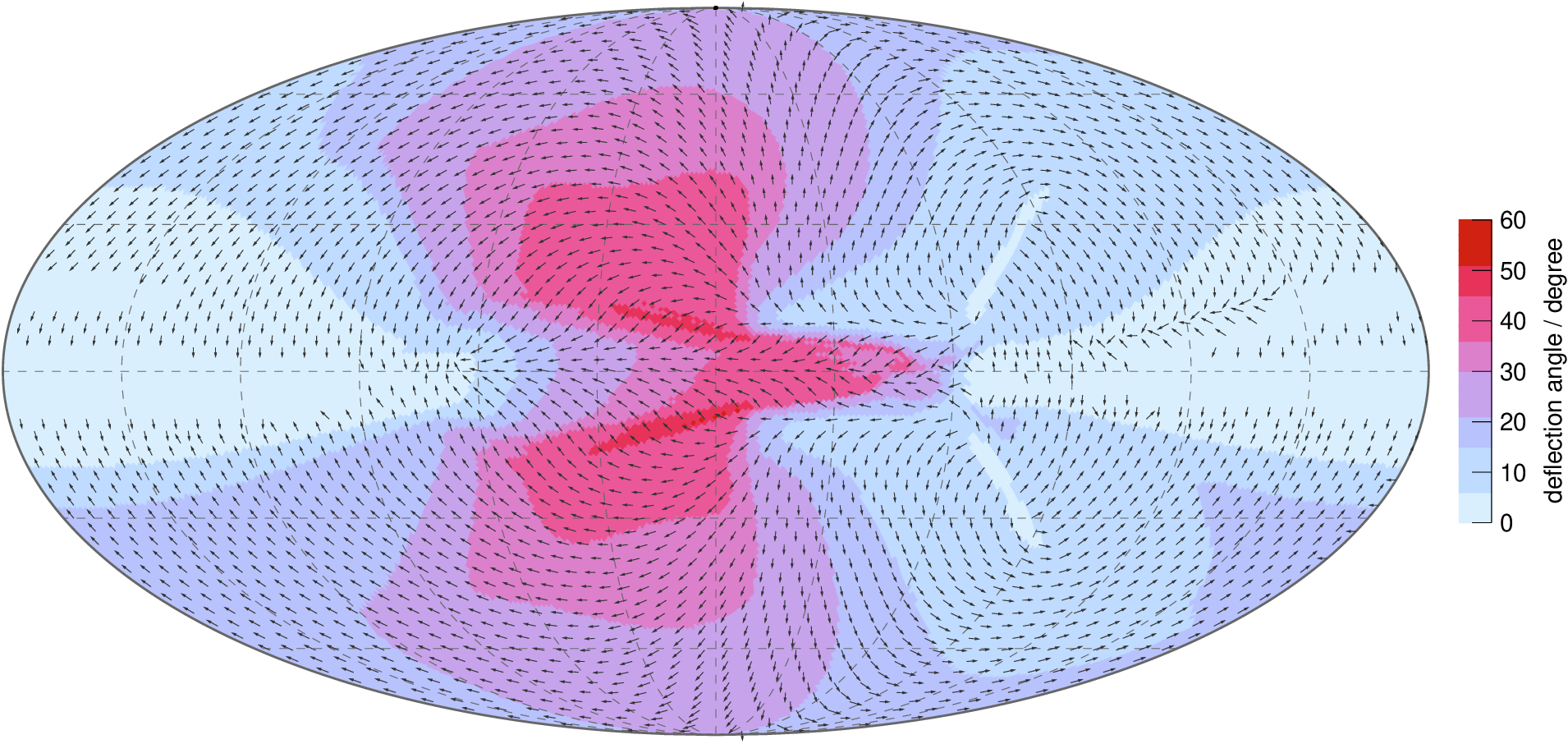}
   \put(\xlabel,\ylabel){\scriptsize \modelBase}
\end{overpic}\qquad
\begin{overpic}[clip,rviewport=-0.02 -0.1 1 1,width=\figw\linewidth]{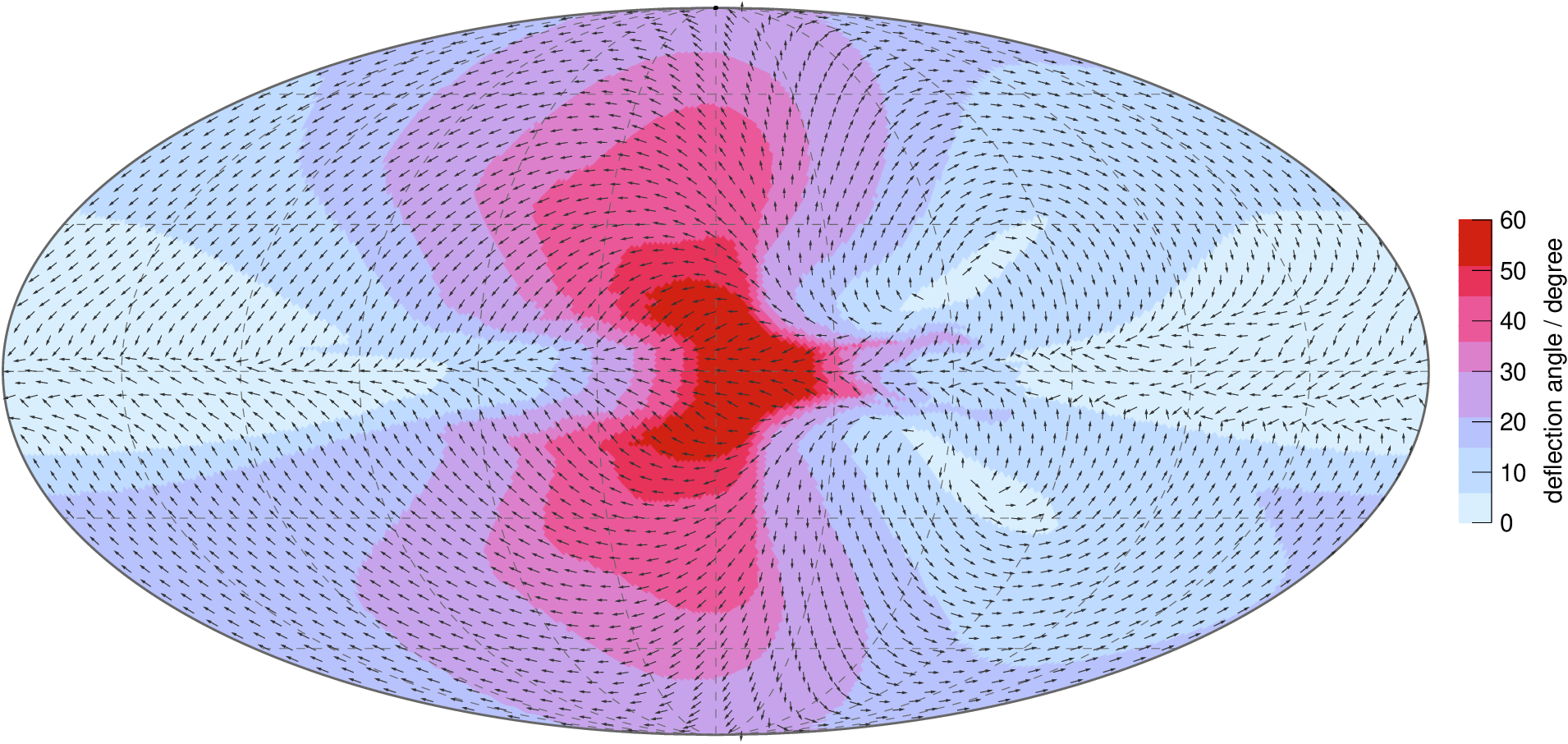}
   \put(\xlabel,\ylabel) {\scriptsize \modelXr}
\end{overpic}
\begin{overpic}[clip,rviewport=-0.02 -0.1 1 1,width=\figw\linewidth]{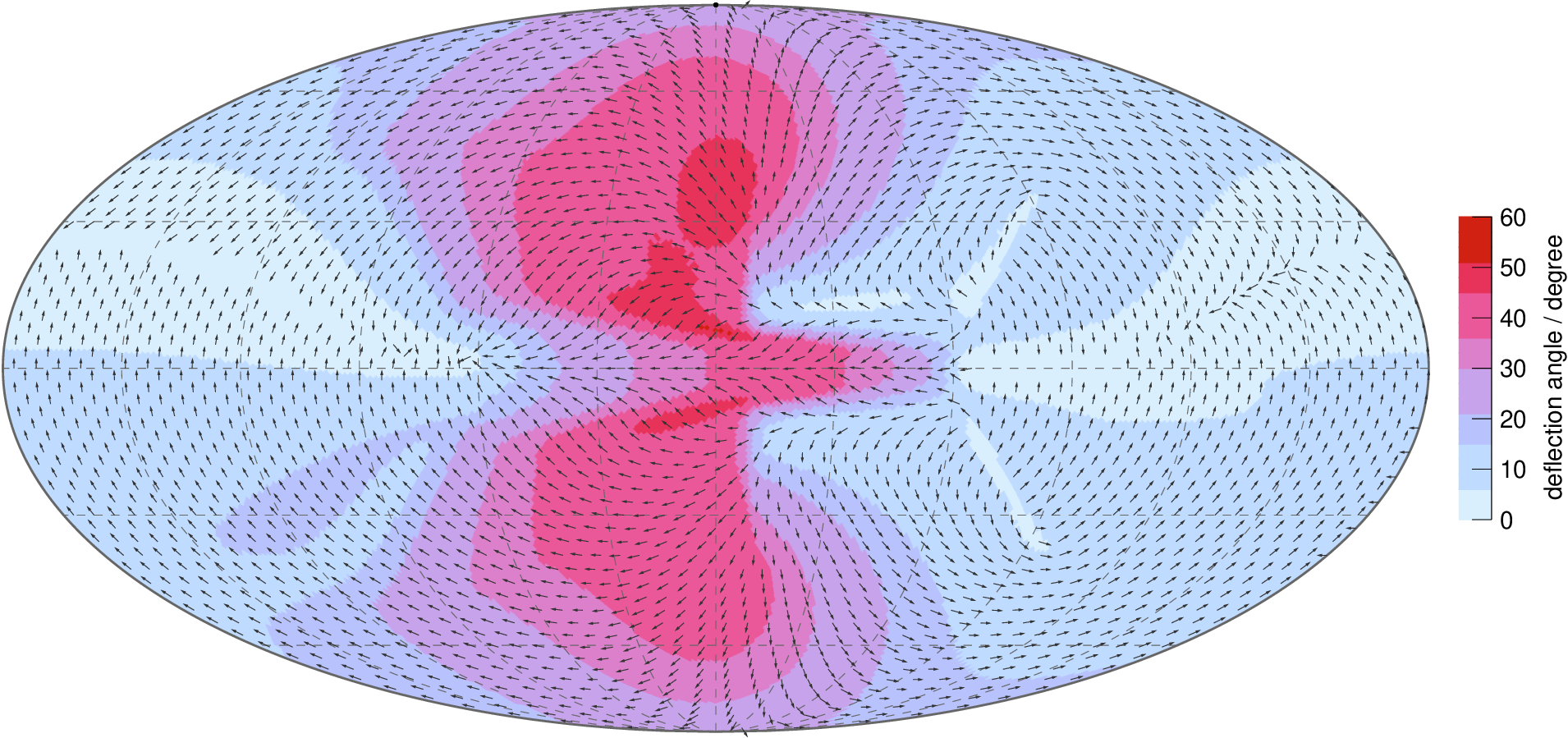}
   \put(\xlabel,\ylabel){\scriptsize \modelSpur}
\end{overpic}\qquad
\begin{overpic}[clip,rviewport=-0.02 -0.1 1 1,width=\figw\linewidth]{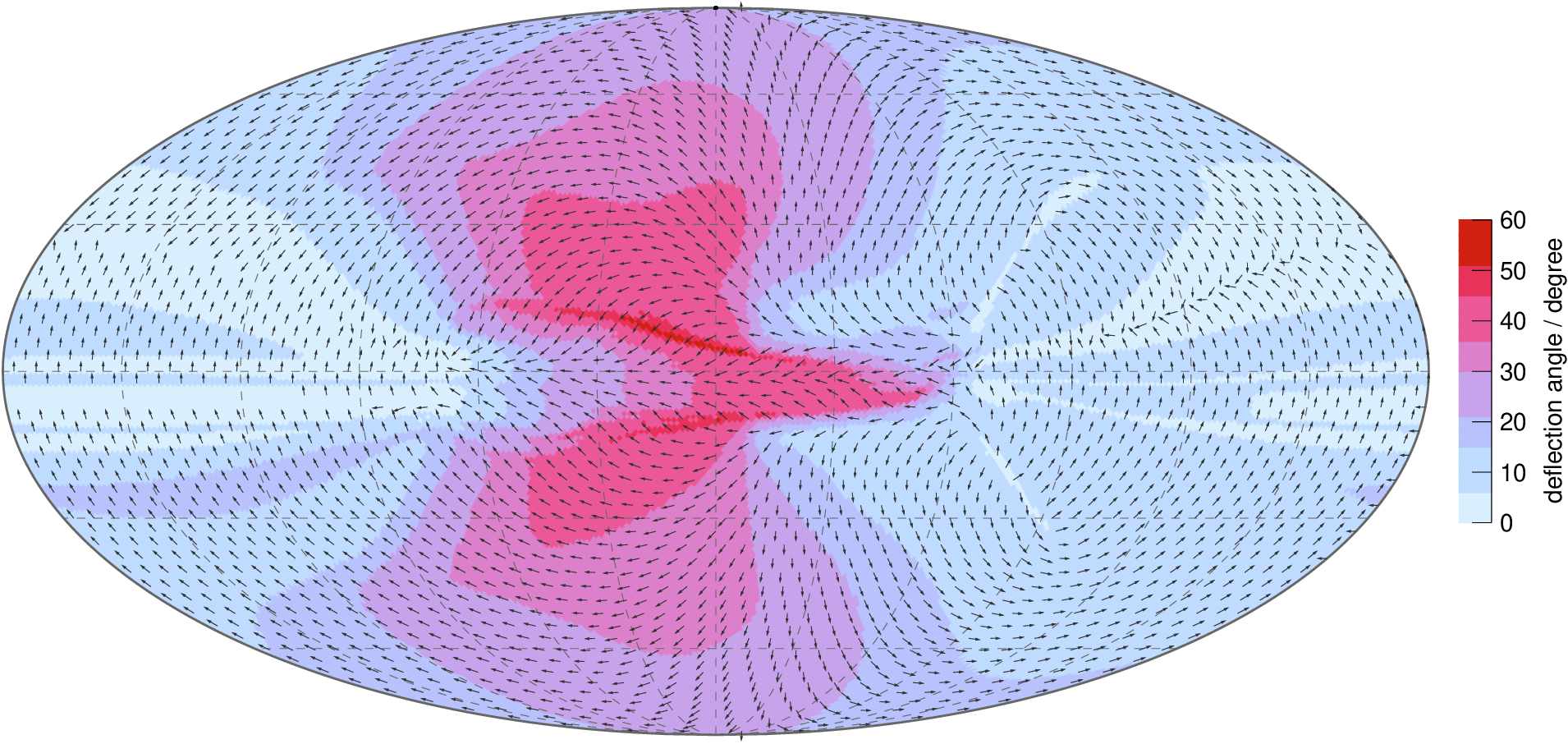}
   \put(\xlabel,\ylabel) {\scriptsize \modelNe}
\end{overpic}
\begin{overpic}[clip,rviewport=-0.02 -0.1 1 1,width=\figw\linewidth]{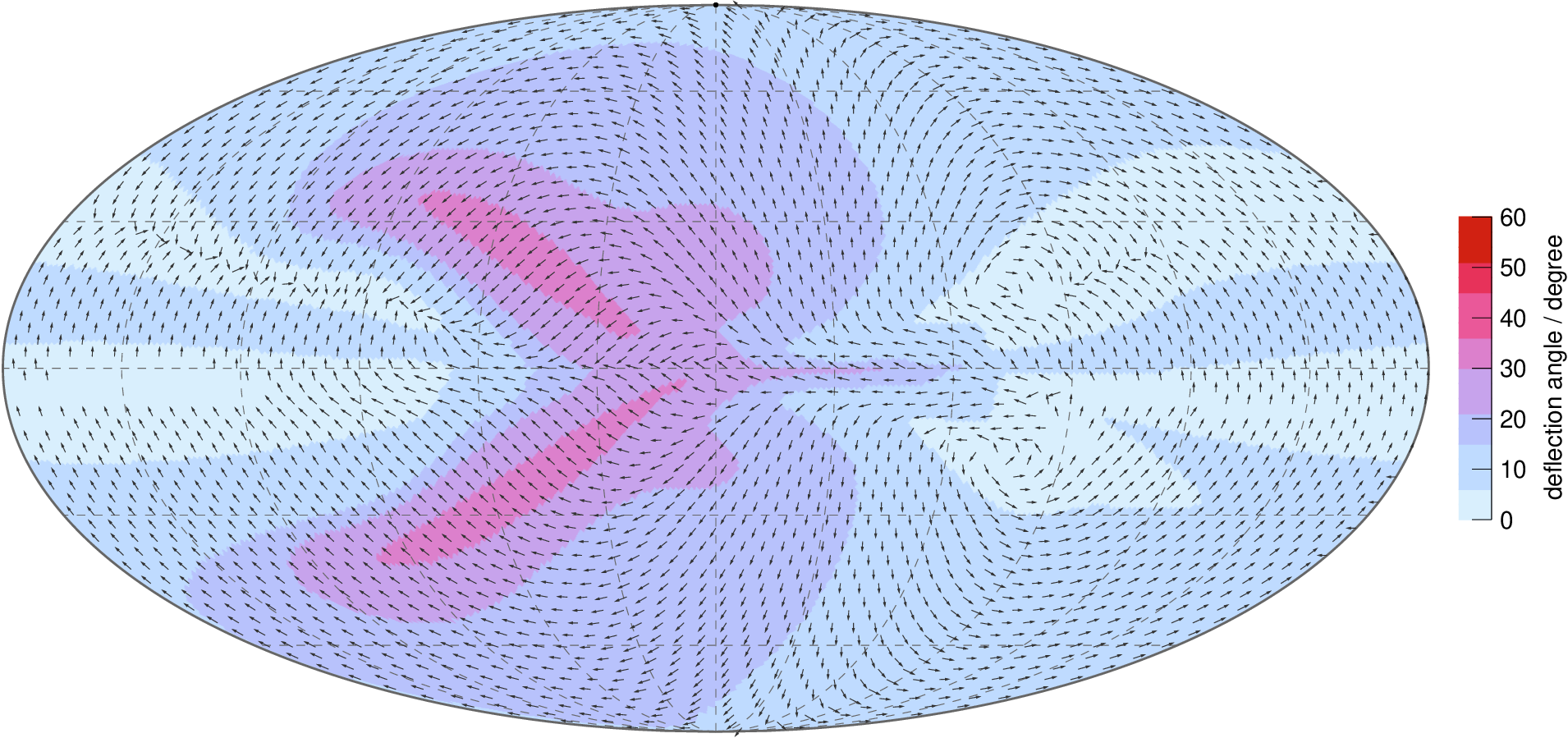}
   \put(\xlabel,\ylabel){\scriptsize \modelTwist}
\end{overpic}\qquad
\begin{overpic}[clip,rviewport=-0.02 -0.1 1 1,width=\figw\linewidth]{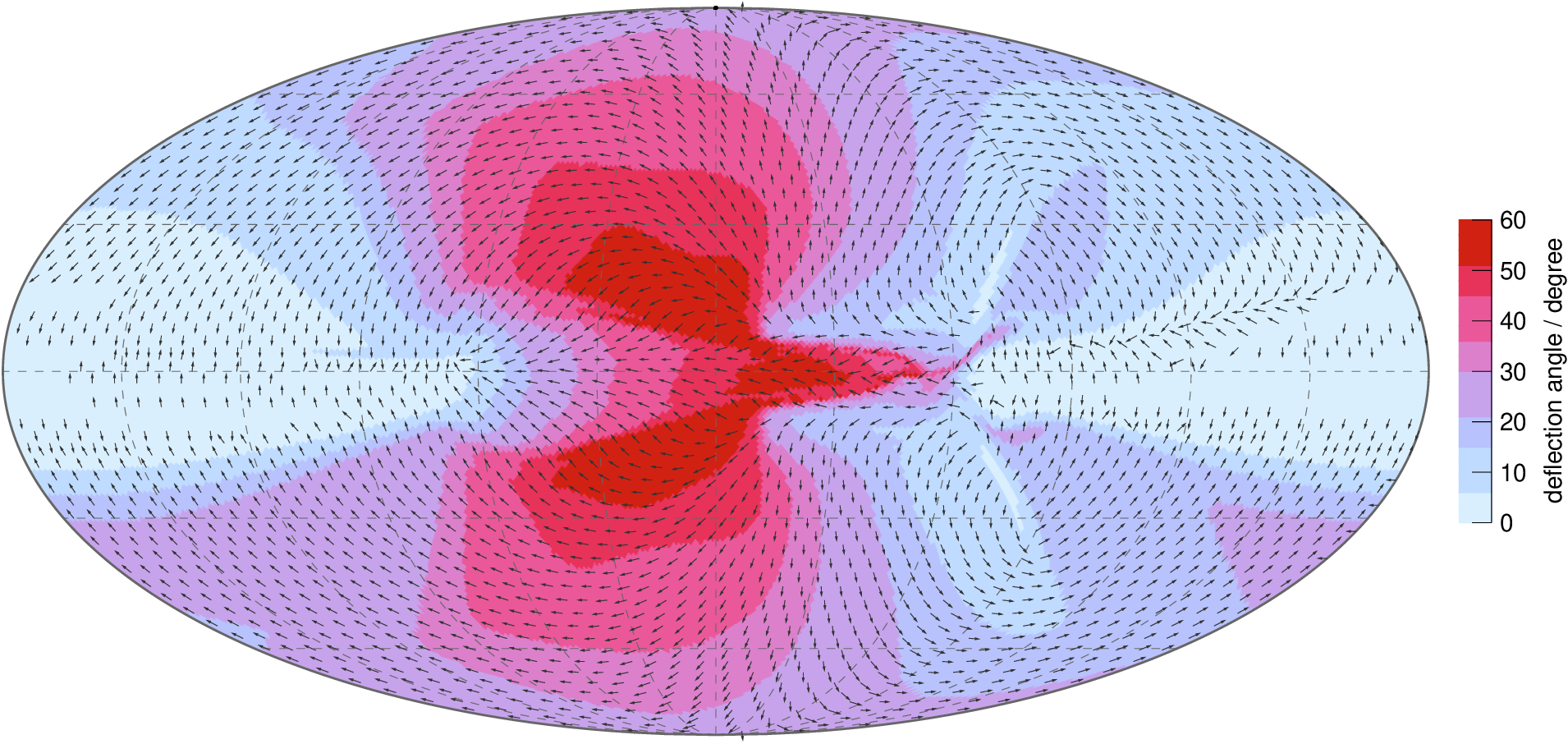}
   \put(\xlabel,\ylabel) {\scriptsize \modelKappa}
\end{overpic}
\begin{overpic}[clip,rviewport=-0.02 -0.1 1 1,width=\figw\linewidth]{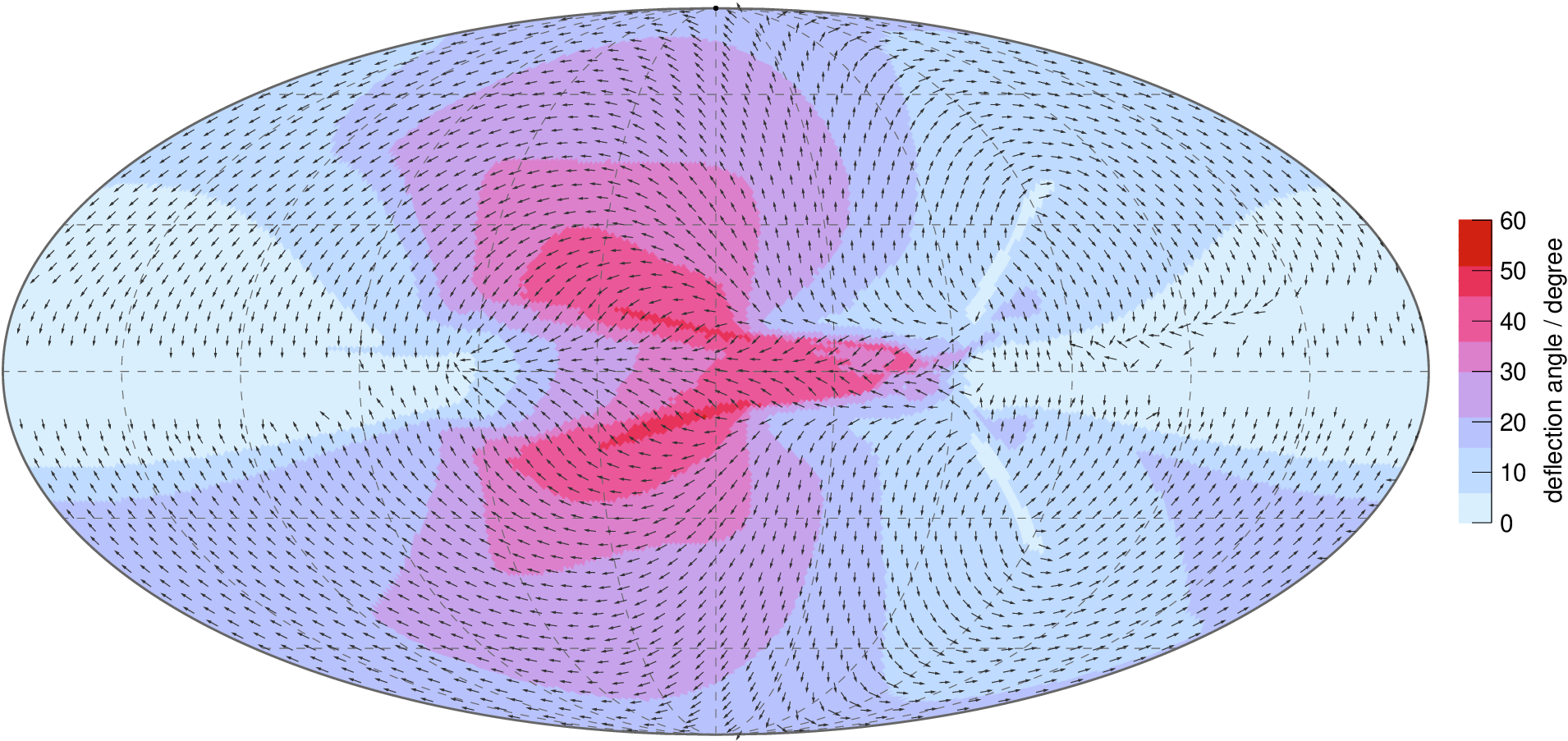}
   \put(\xlabel,\ylabel){\scriptsize \modelCre}
\end{overpic}\qquad
\begin{overpic}[clip,rviewport=-0.02 -0.1 1 1,width=\figw\linewidth]{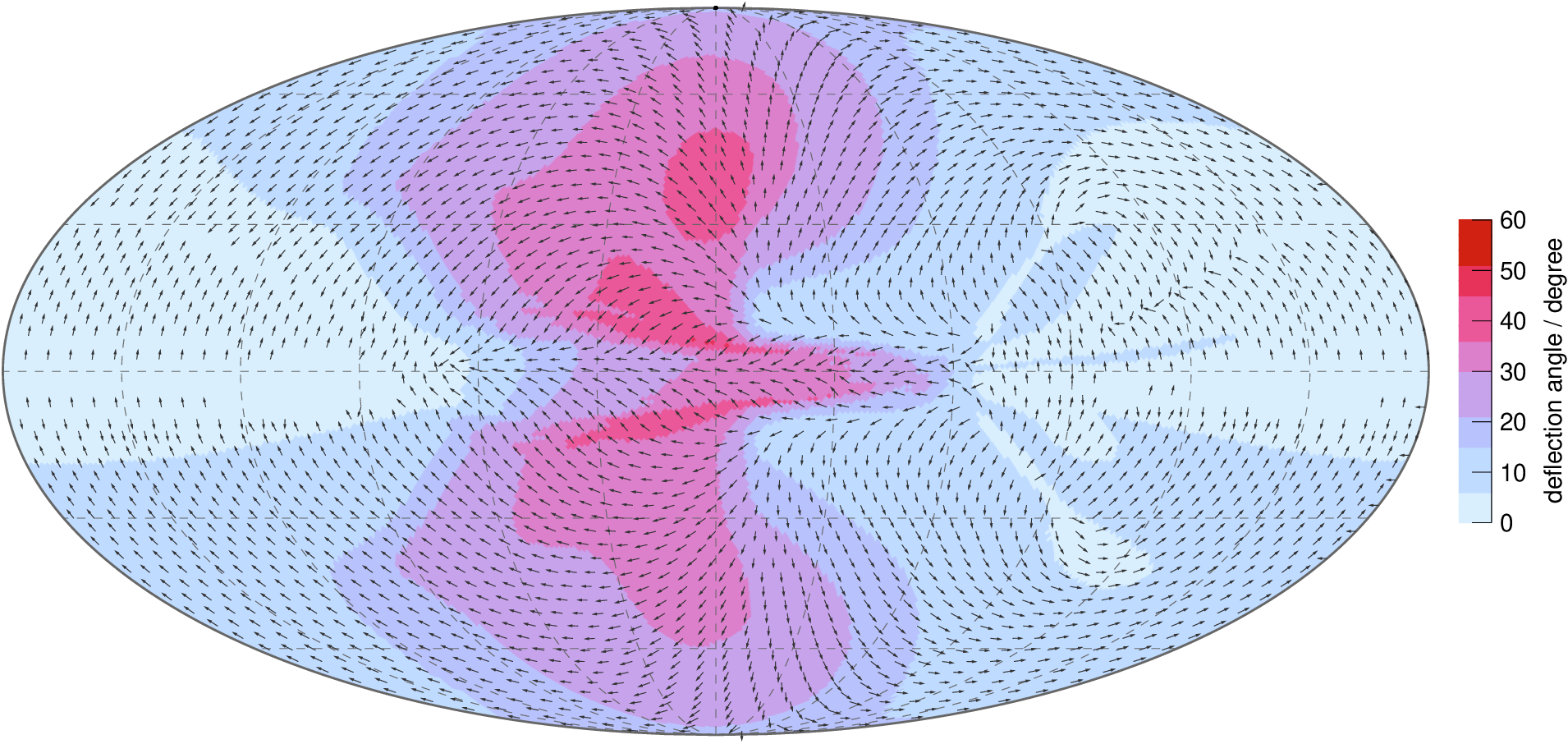}
   \put(\xlabel,\ylabel) {\scriptsize \modelSyn}
\end{overpic}
   \caption{Angular deflections of ultrahigh-energy cosmic rays
     in \JFTwelve model (top) and the eight model variations derived
     in this paper. Colors and arrows denote the size and
     direction of the deflection in the GMF following the particles
     from Earth to the edge of the Galaxy. Positions on the
     skymap denote arrival directions at Earth. The rigidity is $2\times10^{19}$~V.} \label{fig:defl}
\end{figure*}

\subsection{Cosmic-Ray Deflections}

\begin{figure*}[t]
  \centering
  \def\ww{0.85}
  \begin{overpic}[width=\ww\textwidth]{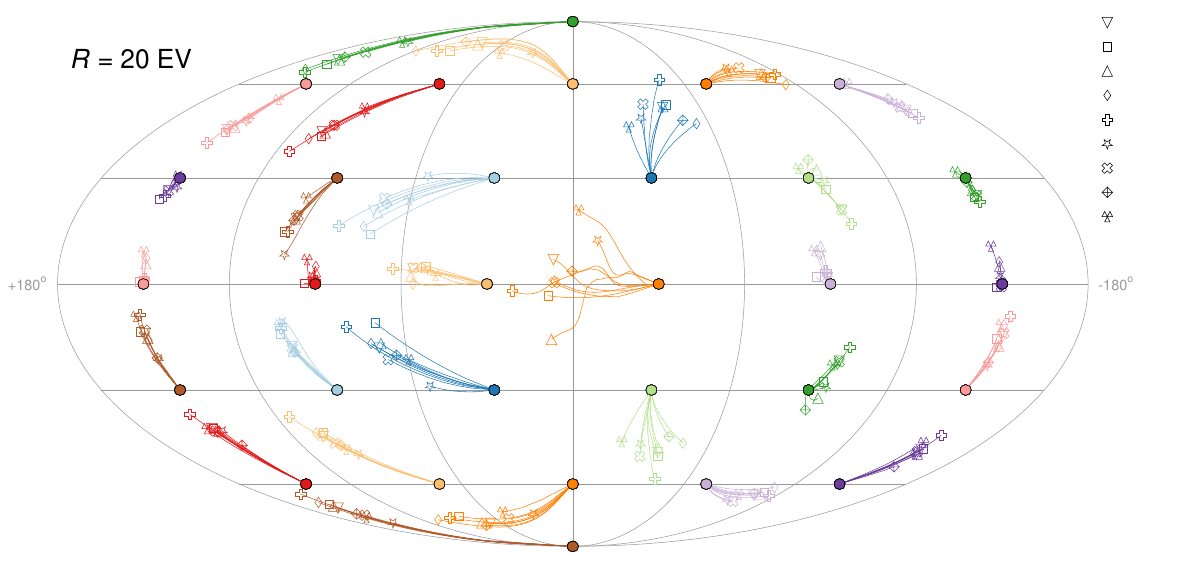}
    \def\xlab{95.3}
    \put(\xlab,45.80000){\scalebox{\ww}{\small\modelBase}}
    \put(\xlab,43.73625){\scalebox{\ww}{\small\modelXr}}
    \put(\xlab,41.67250){\scalebox{\ww}{\small\modelNe}}
    \put(\xlab,39.60875){\scalebox{\ww}{\small\modelSpur}}
    \put(\xlab,37.54500){\scalebox{\ww}{\small\modelKappa}}
    \put(\xlab,35.48125){\scalebox{\ww}{\small\modelTwist}}
    \put(\xlab,33.41750){\scalebox{\ww}{\small\modelCre}}
    \put(\xlab,31.35375){\scalebox{\ww}{\small\modelSyn}}
    \put(\xlab,29.29000){\scalebox{\ww}{\small\JFTwelve}}
\end{overpic}
   \caption{Angular deflections of ultrahigh-energy cosmic rays in the
     eight model variations derived in this paper and \JFTwelve.  The
     cosmic-ray rigidity is \rig{}~EV ($2\times10^{19}$~V). Filled circles denote a grid of
     arrival directions and the open symbols are the back-tracked
     directions at the edge of the Galaxy.} \label{fig:defl2}
\end{figure*}

\begin{figure*}[t]
\def\figh{0.18}
\centering
\includegraphics[clip,rviewport=0 0 0.9 1,height=\figh\textheight]{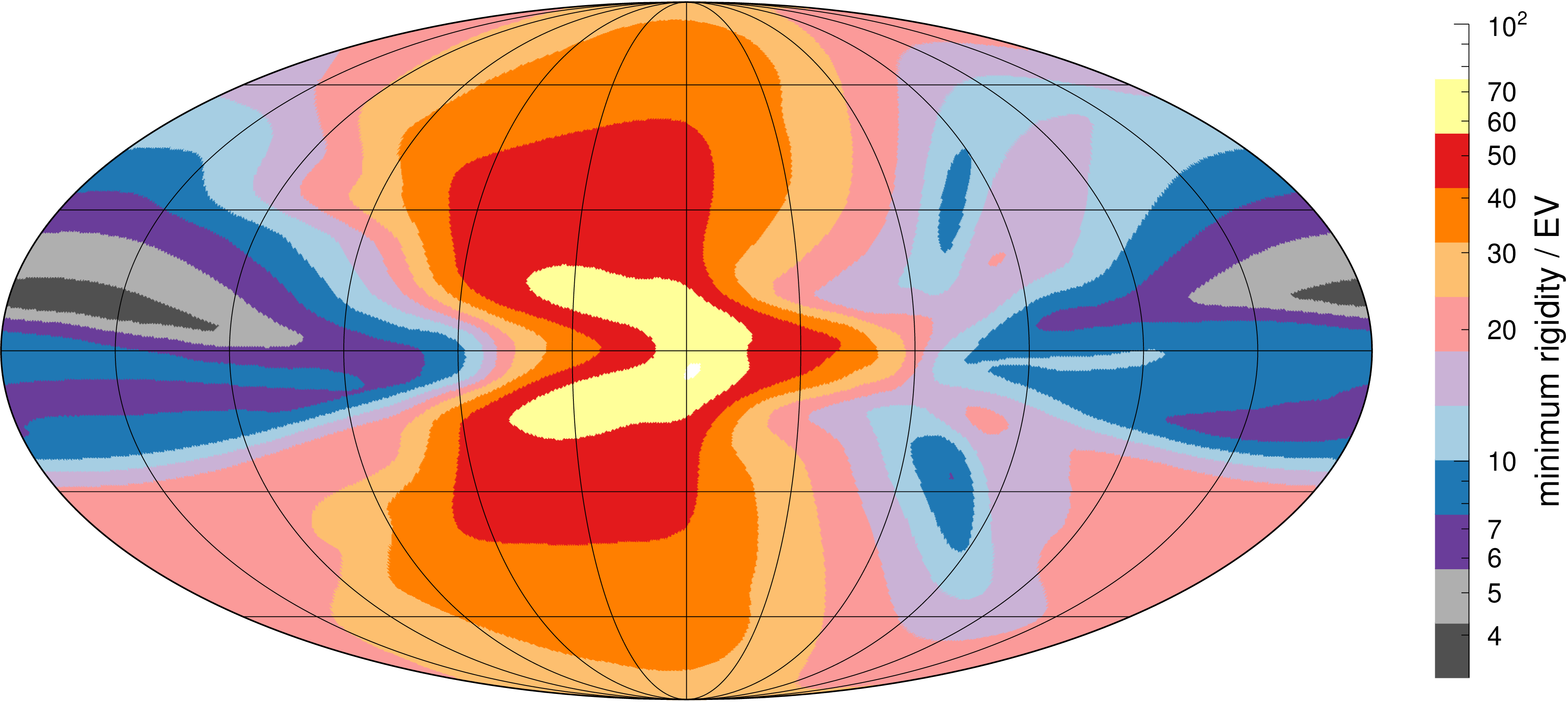}\quad%
\includegraphics[height=\figh\textheight]{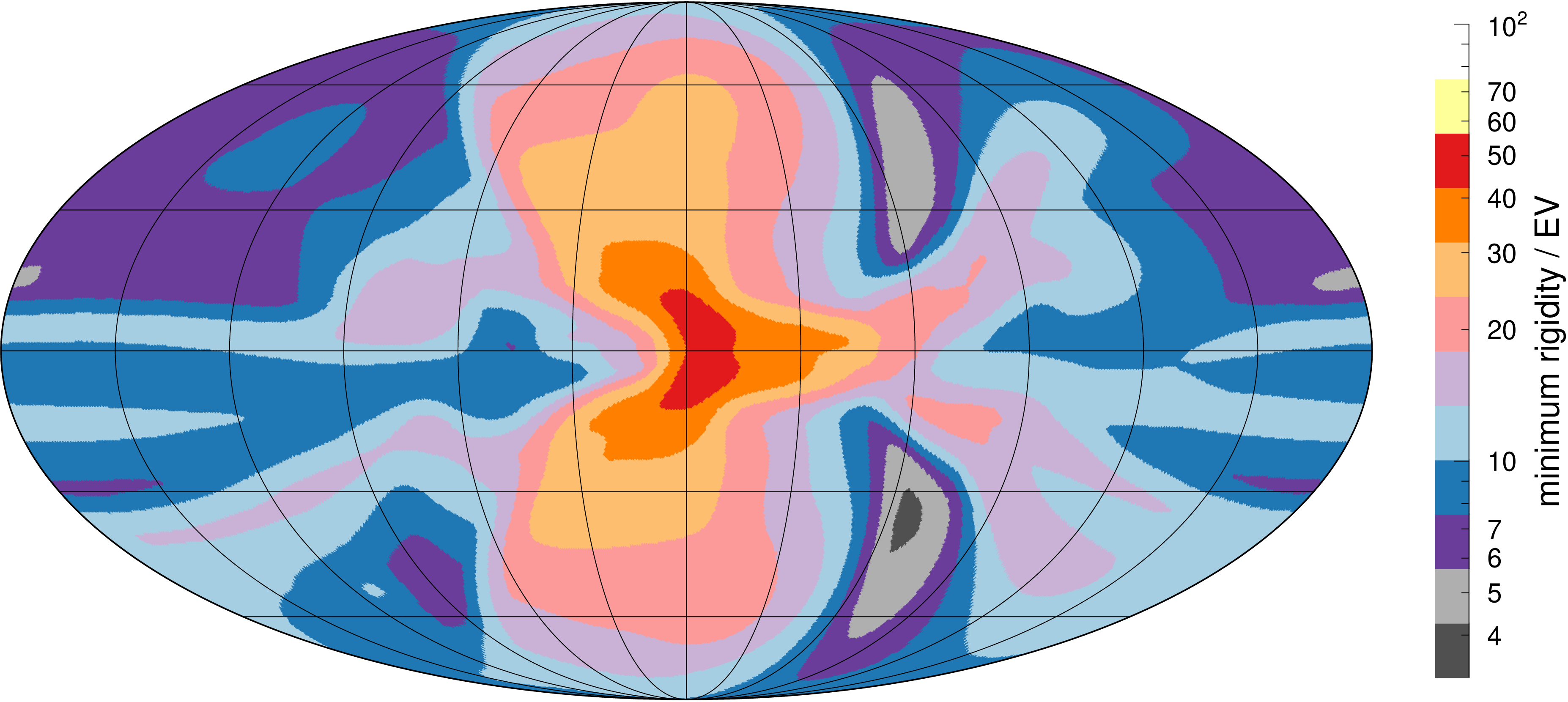}\\
   \caption{Left: Rigidity threshold such that the angular deflection in the given direction is ${\leq}20^\circ$ in all models. Right: Rigidity threshold such that the model predictions of the angular deflection differ by $\leq{20}^\circ$. ($1~\text{EV} = 10^{18}~\text{V}$)} \label{fig:defl3}
\end{figure*}

Given the arrival direction $\bm{v}$ of an ultrahigh-energy cosmic-ray
particle at Earth, the knowledge of the GMF can be used to infer its
arrival direction at the edge of the Galaxy $\bm{u}$. In
\cref{fig:defl} we show sky maps of the {\itshape deflection angle}
$\theta_\text{def} = \arccos(\bm{u}\bm{v})$ for each of the eight
models and the \citetalias{JF12coh} model at a particle rigidity of
\rig{}~EV ($1~\text{EV} = 10^{18}~\text{V}$). These were obtained by
numerically integrating~\citep{cashkarp,Argiro:2007qg} the equation of
motion of a negatively\footnote{Back-tracking a negatively charged
test particle yields the forward trajectory of a positively charged
cosmic ray.}  charged particle in the Galactic magnetic field until it
leaves the Galaxy at a galactocentric radius of
$r_\text{max}=30$~kpc. The magnitude of the deflection angle is
indicated by colors at each of the starting directions $\bm{v}$ on
a \HEALPix grid with $N_\text{side}=64$, and the direction of the
deflection is indicated by an arrow for a subset of directions on a
$N_\text{side}=16$ grid if $\theta_\text{def} > 1^\circ$. As can be
seen, all eight models exhibit qualitatively similar deflection
pattern but with quantitative differences as to be expected from the
results presented in the last section. For instance, the \modelKappa
model has the largest deflections because in this model the magnetic
field scale is set by the polarized synchrotron data, not by the
\RMs. And as expected, the \modelXr model has the largest deflections for
trajectories close to the Galactic center where it has a six times larger
poloidal field strength than the \modelBase model. All models
exhibit a left-right asymmetry with deflections being larger if the
particle is back-tracked towards positive longitudes and smaller for
negative longitudes. This is the consequence of the twisted nature of
the halo field.

For a closer look at the differences between the models, we show the
back-tracked directions $\bm{u}$ in Fig.~\ref{fig:defl2} for a small
set of arrival directions $\bm{v}$. The particle rigidity is again
\rig{}~EV and $\bm{u}$ and $\bm{v}$ are connected by lines
interpolating the back-tracked directions at higher rigidities. This
figure illustrates the similarity of the models, since in many
directions all of them roughly agree on the overall direction of the
deflection, but also shows the model uncertainties, visible as a
scatter in predicted directions for the ensemble of models. It is
worth noting that the deflections predicted by the
widely-used \citetalias{JF12coh} model are generally within the range
of deflections predicted for the GMF models derived in this work.
This is not the case for the deflections calculated with the GMF model
of \citet{2011ApJ...738..192P}, due to the absence of a poloidal
component in that model (c.f., Sec.~\ref{sec:poloidal}).

 \begin{figure*}[ht]
    \def\figw{0.41}
    \def\xlabel{0}
    \def\ylabel{42}
   \centering
\begin{overpic}[clip,rviewport=-0.02 -0.1 1 1,width=\figw\linewidth]{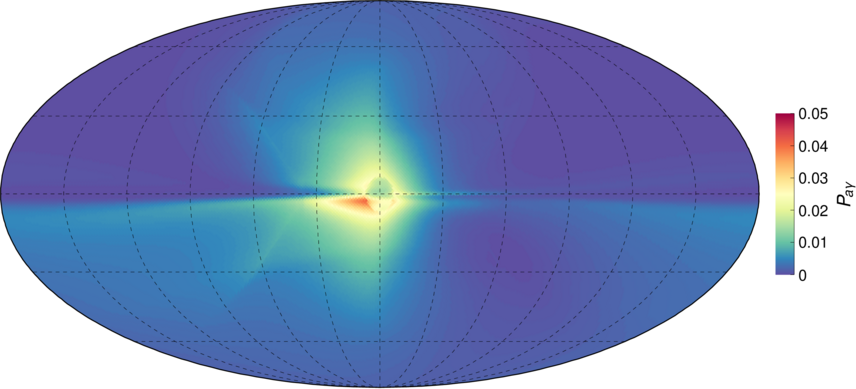}
   \put(\xlabel,\ylabel){\scriptsize \JFTwelve}
 \end{overpic}\\
\begin{overpic}[clip,rviewport=-0.02 -0.1 1 1,width=\figw\linewidth]{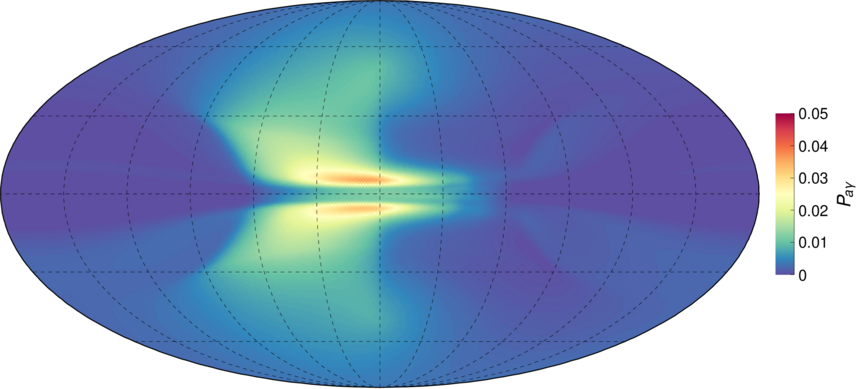}
    \put(\xlabel,\ylabel){\scriptsize \modelBase}
 \end{overpic}\qquad
\begin{overpic}[clip,rviewport=-0.02 -0.1 1 1,width=\figw\linewidth]{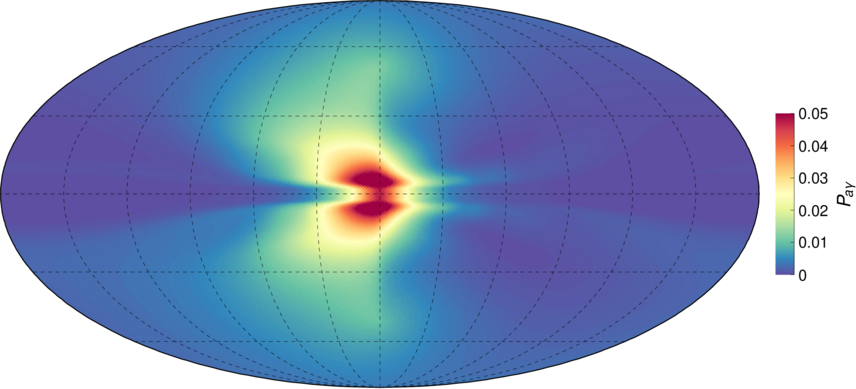}
    \put(\xlabel,\ylabel) {\scriptsize \modelXr}
 \end{overpic}
\begin{overpic}[clip,rviewport=-0.02 -0.1 1 1,width=\figw\linewidth]{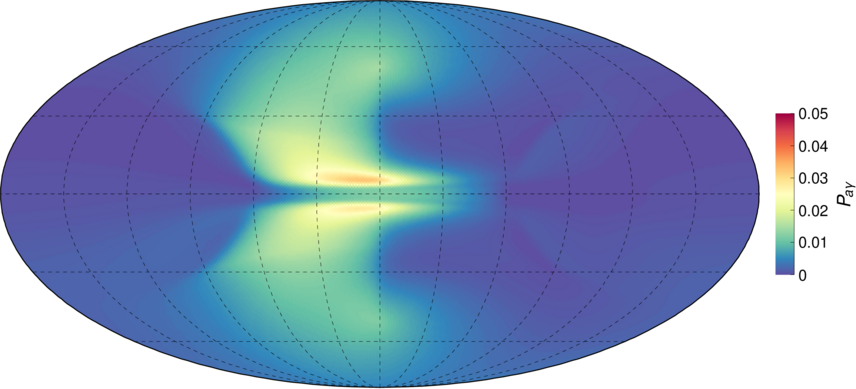}
    \put(\xlabel,\ylabel){\scriptsize \modelSpur}
 \end{overpic}\qquad
\begin{overpic}[clip,rviewport=-0.02 -0.1 1 1,width=\figw\linewidth]{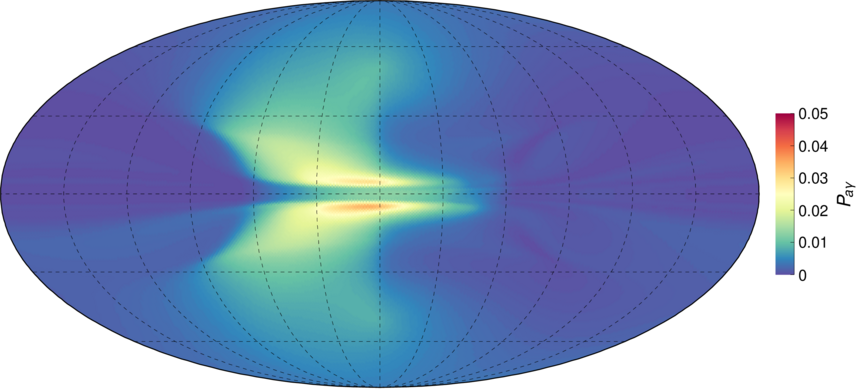}
    \put(\xlabel,\ylabel) {\scriptsize \modelNe}
 \end{overpic}
\begin{overpic}[clip,rviewport=-0.02 -0.1 1 1,width=\figw\linewidth]{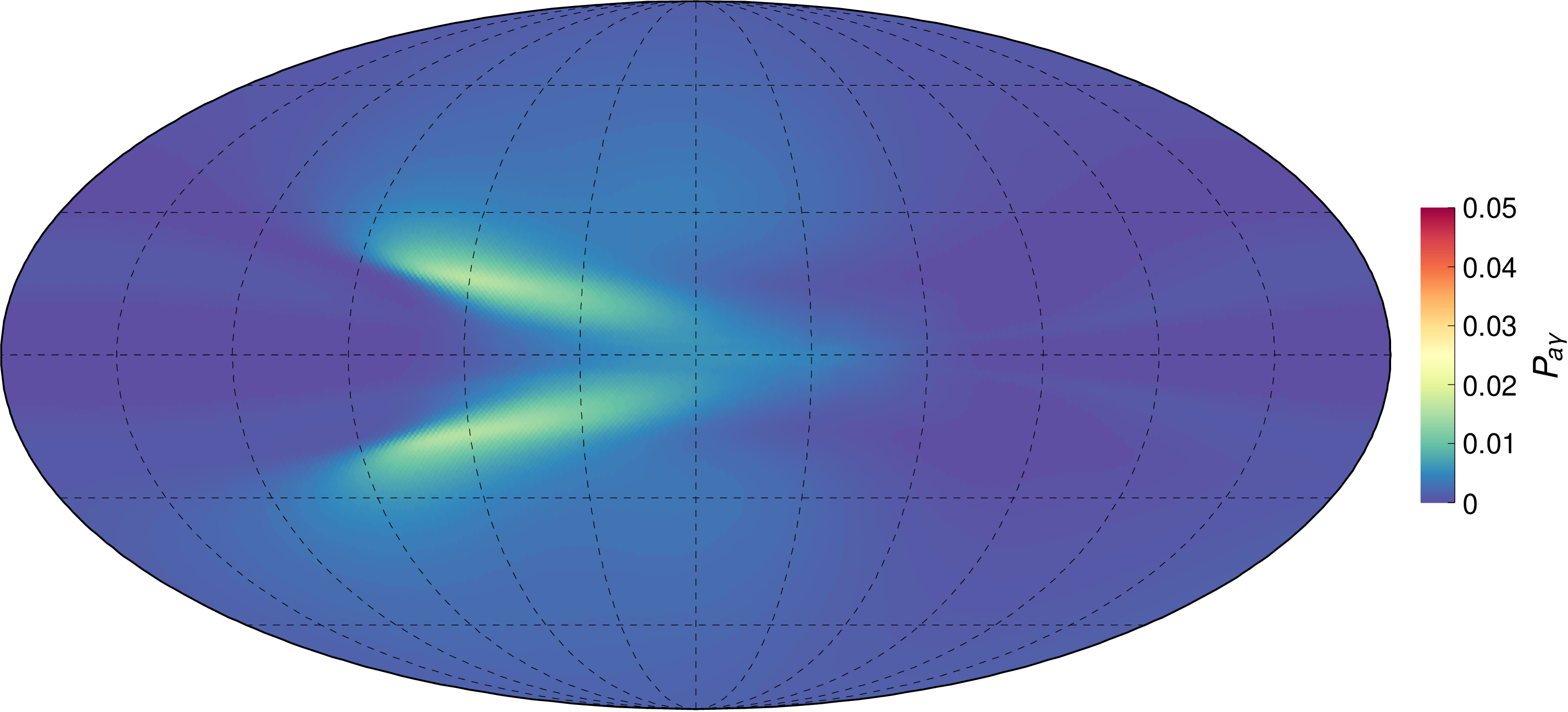}
    \put(\xlabel,\ylabel){\scriptsize \modelTwist}
 \end{overpic}\qquad
\begin{overpic}[clip,rviewport=-0.02 -0.1 1 1,width=\figw\linewidth]{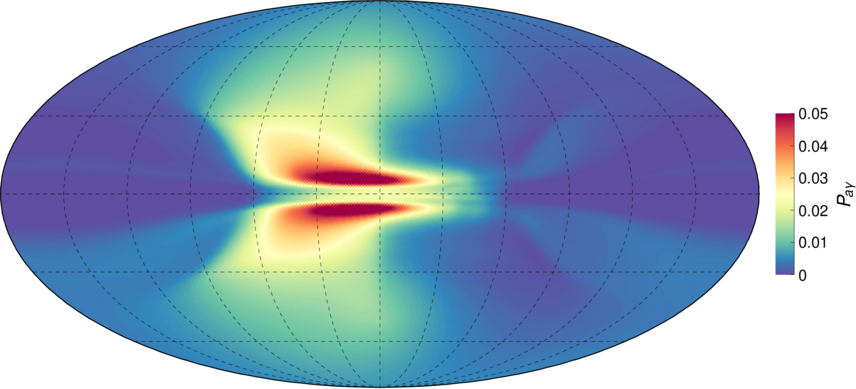}
    \put(\xlabel,\ylabel) {\scriptsize \modelKappa}
 \end{overpic}
\begin{overpic}[clip,rviewport=-0.02 -0.1 1 1,width=\figw\linewidth]{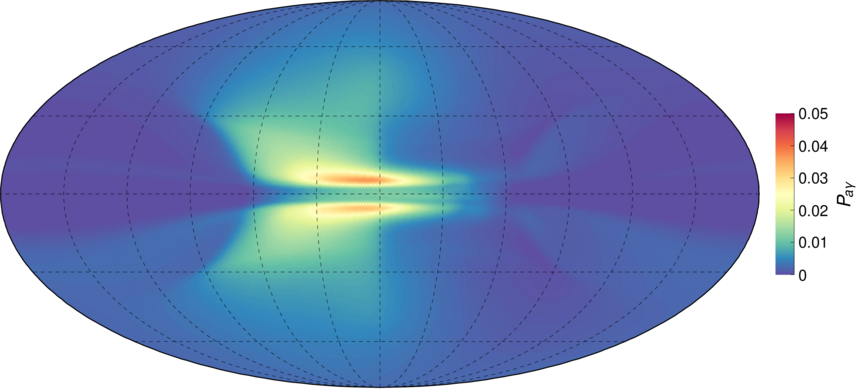}
    \put(\xlabel,\ylabel){\scriptsize \modelCre}
 \end{overpic}\qquad
\begin{overpic}[clip,rviewport=-0.02 -0.1 1 1,width=\figw\linewidth]{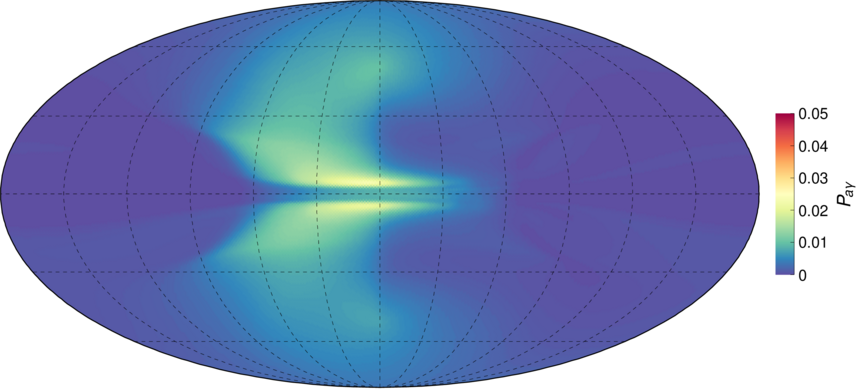}
    \put(\xlabel,\ylabel) {\scriptsize \modelSyn}
 \end{overpic}
   \caption{Axion-photon conversion probabilities, $p_{a\gamma}$, for the
     \JFTwelve model (top), and the eight model variations derived
     in this paper ($g_{a \gamma \gamma} = 5\times 10^{-12}~\text{GeV}^{-1}$, $E_a\gtrsim 1~\text{TeV}$ and $m_a \lesssim
10^{-8}$~eV).}
   \label{fig:paxgam}
 \end{figure*}

Current studies of the anisotropies of ultrahigh-energy cosmic rays
indicate the presence of ``hot spots'' of cosmic-ray clusters at
intermediate angular scales of $20^\circ$
\citep{TelescopeArray:2014tsd, PierreAuger:2022axr}. For the
identification of extragalactic sources related to these
overdensities, a precision in backtracking through the GMF at least as
good as their angular size, $\theta_\text{max}$, is
needed. Figure~\ref{fig:defl3} aims to illustrate this requirement.
In the left panel, we show the minimum rigidity such that the
deflection for a cosmic ray arriving in the given direction is less than
$\theta_\text{max} = 20^\circ$ in all 8 models. Requiring that the
deflections in half of the sky are less than
$\theta_\text{max}=20^\circ$, according to all of these models,
requires the rigidity to be greater than or equal to
$R_{50}^\text{nocorr} = 20$~EV.

The minimum rigidity requirement improves considerably if the arrival
directions are corrected for their expected deflection in the GMF.
The limit on the precision with which we infer the source position
arises from the {\itshape difference} between the models, and not the
overall magnitude of the deflection. The differences of predicted
deflections within the model ensemble are smaller than the deflections
themselves. Therefore, as shown in the right panel
of \cref{fig:defl3}, the required minimum rigidity is lower when the
deflections are corrected for.  With corrections, the rigidity
quantile at which half of the sky can be observed at
$\theta_\text{max}=20^\circ$ or better, decreases to
$\R_{50}^\text{corr} = 11$~EV giving a much greater observational
reach.  Note that this discussion is indicative only, since the
minimal rigidity requirement may change when random fields are
included in the analysis.

\subsection{Axions}
Another important application of the model ensemble presented in this
paper is the prediction of the conversion of astrophysical axion-like
particles \citep[e.g.][]{2010ARNPS..60..405J} to photons in Galactic
magnetic field.  The general expression for axion-photon conversion in
a plasma was derived by \citet{RS88}.  In the limit of small
conversion probability, applicable due to known constraints, we can
use the expression given in Eq.~(S5) of \citep{Safdi+Axion20} for the
axion-to-photon conversion probability between the source, located at
${r}_\text{src}$, and the observer located at the origin,
\begin{multline}
p_{a\gamma} =
\frac{g^2_{a \gamma \gamma}}{4} \!\sum_{i=1,2} \bigg | \int_{ r_{\rm src}}^0 \dd r\, B_i(r)\, {e}^{i \Delta_a \,r - i \int_{ r_{\rm src}}^r dr' \Delta_{||}(r')} \bigg|^2\\
\approx 2.3{\times} 10^{-6} \left(\frac{g_{a \gamma \gamma}}{10^{-12}~\text{GeV}^{-1}}\displaystyle \sum_{i=1,2}\frac{ \int_{ r_{\rm src}}^0 \dd r B_i(r)}{\text{kpc}\,\upmu\text{G}}\right)^2.
\label{eq:convProb}
\end{multline}
The integral is along the line of sight from source to observer and
the subscripts $i=1,2$ refer to the transverse components of the
magnetic field relative to the line of sight. Current limits on the
axion-photon coupling strength are $g_{a \gamma \gamma} \lesssim {\rm
few}{\times} 10^{-12}~\text{GeV}^{-1}$ in the low axion-mass regime
($m_a \lesssim 10^{-6}$~eV)~\citep{ParticleDataGroup:2022pth}.  The
effective wave numbers entering the phase shift are $\Delta_a \equiv
-m_a^2/(2 E) $ and $\Delta_{||}(r') \equiv -\omega_{\rm p}(r)^2/(2
E)$, where the plasma frequency $\omega_{\rm p}
\approx 3.7 {\times} 10^{-12} (\nel/10^{-2} {\rm cm}^{-3})^{-1/2}$ eV.
Thus, in the limit of large axion energies and low axion mass, the
conversion probability simplifies to the last expression that depends
only on the coupling constant and the line-of-sight integral of the
traverse magnetic field as given in the second line
of \cref{eq:convProb}. We checked the validity range of this
approximation by comparing it to the conversion probabilities in
the \JFTwelve magnetic field calculated with the \texttt{gammaALPs}
package~\citep{2022icrc.confE.557M} and found good agreement for
$E_a\gtrsim 1~\text{TeV}$ and $m_a \lesssim 10^{-8}$~eV.

The effect of the new magnetic field models derived in this paper on
the predicted axion-photon conversion probabilities are shown in
Fig.~\ref{fig:paxgam}, where we assumed $g_{a \gamma \gamma} =
5{\times} 10^{-12}~\text{GeV}^{-1}$. These figures look qualitatively
very similar to the ones for the cosmic-ray
deflections, \cref{fig:defl}, since both the conversion probability
and the deflection depend on the perpendicular component of the
magnetic field. The reason for the asymmetry of the conversion
probability in longitude is again the twisted nature of the Galactic
halo field. For a given extragalactic axion source candidate, the
uncertainty of conversion probability can be estimated by finding the
maximum and minimum prediction in the model ensemble in the direction
of the source.

%% file: summary.tex
In this paper, we have developed new, improved models of the coherent
magnetic field of the Milky Way, with better functional descriptions
of the disk and poloidal fields of the Galaxy, as well as the first
step toward a unified model for the toroidal and poloidal magnetic
halo.  We improve on previous analyses by using more and better
variants of auxiliary models for the thermal and cosmic-ray electron
densities.  The parameters of these new models are constrained with
the latest \RM and \PI data, which we have subjected to detailed
scrutiny to resolve or reveal discrepancies.

Equipped with these improvements, and with a fast optimization
framework that made it computationally feasible to investigate many
model variations, we can infer the following insights about the
coherent magnetic field of the Milky Way:
\begin{itemize}[nosep,leftmargin=0em,labelwidth=*,align=left]
  \item The local pitch angle of the disk field is $(11\pm1)^\circ$.
  \item A magnetic field arranged as a grand-design spiral in the
  Galactic disk fits the data well; interestingly, the inferred field
  reversals occur at the position of tracers of the spiral arms of the
  matter density. However, a grand-design spiral is not needed: a model
  with only a coherent local spur describes the \RM and \PI data
  equally well.

  \item Within the 10\% statistical precision of the data, the
  magnitude of the toroidal halo field is the same below and above the
  disk.
  \item The vertical scale height of the inferred toroidal halo is
  anticorrelated to the assumed half-height of the cosmic-ray
  diffusion volume, with a lower 5-$\sigma$ limit of $\hDiff \geq
  2.9$~kpc.
  \item Compatibility of the magnetic field strengths inferred from
  the \RM and \PI data can be achieved either through a striated
  random field having about the same energy density as the coherent
  field, or through an anticorrelation between the thermal electrons and
  magnetic field strength with a coefficient of $-0.4$.
  \item The new data corroborates the existence of the poloidal halo
  field introduced in \citetalias{JF12coh} with high significance.
  The strength of the poloidal field in the inner Galaxy is currently
  not well constrained due to the need for masking local structures.
  \item A simple unified halo model, in which the toroidal field is
  generated by the shearing of the poloidal field due to the Galactic
  rotation, fits the data well with only 6 instead of the usual 10
  free halo parameters.
\end{itemize}

We selected an ensemble of eight model variations\footnote{A C++
implementation of these GMF models can be found
at \cite{zenodo_10627091}.} that encompass a large range of
assumptions that are compatible with the data, see
Table~\ref{tab:allmodels}. As a first application, we employed these
variations to study the deflections of ultrahigh-energy cosmic rays in
the GMF. We find that the deflections predicted by the
widely-used \citetalias{JF12coh} model are close to the ones from the
new model ensemble.  An important conclusion of our work is that the
UHECR deflection uncertainties derived from the model ensemble are
smaller than the deflection itself for most of the sky at ultrahigh
energies, and we can localize the regions of greatest uncertainty.

The comprehensive study presented here significantly improves our
knowledge of the global structure of the coherent magnetic field of
the Milky Way, and for the first time provides a range of model
possibilities to explore the impact of GMF uncertainties on
GMF-sensitive science. However, it is just a step in the journey to
achieving accurate knowledge of the magnetic field of the
Galaxy. Directions for future improvements of the GMF modeling have
been noted throughout this paper. These include a better understanding
of Galactic foregrounds, an investigation of a possible
position-dependence of the striation factor and a self-consistent
modeling of the diffusion of cosmic-ray electrons in the GMF.

%% file: ackn.tex
\section*{Acknowledgements}
We would like to thank Denis Allard, Rainer Beck, Katia Ferri{\`e}re,
Michael Kachelriess, Jens Kleimann, Alexander Korochkin and Darko
Veberi\v{c} for useful comments on this manuscript.  MU acknowledges
the support from the EU-funded Marie Curie Outgoing Fellowship, Grant
PIOF-GA-2013-624803 for part of this work and the hospitality of
CCPP/NYU. This research of GRF has been supported by NASA grant
NNX10AC96G and NSF grants NSF-PHY-1212538, NSF-PHY-1517319,
NSF-PHY-2013199, and by the Simons Foundation.

We acknowledge the use of the Legacy Archive for
Microwave Background Data Analysis (LAMBDA), part of the High Energy
Astrophysics Science Archive Center (HEASARC). HEASARC/LAMBDA is a
service of the Astrophysics Science Division at the NASA Goddard Space
Flight Center. This work has benefited from discussions during the
program "Towards a Comprehensive Model of the Galactic Magnetic Field"
at NORDITA in April 2023, which is partly supported by NordForsk and
Royal Astronomical Society.

\software
{The results of this paper have been derived using the software
  packages \HEALPix~\citep{2005ApJ...622..759G},
  \Offline~\citep{Argiro:2007qg}, and \ROOT~\citep{Brun:1997pa}.}

%% file: appendix.tex
\section{Comparison of Synchrotron Data Products}
\label{sec:synappendix}
Sky maps of the Stokes \Q and \U parameters of the polarized
synchrotron emission are shown in the sub-panels along the diagonal of
Figs.~\ref{fig:syndatadiff}. The values
are from the 9-year \WMAP ``base'' model~\citep{2013ApJS..208...20B},
the third \Planck data release (DR3.0)~\citep{2020A&A...641A...4P}, our
simple arithmetic average of \WMAP and \Planck, and the CosmoGlobe
results derived from a combined analysis of \WMAP and \Planck
data~\citep{2023arXiv230308095W}. The difference between each data set
is shown as sky maps in the off-diagonal panels. As can be seen, the
\WMAP and \Planck maps exhibit large differences in the Galactic plane,
that could be related to different levels of
temperature-to-polarization leakage in the two data
sets~\citep{Svalheim:2022mks}. This region of the sky is, however, not
relevant for our analysis as it is masked in fit. Moreover, there are
large-scale systematic differences between the \Q and \U parameters
derived by the two collaborations. Our arithmetic average is by
construction in between the two data products and the values from the
global reanalysis from CosmoGlobe show yet another large-scale
distribution of \Q and \U differences.
\def\figw{0.236}
\begin{figure*}[!ht]
  \centering
\includegraphics[width=\figw\linewidth]{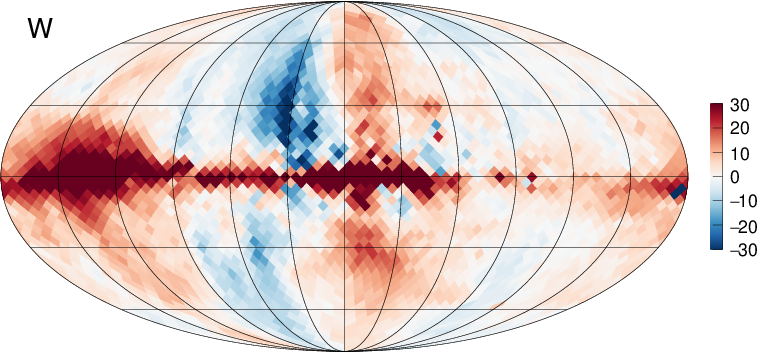}\quad%
\includegraphics[width=\figw\linewidth]{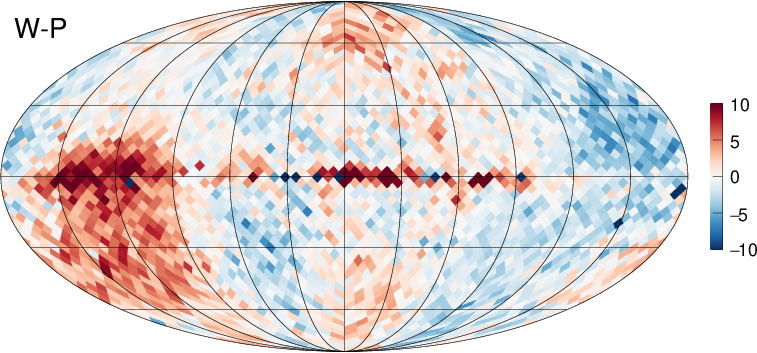}\quad%
\includegraphics[width=\figw\linewidth]{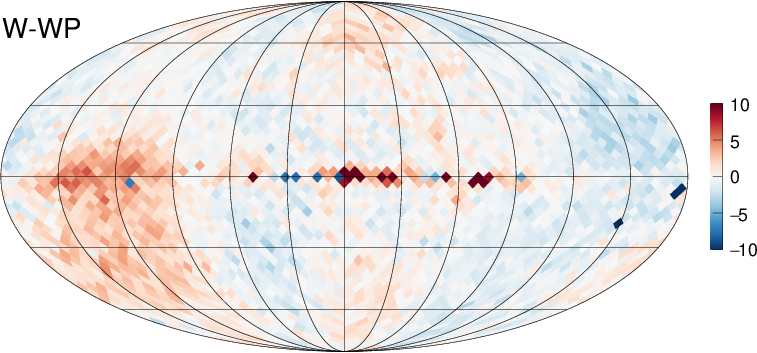}\quad%
\includegraphics[width=\figw\linewidth]{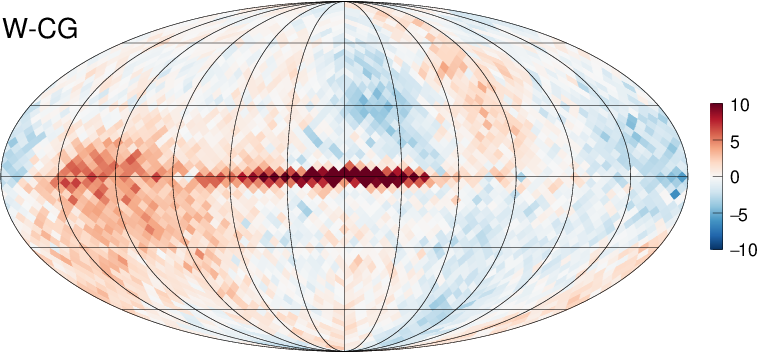}\\[\baselineskip]
\includegraphics[width=\figw\linewidth]{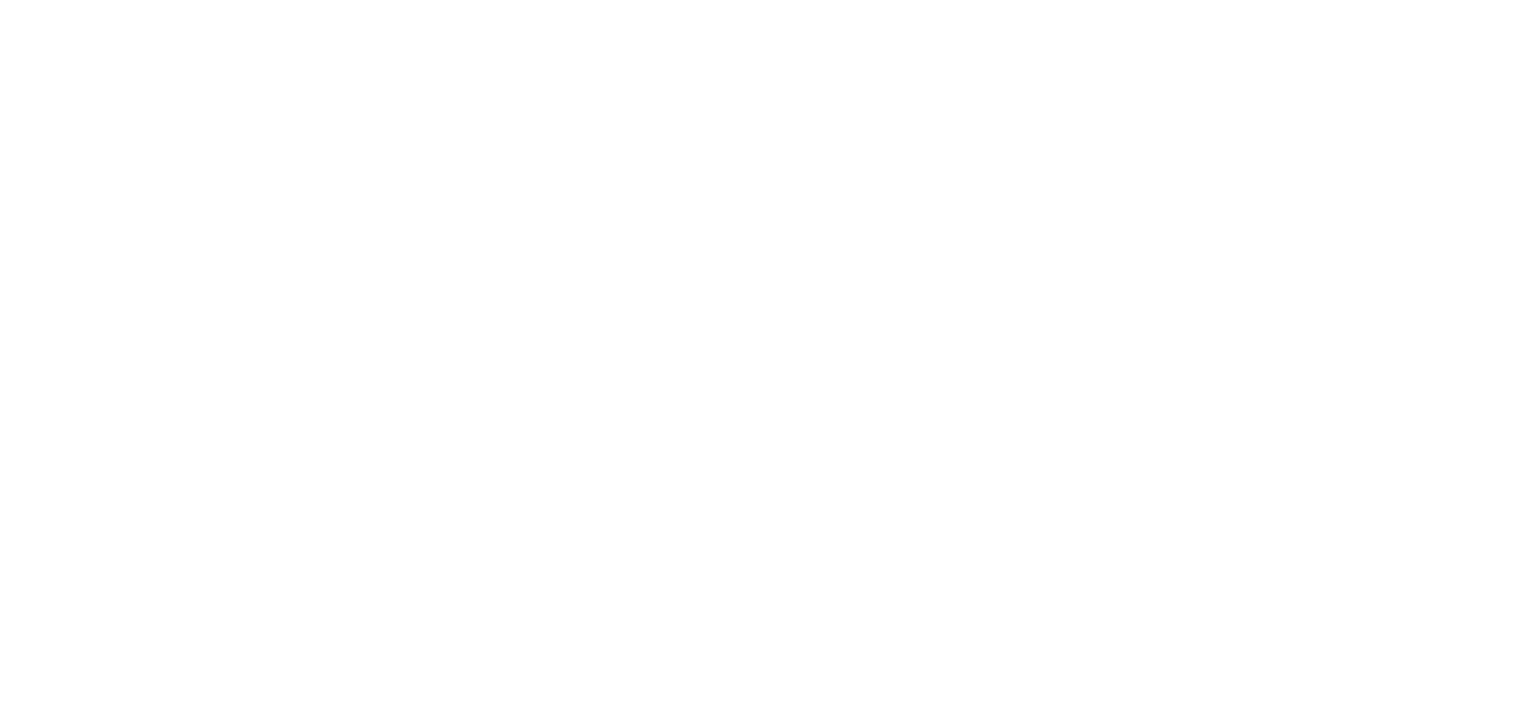}\quad%
\includegraphics[width=\figw\linewidth]{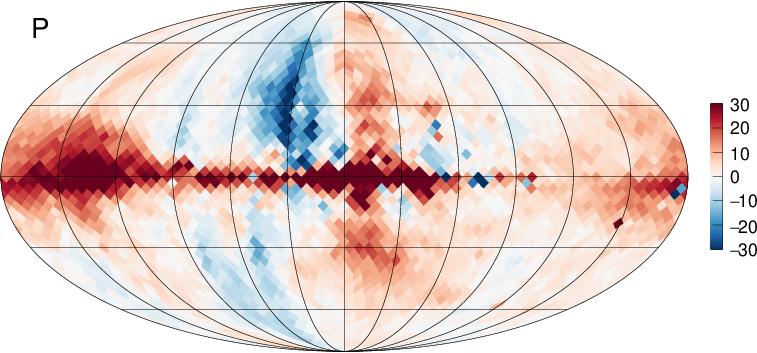}\quad%
\includegraphics[width=\figw\linewidth]{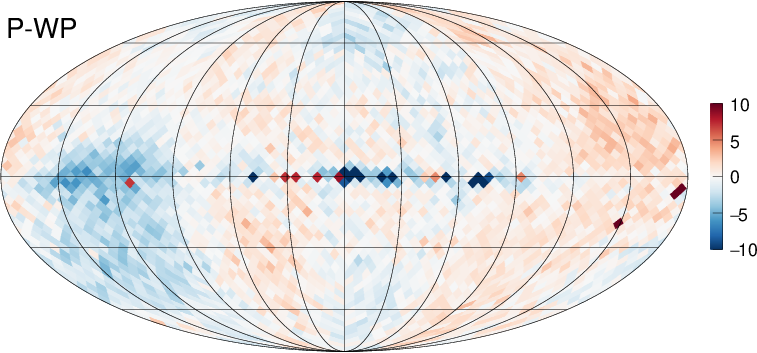}\quad%
\includegraphics[width=\figw\linewidth]{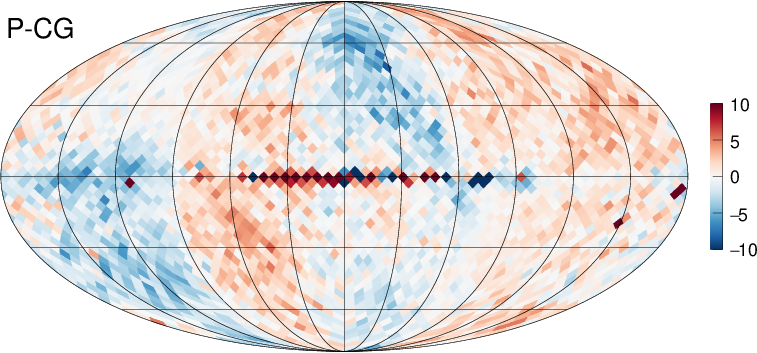}\\[\baselineskip]
\includegraphics[width=\figw\linewidth]{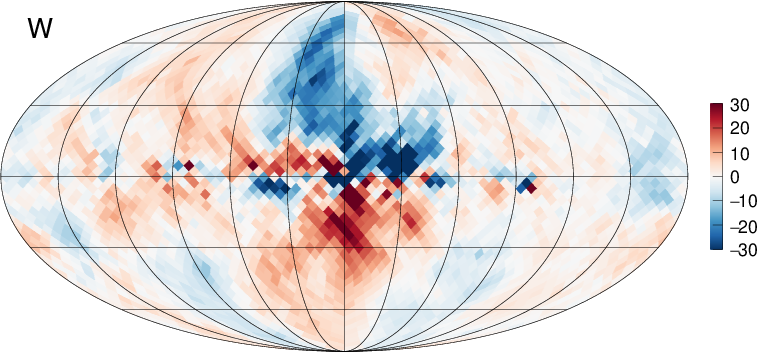}\quad%
\includegraphics[width=\figw\linewidth]{pics/matrix_empty.png}\quad%
\includegraphics[width=\figw\linewidth]{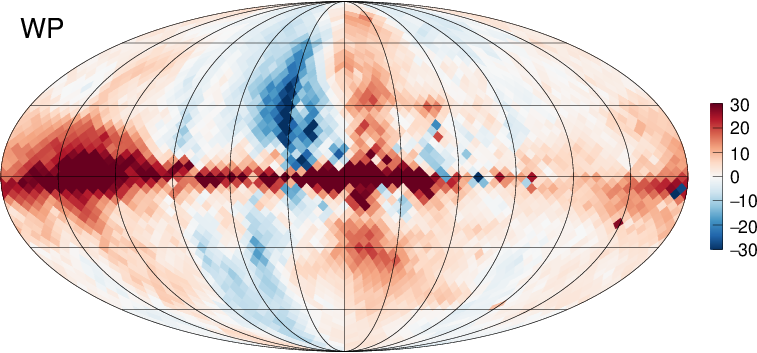}\quad%
\includegraphics[width=\figw\linewidth]{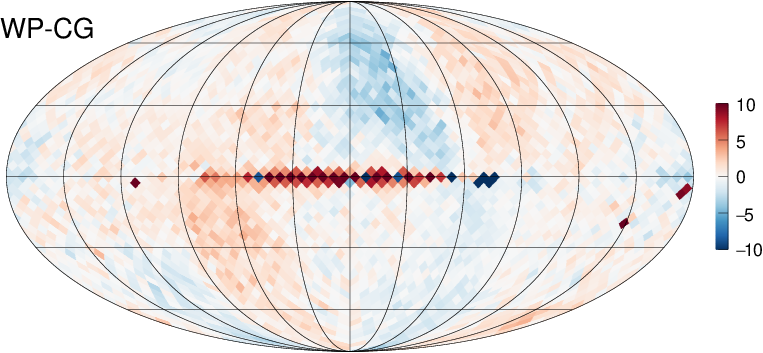}\\[\baselineskip]
\includegraphics[width=\figw\linewidth]{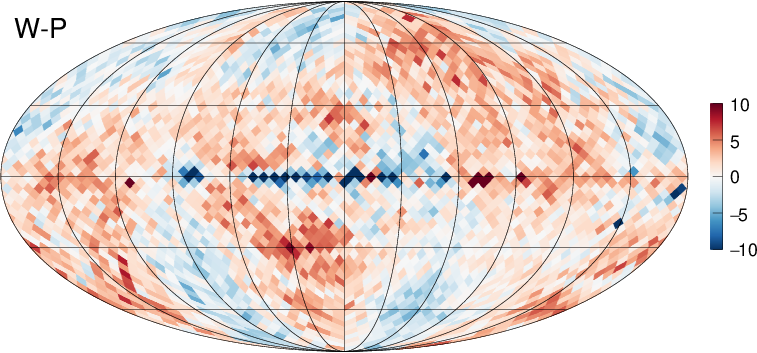}\quad%
\includegraphics[width=\figw\linewidth]{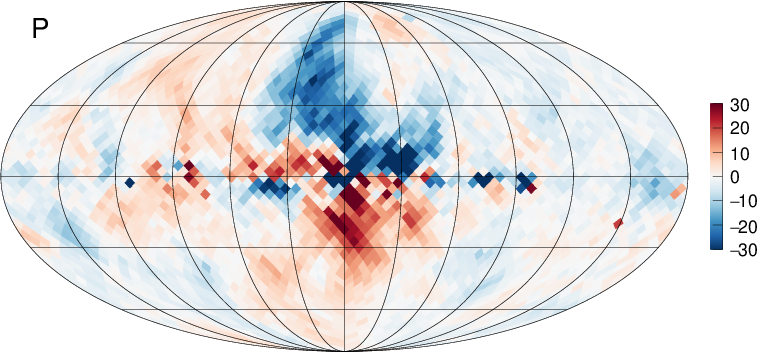}\quad%
\includegraphics[width=\figw\linewidth]{pics/matrix_empty.png}\quad%
\includegraphics[width=\figw\linewidth]{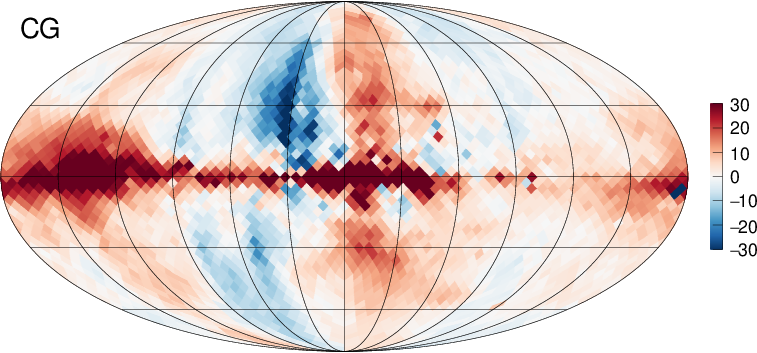}\\[\baselineskip]
\includegraphics[width=\figw\linewidth]{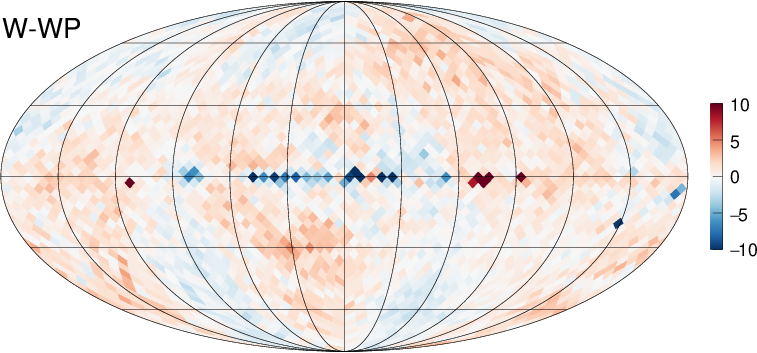}\quad%
\includegraphics[width=\figw\linewidth]{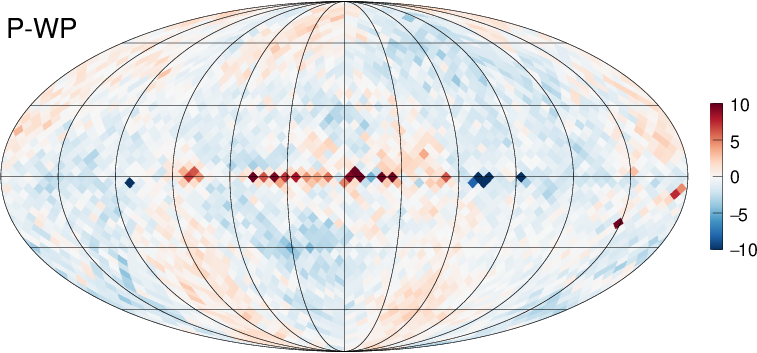}\quad%
\includegraphics[width=\figw\linewidth]{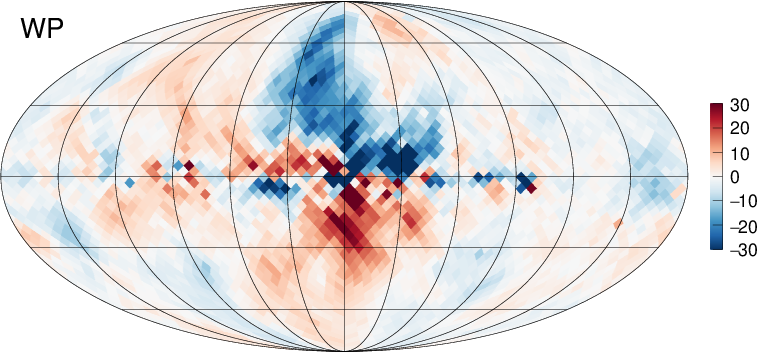}\quad%
\includegraphics[width=\figw\linewidth]{pics/matrix_empty.png}\\[\baselineskip]
\includegraphics[width=\figw\linewidth]{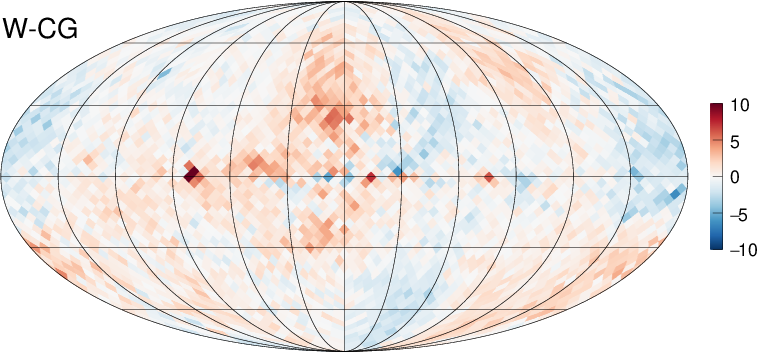}\quad%
\includegraphics[width=\figw\linewidth]{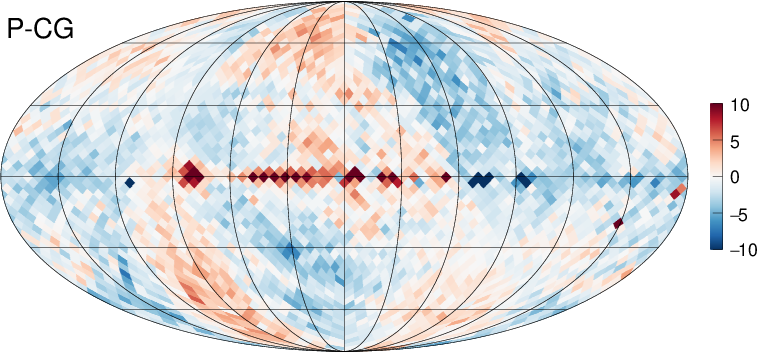}\quad%
\includegraphics[width=\figw\linewidth]{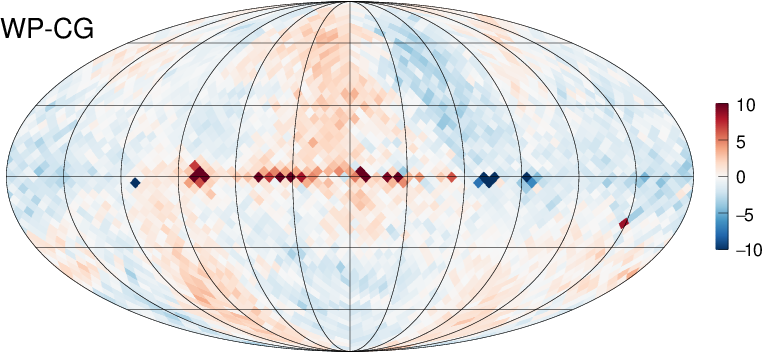}\quad%
\includegraphics[width=\figw\linewidth]{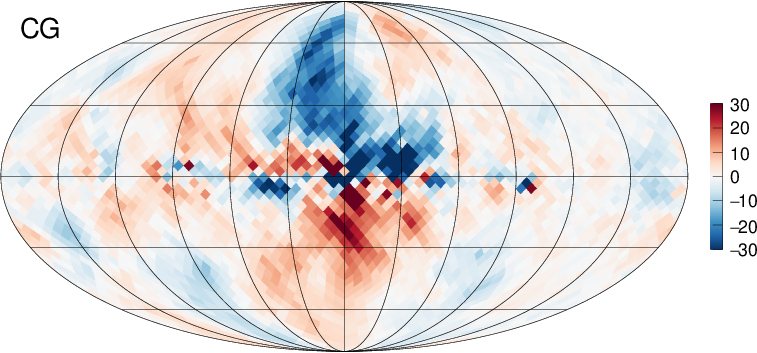}\\

\caption{Values (diagonal) and differences (off-diagonal) of Stokes
  parameters at 30~GHz in $\upmu$K of \WMAP (W), \Planck (P), our
  arithmetic average of \WMAP and \Planck (WP), and the \WMAP-\Planck
  combination of the CosmoGlobe collaboration (CG). Stokes \Q parameters
  are shown as an upper triangular matrix of plots at the top, Stokes
  \U parameters are displayed as a lower triangular matrix of plots at the bottom. Values are given in $\upmu$K.}
\label{fig:syndatadiff}
\end{figure*}

\begin{figure}[!t]
  \def\xlabel{0}
  \def\ylabel{48}
  \centering
  \begin{overpic}[width=0.49\linewidth]{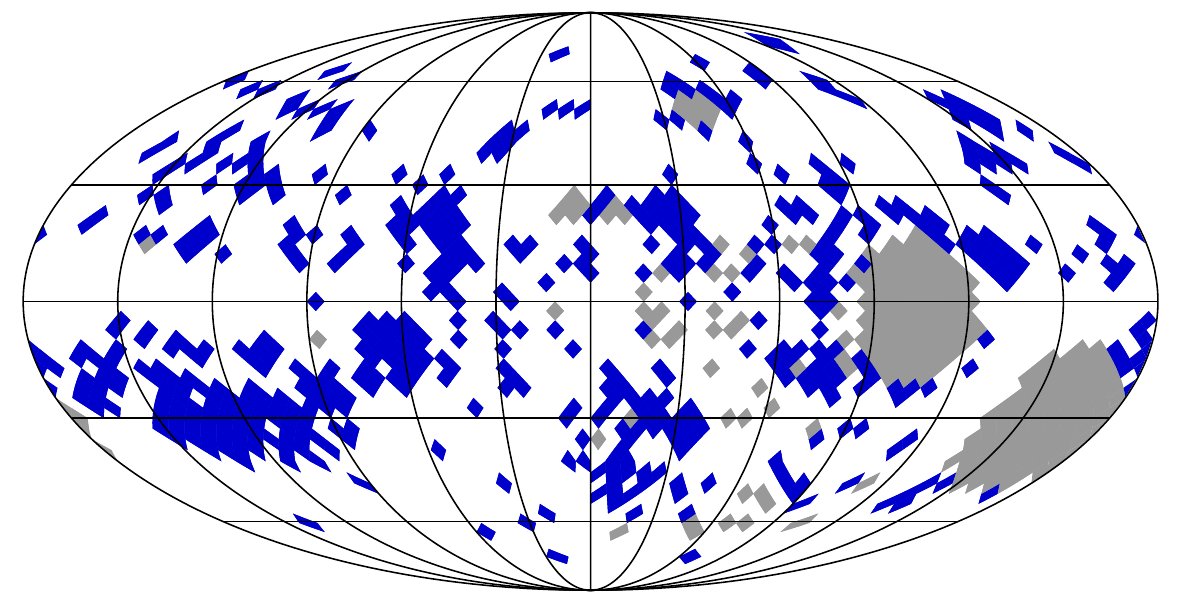}
  \put(\xlabel,\ylabel){(a) \RM mask}\end{overpic}%
  \begin{overpic}[width=0.49\linewidth]{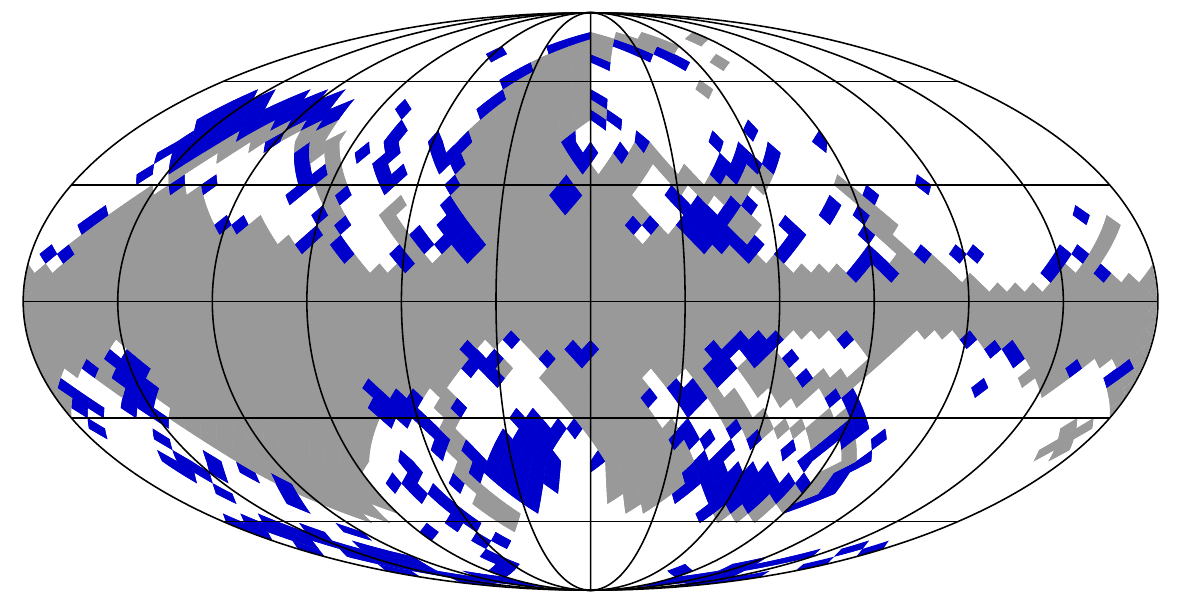}
    \put(\xlabel,\ylabel){(b) \Q and \U mask}\end{overpic}
  \caption{Pixel masks in Galactic coordinates. Left: \RM, right: polarized
    synchrotron intensity (Stokes \Q and \U parameters). The gray pixels
    are masked in the standard analysis (see Secs. \ref{sec:rmData} and \ref{sec:syndata}) and the additionally masked pixels based on the ``pull'' are shown in blue (see text). \label{fig:pullmask}}
\end{figure}
\begin{figure}[!h]
  \centering
  \begin{overpic}[clip,rviewport=0 0 1 1,
      width=0.5\linewidth]{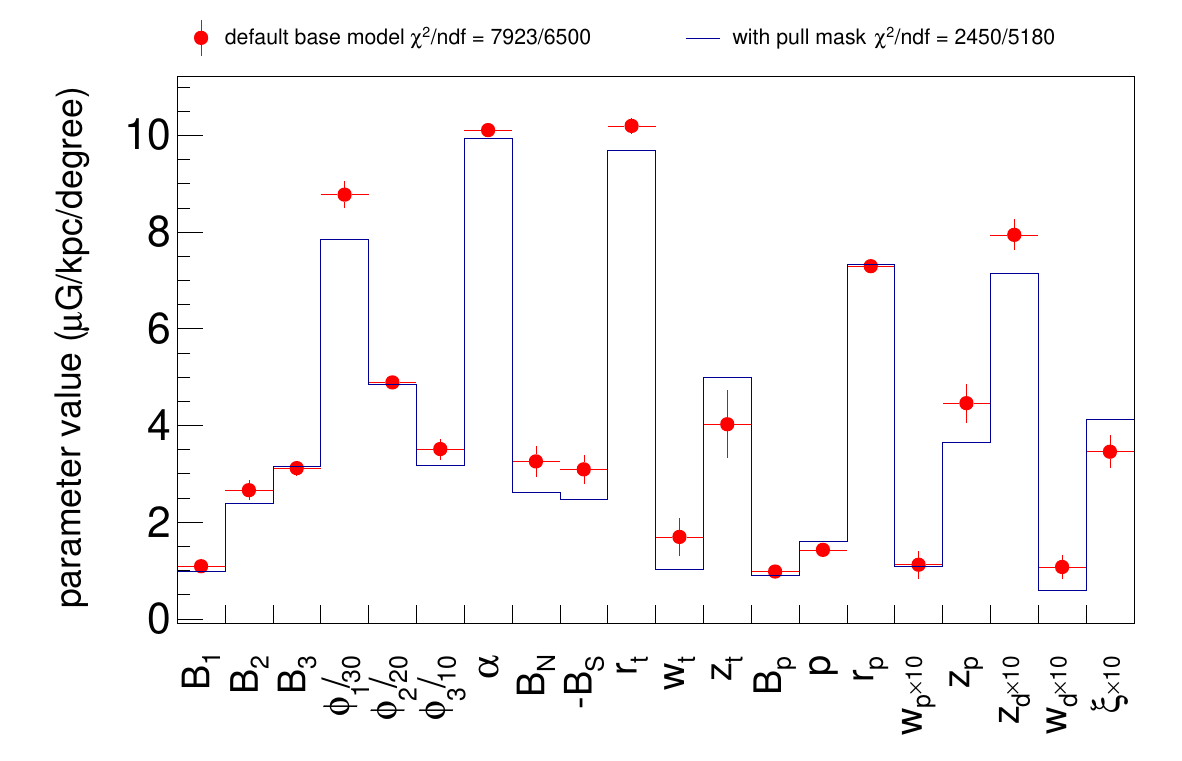}
   \end{overpic}
   \caption{Fit parameters of the \modelBase model obtained with the standard
      pixel mask (red points with uncertainties) and with the conservative
      ``pull'' mask shown in Fig.~\ref{fig:pullmask}\label{fit:pullpars}}
\end{figure}

\section{Foreground Contamination}
\label{ref:fg}
An important challenge for the inference of the global structure of
the large scale of the GMF is the contamination of the observables
with foreground structures~\citep[e.g.\ Sec.\ 5.2
  in][]{Jaffe:2019iuk}.  If these structures are sufficiently close to
the observer they can appear as features of large angular extent and
could either be mistaken for large-scale global features of the GMF
and/or bias the fitted parameters.

A foreground of particular importance are the known loops and spurs of
nearby supernova remnants. We performed a few initial tests with a
magnetic field compressed at the edge of a local foreground bubble
\citep[e.g.][]{1991ApJ...375..239F} using a simple spherical model
with magnetic flux conservation, but could not find a configuration
that described the data. More systematic studies are needed to
definitively exclude this possibility, e.g.\ including the known loops
into the polarized intensity. It should, however, be noted
that we have masked the edges of the classical loops I to IV from our
fit, see Sec.~\ref{sec:syndata}, excluding also the possible imprint
of local filamentary magnetic structures suggested
by~\citet{2021ApJ...923...58W}.  Furthermore, given the angular
positions of the loops, it is not obvious how they could conspire to
create a foreground that appears as a North-South-symmetric poloidal
field with $\Delta \BP = (0.12\pm 0.07)~\muG$, see Sec.~\ref{sec:poloidal}.

A similar argument can be brought forward regarding the possibility
that the butterfly pattern of the \RMs, \cref{eq:rmantisym}, is the
imprint of the local environment as e.g.\ suggested by the
identification of visually similar features identified in \RM sky maps
derived from simulated galaxies
\citep[e.g.][]{2018MNRAS.481.4410P,2023NatAs.tmp..176R}. Our finding
that for all model variations the magnitudes of the Northern and
Southern magnetic field strength of the toroidal component are
compatible within a few tenths of \muG (see Tab.~\ref{tab:modelpars})
is difficult to reconcile with an origin of ordered fields on scales
of 1-2 kpc.\\

For a data-driven test of the influence of unmodeled local structures
on the model parameters, we repeat the fit of the \modelBase model
excluding pixels with a large deviation of the data from the model.
We exclude pixels if the magnitude of the ``pull'' =
(data-model)/uncertainty) exceeds 1.5, see
Fig.~\ref{fig:basemodel}. The resulting masks for the rotation
measures and polarized intensity are displayed in
Fig.~\ref{fig:pullmask}, where the additionally masked pixels are
shown in blue. As can be seen, the new mask excludes the pixels below
the fan region with large \RM residuals centered at $(\ell,b)\sim
(120^\circ,-30^\circ)$ and pixels around the already masked Gum nebula
centered at $(\ell,b)\sim (105^\circ,-9^\circ)$. The additionally
masked pixels in polarized intensity are mainly at the edge of the
existing mask close to the North Polar Spur (or loop I) and at the
Northern edge of loop III centered at $(\ell,b)\sim
(124^\circ,15^\circ)$. Excluding these pixels removes 1320 data points
from the fit, i.e.\ a reduction of about 20\%. After re-optimizing the
parameters, we find a drastically changed fit quality which improves
from a reduced $\chi^2$ of 1.20 for the default fit to 0.47. Note that
a reduced $\chi^2$ as low as this value would usually be considered
suspicious (``too good'') and could indicate that the pull cut is too
tight removing too many pixels. On the other hand, it could result
from a small coherence length of the magnetic turbulence as indicated
by our analysis of simulated skymaps discussed in
Sec.~\ref{sec:modelopt}.

The parameters from the fit with the data-driven mask are compared to the default ones in
Fig.~\ref{fit:pullpars}. As can be seen, even after the removal of
20\% of the data points, the fit parameters are qualitatively
similar. The parameters of the poloidal field are impacted the least
by the change of the data set and also the field strengths and the
pitch angle of the disk field component change very little. The most
significant changes are observed for the toroidal field for which the
magnitude of the Northern and Southern field strengths are reduced by
about 0.6~\muG. We verified that this change is driven by the \RM
pixels below the fan region. Since similarly low values of the
toroidal field strengths were obtained for the \modelSyn model, these
changes can be considered to be within the overall modeling
uncertainties bracketed by the model ensemble presented in this paper.

These studies with a more restrictive mask make it plausible that our modeling is not
largely driven by local features. However, the changes of the
model parameters in Fig.~\ref{fit:pullpars} also demonstrate the gain in
precision that could be achieved after including a description of the local magnetized environment to the fit, in particular modeling the Local Bubble and loops along the lines of e.g.~\citep{2018A&A...611L...5A} and ~\citep{Mertsch:2013pua}.

\section{Parameter Correlations}
\label{sec:cova}
We estimate the covariance matrix $\bm{V}$ of the best-fit parameters $\hat{\bm{p}}$ from the second derivatives of the $\chi^2$ at the minimum,
\begin{equation}
  V_{ij}^{-1}(\hat{\bm{p}}) = \frac{1}{2} \left( \frac{\partial^2 \chi^2}{\partial p_i \partial p_j} \right)_{\bm{p} = \hat{\bm{p}}}
  \label{eq:cova}
  \end{equation}\citep[e.g.][]{Frodesen:1979fy}.
We evaluate \cref{eq:cova} using central
differences with one Richardson extrapolation
\citep[eg][]{2002nrca.book.....P} because for our 20-parameter fits {\scshape Minuit}'s
{\scshape Hesse} algorithm often resulted in numerically
unsatisfactory results. The step size for our numerical
differentiation is $\Delta x_i = \alpha\, \sigma_i$, where $\sigma_i$
denotes the uncertainty of parameter $i$ estimated with {\scshape
  Minos} (see Sec.~\ref{sec:modelopt}). The common scale parameter
$\alpha$ is decreased until the differences $||\bm{V}(\alpha_{k+1})-\bm{V}(\alpha_k)||$  between iterations $k$ increase, indicating the
onset of numerical noise, typically $\alpha = 0.05$.

\begin{figure}[!ht]
  \centering
  \includegraphics[width=0.5\linewidth]{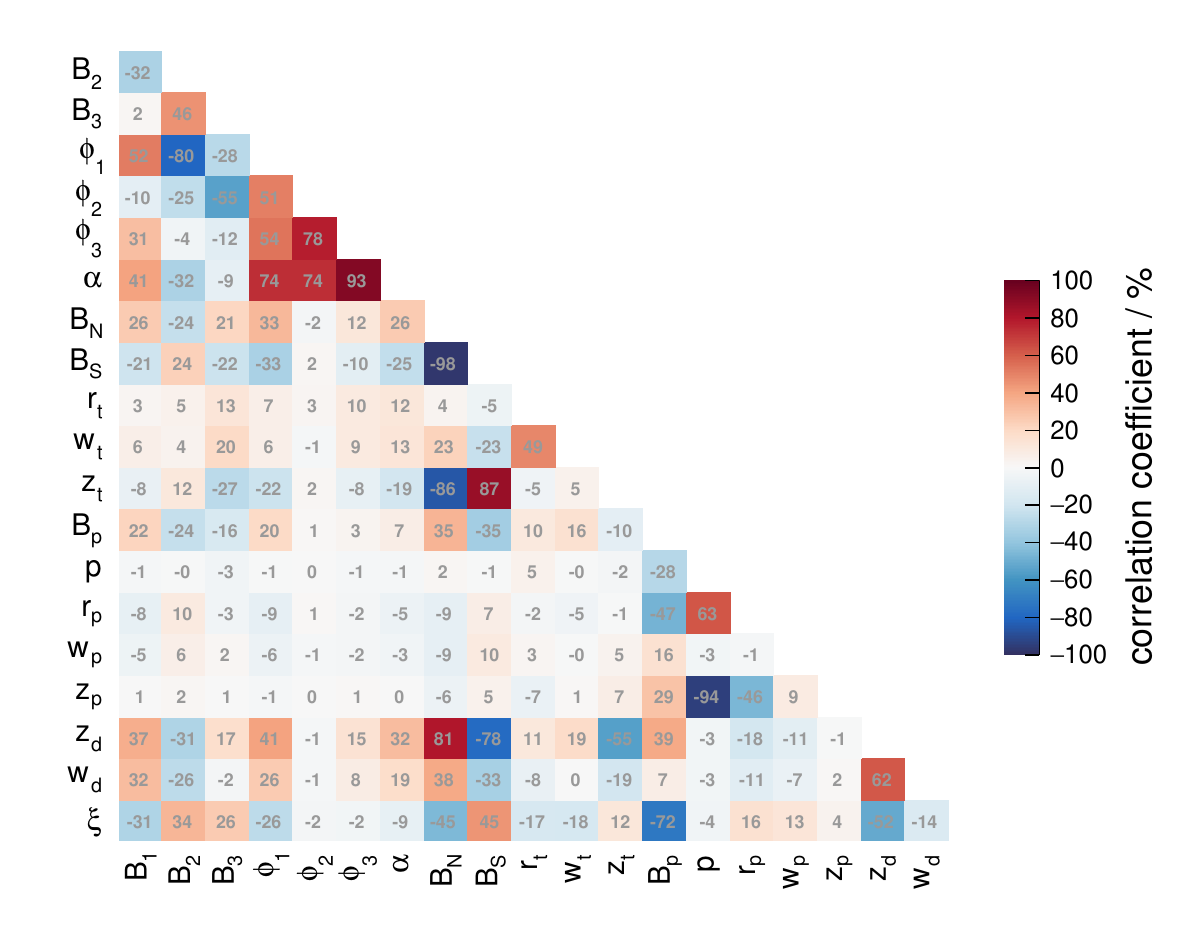}
  \caption{Visualization of the correlation matrix of the 20 parameters
    of the \modelBase model.}
\label{fig:baseCorr}
\end{figure}

\begin{figure*}[!ht]
  \def\figh{0.15}
  \begin{minipage}[b]{0.9\textwidth}
  \includegraphics[clip,rviewport=0 0 0.818 1,height=\figh\textheight]{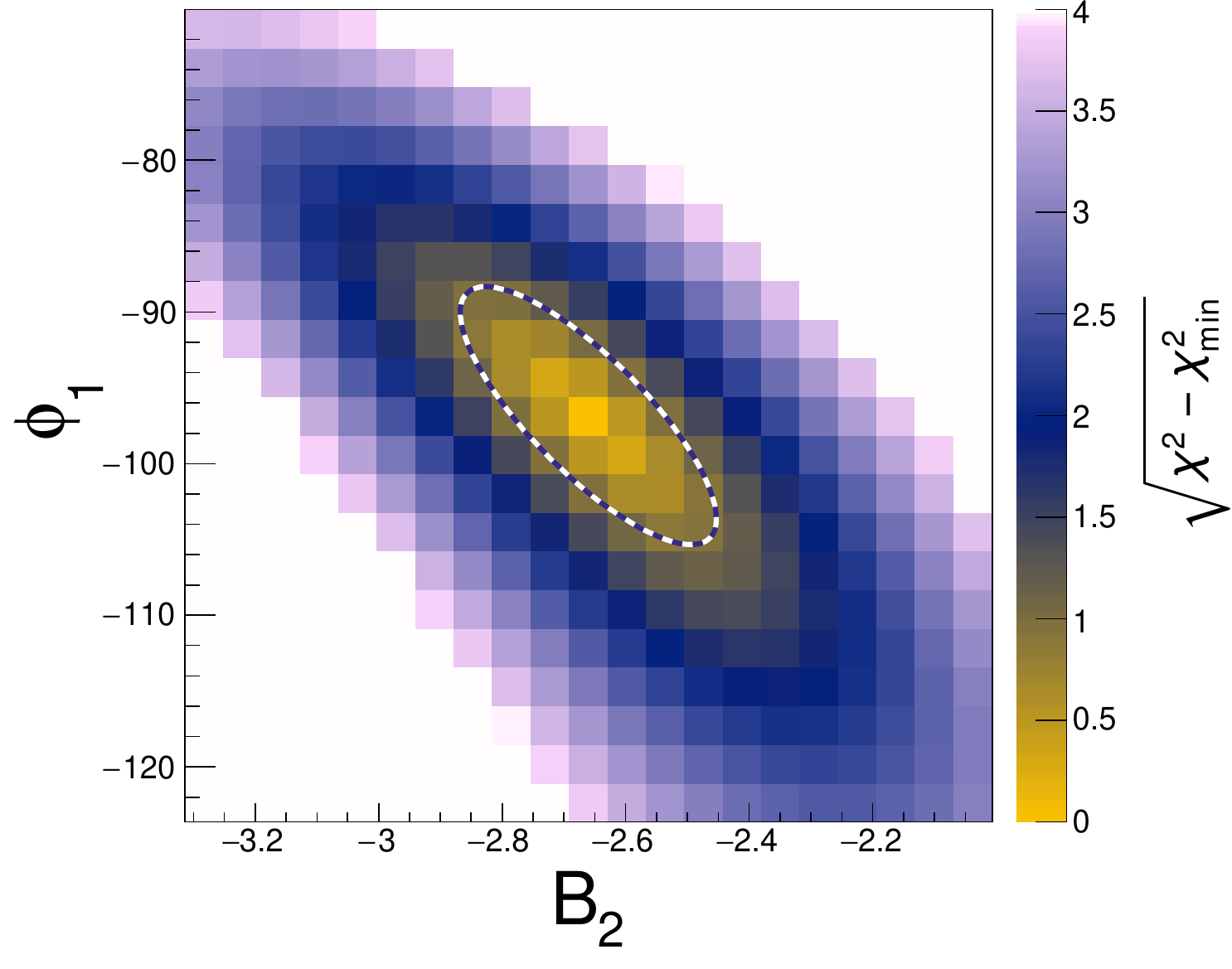}\quad\includegraphics[clip,rviewport=0 0 0.818 1,height=\figh\textheight]{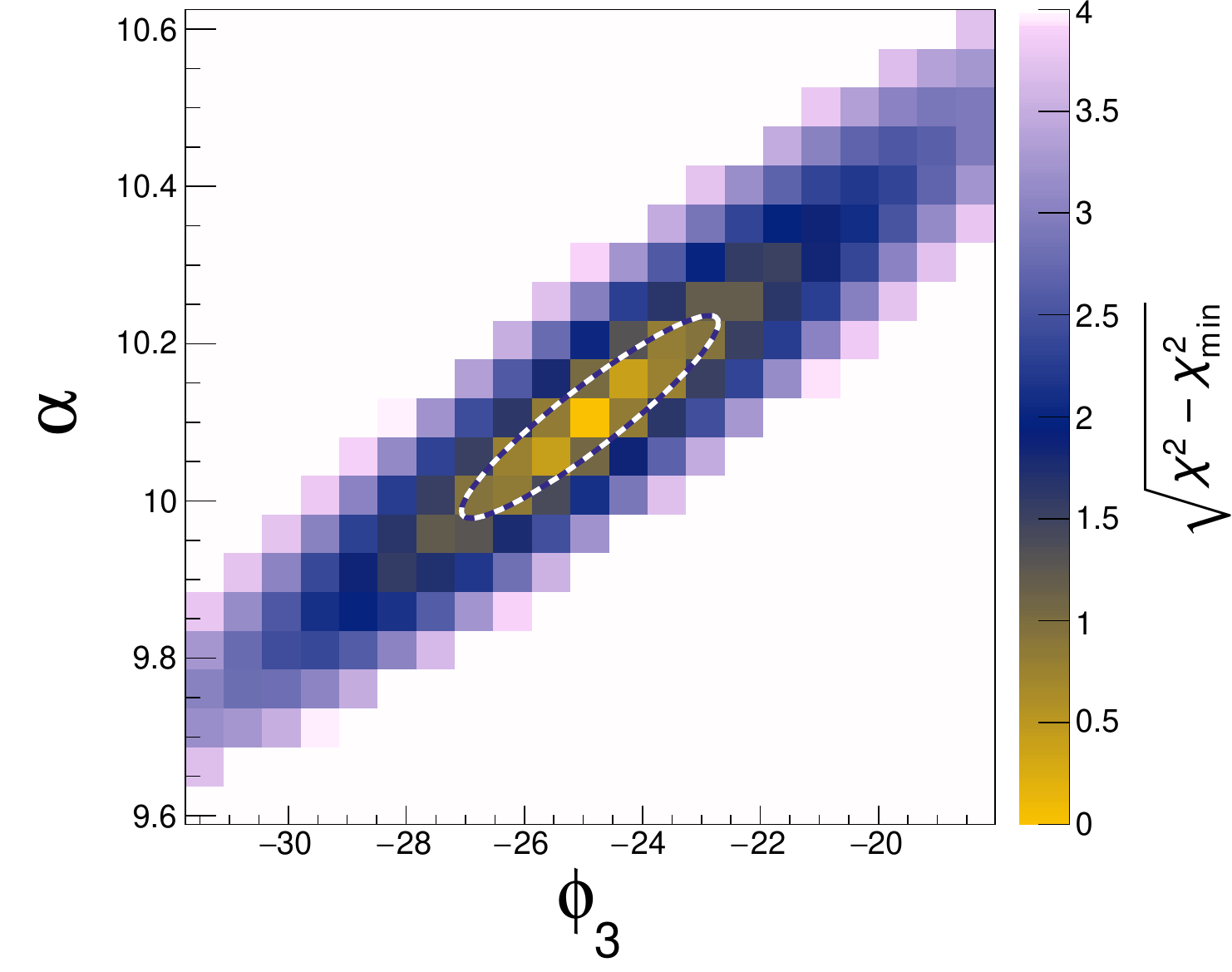}\quad\includegraphics[clip,rviewport=0 0 0.818 1,height=\figh\textheight]{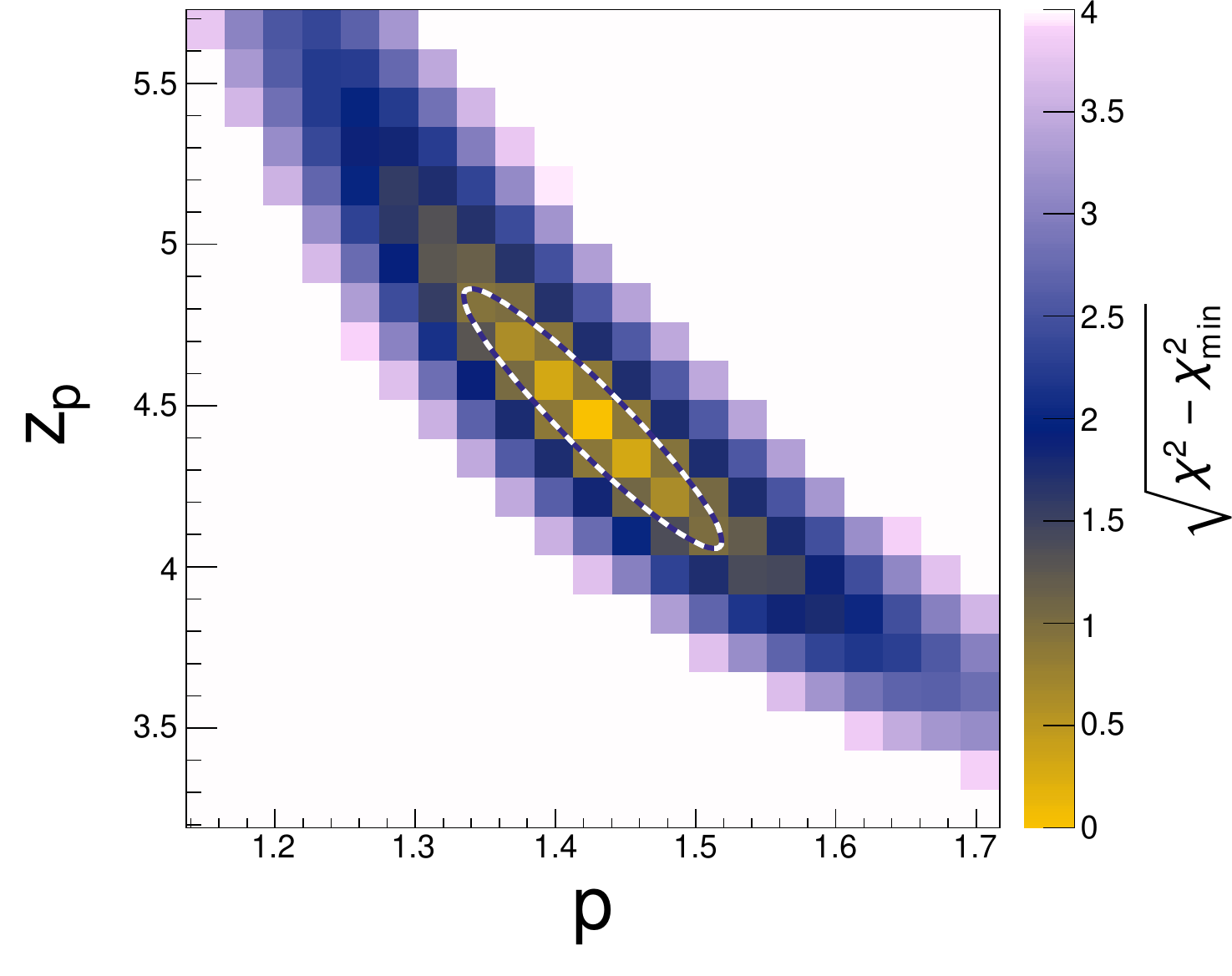}\quad\includegraphics[clip,rviewport=0 0 0.818 1,height=\figh\textheight]{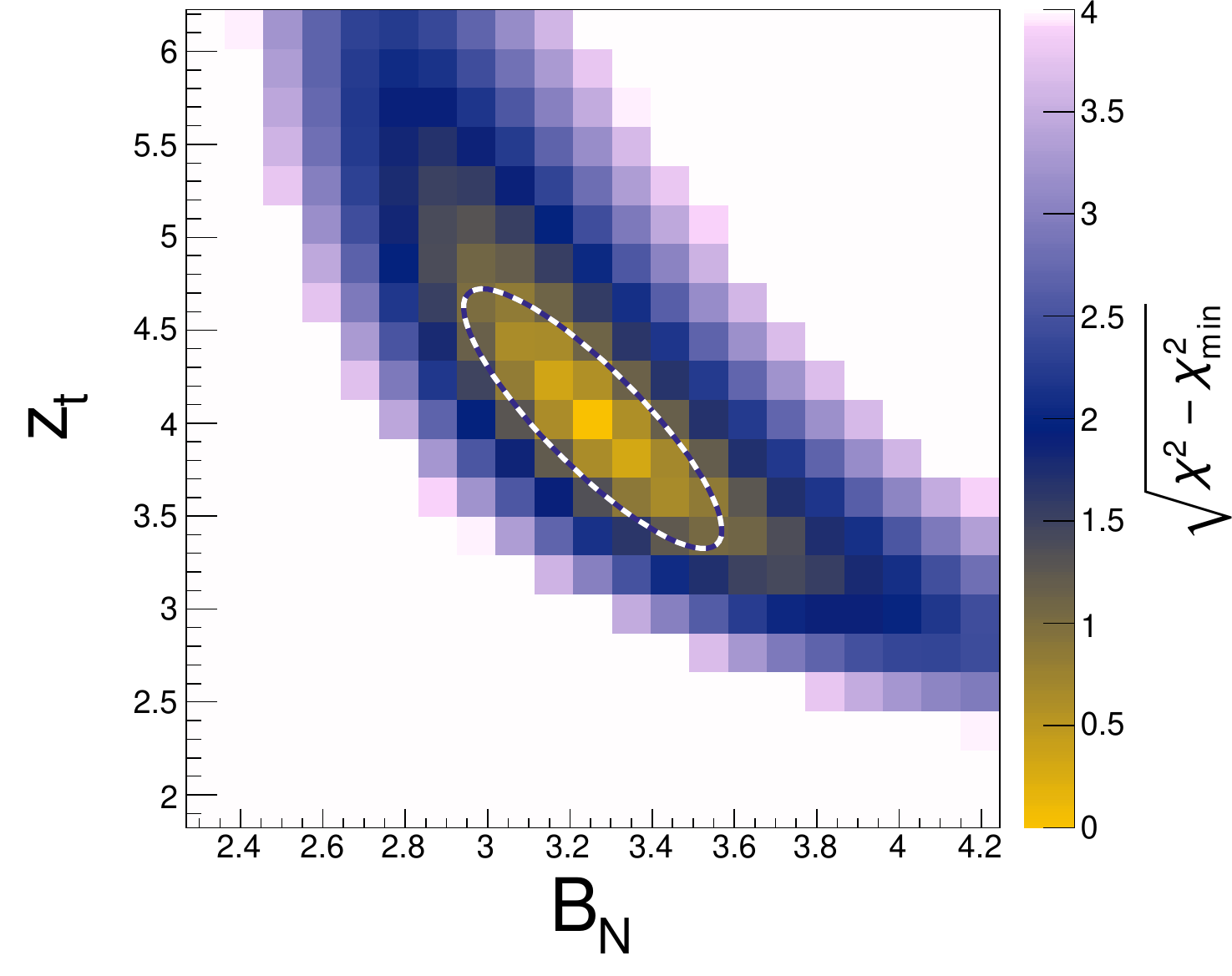}\\
  \includegraphics[clip,rviewport=0 0 0.818 1,height=\figh\textheight]{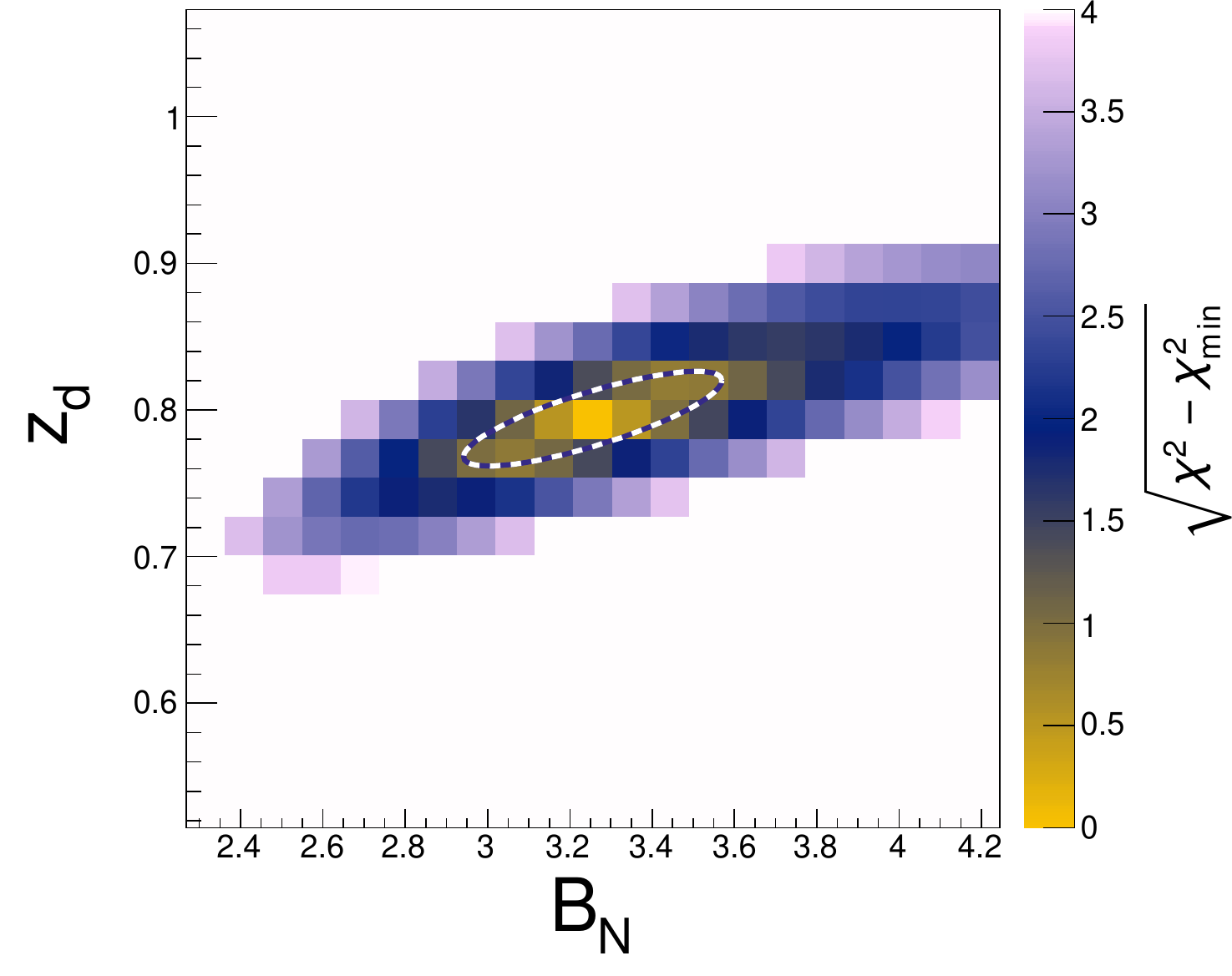}\quad\includegraphics[clip,rviewport=0 0 0.818 1,height=\figh\textheight]{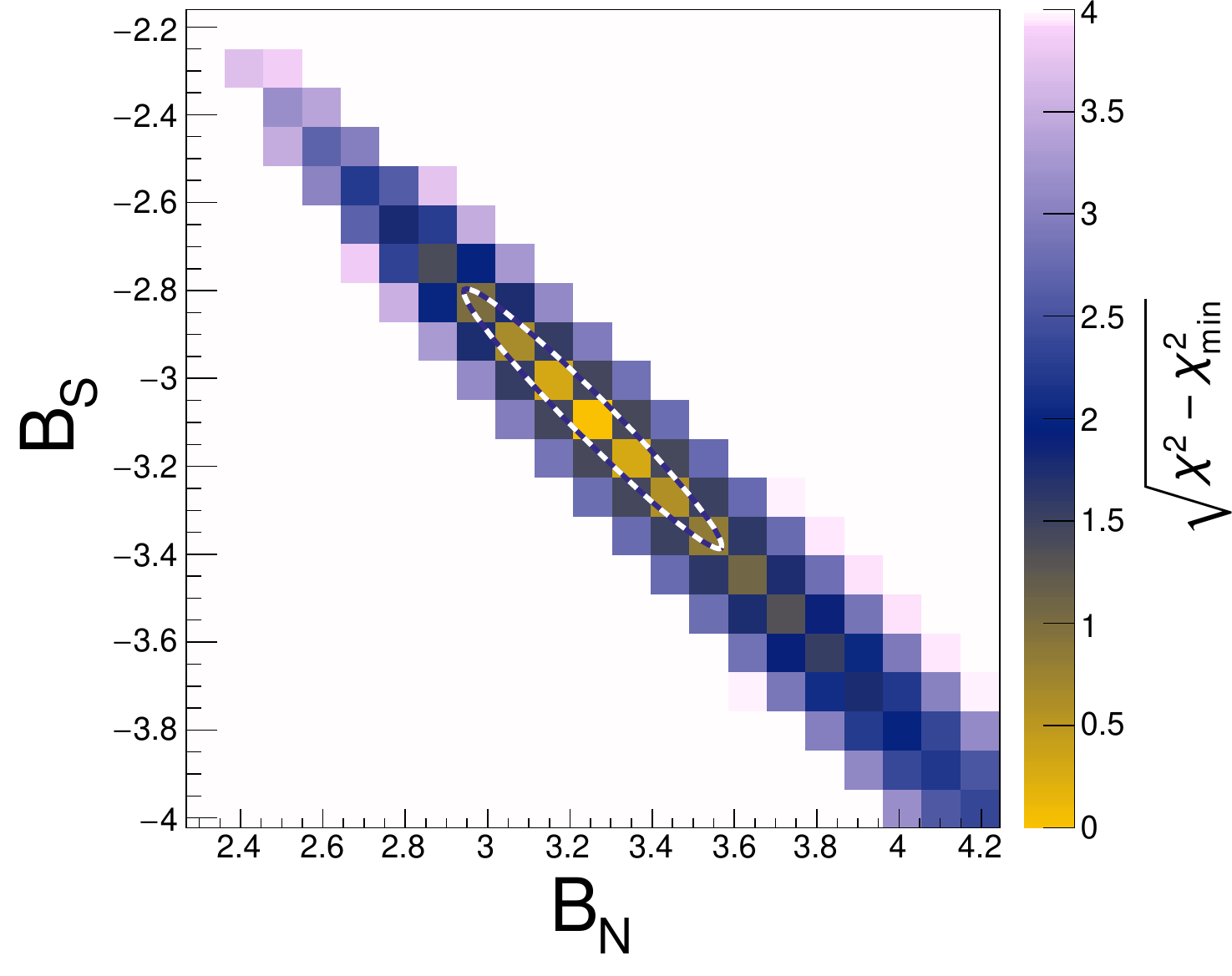}\quad\includegraphics[clip,rviewport=0 0 0.818 1,height=\figh\textheight]{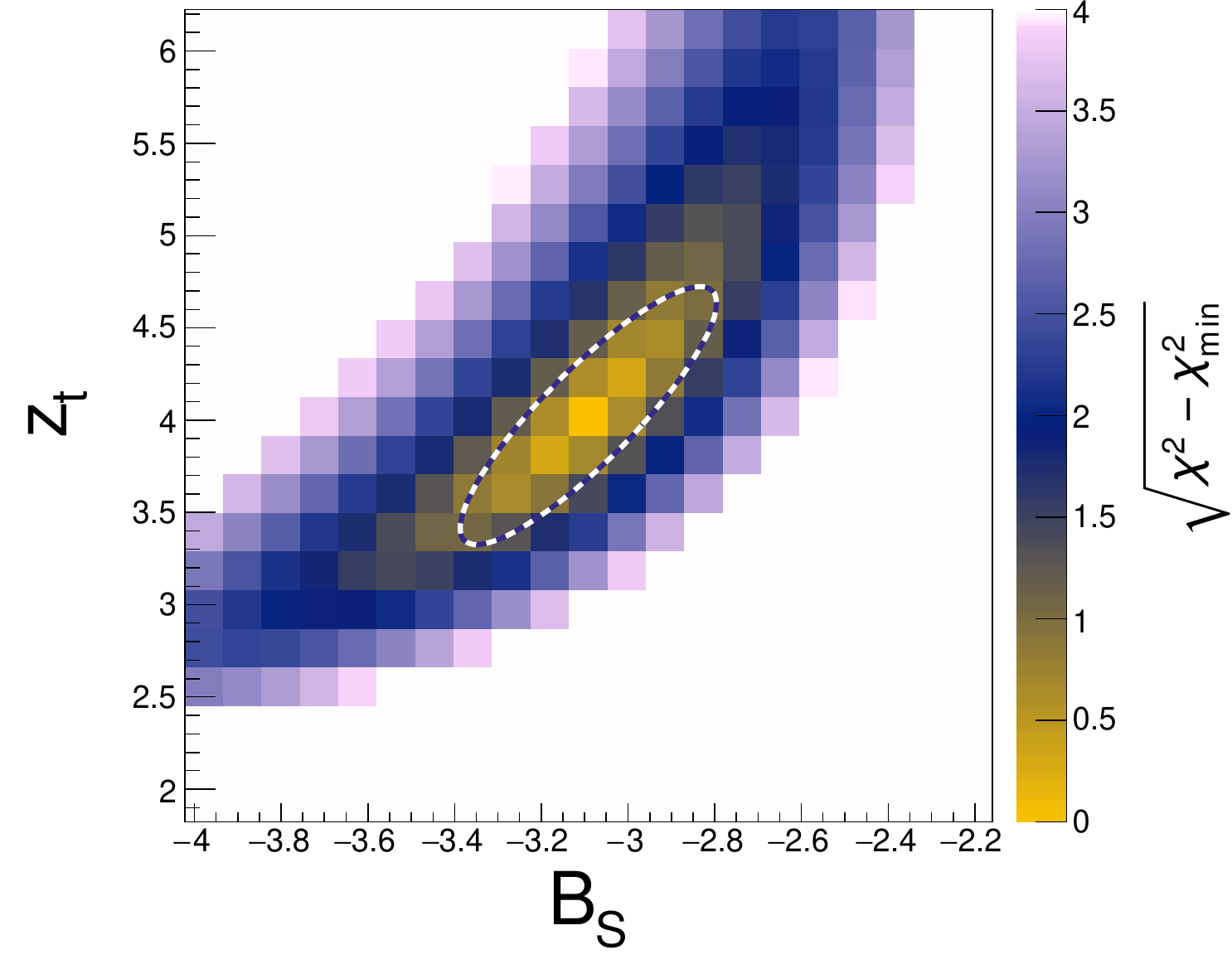}\quad\includegraphics[clip,rviewport=0 0 0.818 1,height=\figh\textheight]{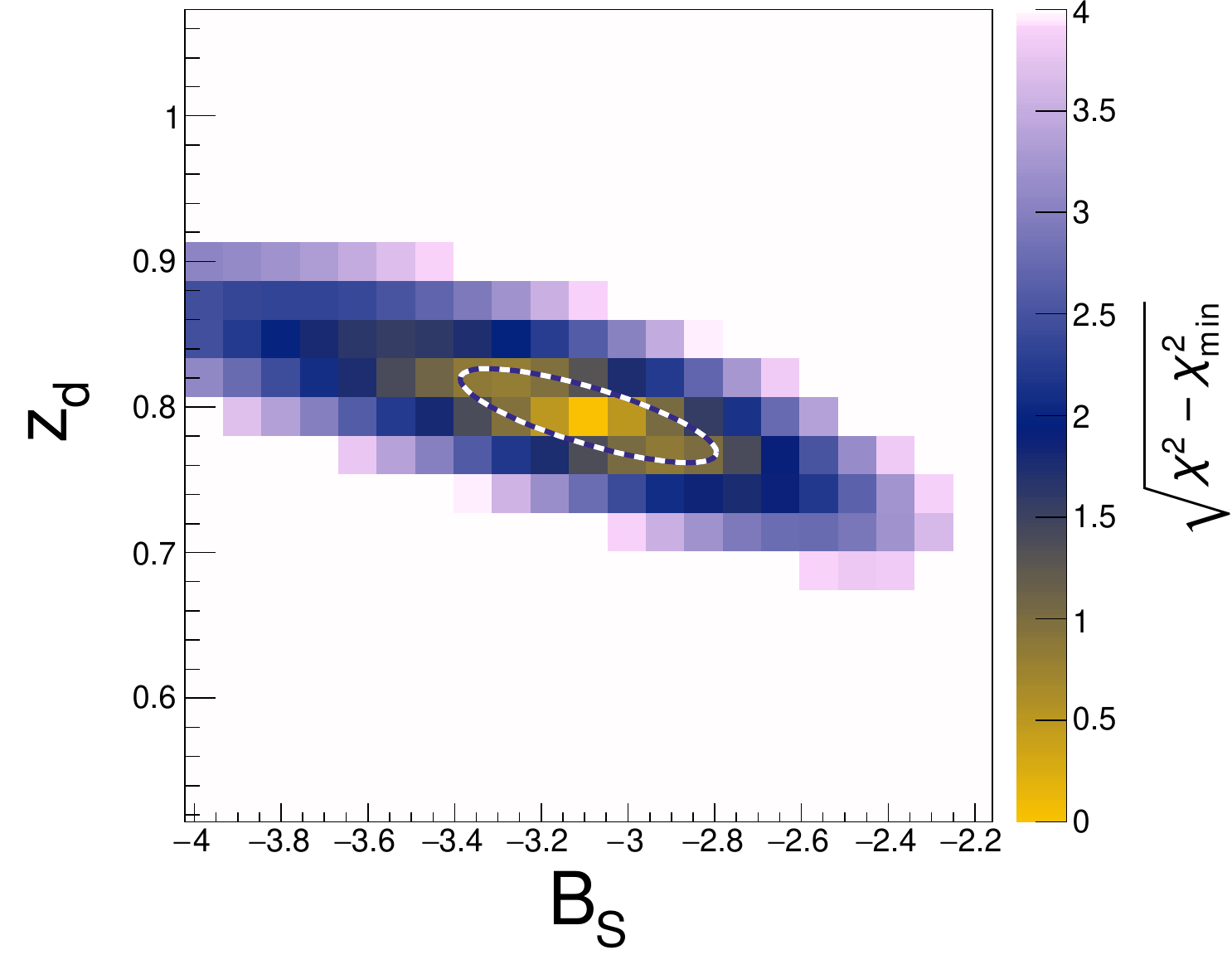}
  \end{minipage}
  \begin{minipage}[b]{0.09\textwidth}
\includegraphics[clip,rviewport=0.818 0.05 1 1,height=0.29\textheight]{pics/uf23_base_8_17_profLike_e.pdf}
  \end{minipage}
  \caption{Profile likelihood scan of the eight parameter pairs of the
    \modelBase model with the largest (anti-) correlation (units are
    $\upmu$G, degree and kpc as in Table~\ref{tab:glossary}); the
    color scale denotes the deviation of the fit quality from the
    optimum. The dashed ellipses show the approximate 1-$\sigma$ contour derived from the
    covariance matrix.}
 \label{fig:parCorr}
\end{figure*}

As an example, we show the correlation matrix ($\rho_{ij} = V_{ij} /
\sqrt{V_{ii} V_{jj}}$) of the parameters of the \modelBase model in
Fig.~\ref{fig:baseCorr}. As can be seen, most parameters have low
correlation coefficients indicating, e.g., a good factorization of the
shape of the toroidal GMF from the other components of the model.
However, some parameters are highly (anti-)correlated. The
two-dimensional profile likelihood contour of the eight parameter
combinations with the highest (anti-)correlation is shown in
Fig.~\ref{fig:parCorr}. Here we also superimposed the approximation
derived from \cref{eq:cova} demonstrating good agreement with
the numerical scan.

Particularly interesting is the large anti-correlation of the field
strengths of the toroidal halo, \BNT and \BST. This anti-correlation
is introduced by the common scale height \zT. We tested a fit of
separate scale heights in the North and South, but found only a minor
improvement in fit quality when leaving them separately free in the
fit ($\delta\chi^2=-22$).\\

In addition to being a valuable diagnostic for the correlation of fit
parameters, the covariance matrix is essential to propagate the
uncertainties of the parameters. For numerical applications,
the simplest method for the error propagation is to draw samples
$\tilde{\bm{p}}$ of the parameter vector $\bm{p}$ that are distributed
according to the covariance matrix. This can for instance be achieved
via
  \begin{equation}
    \tilde{\bm{p}} = \hat{\bm{p}} + \bm{L}\bm{n},
    \label{eq:chol}
  \end{equation}
  where $\bm{L}$ is the lower triangular matrix of the Cholesky
  decomposition of the covariance matrix, $\bm{V} =
  \bm{L}\bm{L}^\text{T}$, and $\bm{n}$ denotes a vector of
  standard-normally distributed random numbers.

\begin{figure*}[!ht]
  \centering
  \includegraphics[width=0.8\linewidth]{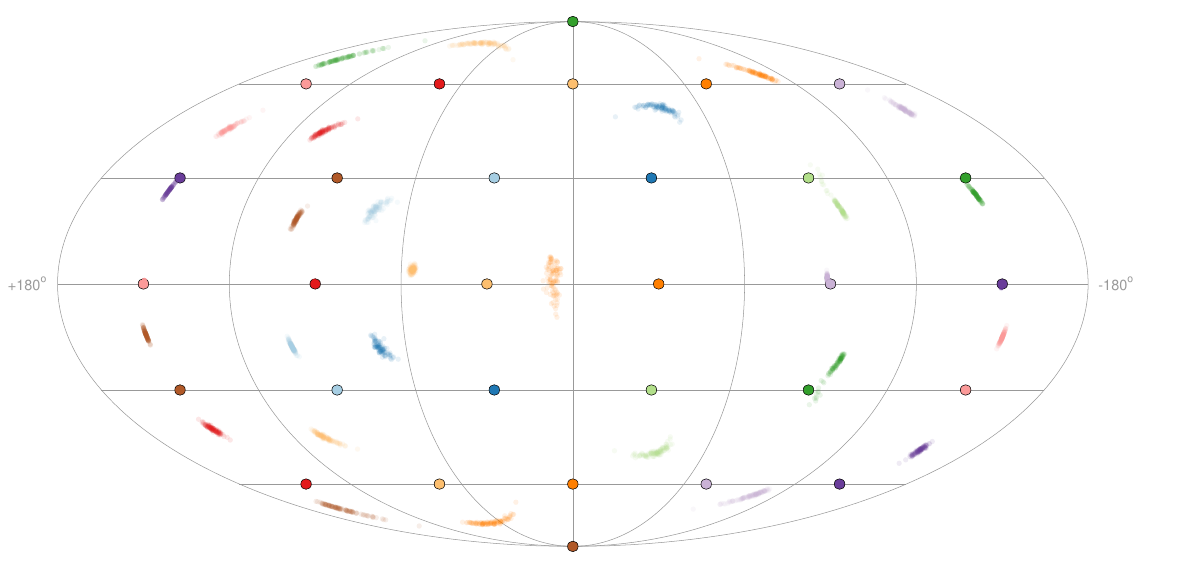}\caption{Propagation
    of parameter uncertainties to deflection uncertainties. Large
    filled circles denote a grid of arrival directions at Earth. Small
    transparent points of the same color show the backtracked
    directions at the edge of the Galaxy using the \modelBase model
    and a rigidity of $\R=20$~EV. For every arrival direction, 100
    sets of model parameters are drawn
    randomly from the covariance matrix and a cosmic ray is backtracked in the corresponding field.}
 \label{fig:errProp}
\end{figure*}

  How these samples can be used for error propagation is illustrated in
  Fig.~\ref{fig:errProp}. Here we show the backtracked directions of
  particles starting on a grid of arrival directions at Earth similar
  to Fig.~\ref{fig:defl2}. For each grid point, we backtracked
  100 particles, each through a different version of the \modelBase
  model using a different draw of the parameter vector. The point density
  is a direct measure of the probability of a particular direction given
  the covariance of the fitted parameters.  For most of the grid points, the standard deviation of points is
  smaller than the difference between models in Fig.~\ref{fig:defl2},
  showing that the systematic differences between models dominate over the
  parameter uncertainties. The sampling of the parameters from the
  covariance matrix can nevertheless be useful, for instance, if
  instead of a discrete 8-fold uncertainty mapping a smoother
  distribution is needed, which may be obtained by sampling the
  parameters of all eight model variants. The numerical values of the
  elements of $V$ for the eight GMF models as well as an example for
  the sampling according to \cref{eq:chol} are available in
  \cite{zenodo_10627091}.

